\documentclass[
 superscriptaddress,
 reprint, 
 amsmath,amssymb,
 prc,aps,nofootinbib
]{revtex4-2}
\usepackage[utf8]{inputenc}
\usepackage{graphics}
\usepackage{graphicx}% Include figure files
\usepackage{dcolumn}% Align table columns on decimal point
\usepackage{bm}% bold math
\usepackage{braket}
\usepackage{textcomp}
\usepackage{natbib}
\usepackage[dvipsnames]{xcolor}
\usepackage[colorlinks]{hyperref} 
\hypersetup{
      colorlinks=true,
      linkcolor=blue,
      filecolor=blue,
      citecolor = teal,      
      urlcolor=blue,
      }
\begin{document}
\title{Model-independent measurement of isospin diffusion in Ni-Ni systems at intermediate~energy}

\author{C.~Ciampi}
 \email{caterina.ciampi@ganil.fr}
 \affiliation{Grand Accélérateur National d'Ions Lourds (GANIL), CEA/DRF–CNRS/IN2P3, Boulevard Henri Becquerel, F-14076 Caen, France}
\author{J.~D.~Frankland}
 \affiliation{Grand Accélérateur National d'Ions Lourds (GANIL), CEA/DRF–CNRS/IN2P3, Boulevard Henri Becquerel, F-14076 Caen, France} 
\author{D.~Gruyer}
\affiliation{Université de Caen Normandie, ENSICAEN, CNRS/IN2P3, LPC Caen UMR6534, F-14000 Caen, France}
\affiliation{Grand Accélérateur National d'Ions Lourds (GANIL), CEA/DRF–CNRS/IN2P3, Boulevard Henri Becquerel, F-14076 Caen, France}
\author{N.~Le~Neindre}
 \affiliation{Université de Caen Normandie, ENSICAEN, CNRS/IN2P3, LPC Caen UMR6534, F-14000 Caen, France}
\author{S.~Mallik}
 \affiliation{Physics Group, Variable Energy Cyclotron Centre, 1/AF Bidhan Nagar, Kolkata 700064, India}
\author{R.~Bougault}
 \affiliation{Université de Caen Normandie, ENSICAEN, CNRS/IN2P3, LPC Caen UMR6534, F-14000 Caen, France}
\author{A.~Chbihi}
\affiliation{Grand Accélérateur National d'Ions Lourds (GANIL), CEA/DRF–CNRS/IN2P3, Boulevard Henri Becquerel, F-14076 Caen, France}
\author{L.~Baldesi}
\affiliation{INFN - Sezione di Firenze, 50019 Sesto Fiorentino, Italy}
\affiliation{Dipartimento di Fisica e Astronomia, Universit\`{a} di Firenze, 50019 Sesto Fiorentino, Italy}
\author{S.~Barlini}
\affiliation{INFN - Sezione di Firenze, 50019 Sesto Fiorentino, Italy}
\affiliation{Dipartimento di Fisica e Astronomia, Universit\`{a} di Firenze, 50019 Sesto Fiorentino, Italy}
\author{E.~Bonnet}
\affiliation{SUBATECH UMR 6457, IMT Atlantique, Université de Nantes, CNRS-IN2P3, F-44300 Nantes, France}
\author{B.~Borderie}
\affiliation{Université Paris-Saclay, CNRS/IN2P3, IJCLab, 91405 Orsay, France}
\author{A.~Camaiani}
 \affiliation{INFN - Sezione di Firenze, 50019 Sesto Fiorentino, Italy}
 \affiliation{Dipartimento di Fisica e Astronomia, Universit\`{a} di Firenze, 50019 Sesto Fiorentino, Italy}
\author{G.~Casini}
 \affiliation{INFN - Sezione di Firenze, 50019 Sesto Fiorentino, Italy}
\author{I.~Dekhissi}
 \affiliation{Université de Caen Normandie, ENSICAEN, CNRS/IN2P3, LPC Caen UMR6534, F-14000 Caen, France}
 \affiliation{Grand Accélérateur National d'Ions Lourds (GANIL), CEA/DRF–CNRS/IN2P3, Boulevard Henri Becquerel, F-14076 Caen, France}
\author{D.~Dell'Aquila}
\affiliation{Dipartimento di Fisica ``Ettore Pancini'', University of Napoli Federico II, 80126 Napoli, Italy}
\affiliation{INFN - Sezione di Napoli, 80126 Napoli, Italy}
\author{J.~A.~Dueñas}
\affiliation{Departamento de Ingeniería Eléctrica y Centro de Estudios Avanzados en Física, Matemáticas y Computación, Universidad de Huelva, 21007 Huelva, Spain}
\author{Q.~Fable}
\affiliation{Grand Accélérateur National d'Ions Lourds (GANIL), CEA/DRF–CNRS/IN2P3, Boulevard Henri Becquerel, F-14076 Caen, France}
\author{F.~Gramegna}
\affiliation{INFN - Laboratori Nazionali di Legnaro, 35020 Legnaro, Italy}
\author{C.~Gouyet}
\affiliation{Grand Accélérateur National d'Ions Lourds (GANIL), CEA/DRF–CNRS/IN2P3, Boulevard Henri Becquerel, F-14076 Caen, France}
\author{M.~Henri}
\affiliation{CEA/INSTN/UECC, 143 Chemin de la Crespinière – ZA Les Vindits, F-50130 Cherbourg en Cotentin, France}
\author{B.~Hong}
\affiliation{Center for Extreme Nuclear Matters (CENuM), Korea University, Seoul 02841, Republic of Korea}
\affiliation{Department of Physics, Korea University, Seoul 02841, Republic of Korea}
\author{S.~Kim}
 \affiliation{Center for Exotic Nuclear Studies, Institute for Basic Science, Daejeon 34126, Republic of Korea}
\author{A.~Kordyasz}
\affiliation{Heavy Ion Laboratory, University of Warsaw, 02-093 Warszawa, Poland}
\author{T.~Kozik}
\affiliation{Faculty of Physics, Astronomy and Applied Computer Science, Jagiellonian University, S. Łojasiewicza 11, 30-348 Krakow, Poland}
\author{M.~J.~Kweon}
\affiliation{Center for Extreme Nuclear Matters (CENuM), Korea University, Seoul 02841, Republic of Korea}
\affiliation{Department of Physics, Inha University, Incheon 22212, Republic of Korea}
\author{I.~Lombardo}
\affiliation{INFN - Sezione di Catania, 95123 Catania, Italy}
\affiliation{Dipartimento di Fisica e Astronomia, Universit\`a di Catania, via S. Sofia 64, 95123 Catania, Italy}
\author{O.~Lopez}
\affiliation{Université de Caen Normandie, ENSICAEN, CNRS/IN2P3, LPC Caen UMR6534, F-14000 Caen, France}
\author{L.~Manduci}
\affiliation{Ecole des Applications Militaires de l’Energie Atomique (EAMEA), BP 19, F-50115 Cherbourg Armées, France}
\affiliation{Université de Caen Normandie, ENSICAEN, CNRS/IN2P3, LPC Caen UMR6534, F-14000 Caen, France}
\author{T.~Marchi}
 \affiliation{INFN - Laboratori Nazionali di Legnaro, 35020 Legnaro, Italy}
\author{K.~Mazurek}
 \affiliation{Institute of Nuclear Physics Polish Academy of Sciences, PL-31342 Krakow, Poland}
\author{S.~H.~Nam}
\affiliation{Center for Extreme Nuclear Matters (CENuM), Korea University, Seoul 02841, Republic of Korea}
\affiliation{Department of Physics, Korea University, Seoul 02841, Republic of Korea}
\author{J.~Park}
\affiliation{Center for Extreme Nuclear Matters (CENuM), Korea University, Seoul 02841, Republic of Korea}
\affiliation{Department of Physics, Korea University, Seoul 02841, Republic of Korea}
\author{M.~P\^{a}rlog}
\affiliation{Université de Caen Normandie, ENSICAEN, CNRS/IN2P3, LPC Caen UMR6534, F-14000 Caen, France}
\affiliation{``Horia Hulubei'' National Institute for R\&D in Physics and Nuclear Engineering (IFIN-HH), P.~O.~Box MG-6, Bucharest Magurele, Romania}
\author{G.~Pasquali}
 \affiliation{INFN - Sezione di Firenze, 50019 Sesto Fiorentino, Italy}
 \affiliation{Dipartimento di Fisica e Astronomia, Universit\`{a} di Firenze, 50019 Sesto Fiorentino, Italy}
\author{S.~Piantelli}
 \affiliation{INFN - Sezione di Firenze, 50019 Sesto Fiorentino, Italy}
\author{G.~Poggi}
 \affiliation{INFN - Sezione di Firenze, 50019 Sesto Fiorentino, Italy}
 \affiliation{Dipartimento di Fisica e Astronomia, Universit\`{a} di Firenze, 50019 Sesto Fiorentino, Italy}
\author{A.~Rebillard-Soulié}
\affiliation{Université de Caen Normandie, ENSICAEN, CNRS/IN2P3, LPC Caen UMR6534, F-14000 Caen, France}
\author{R.~Revenko}
\affiliation{Grand Accélérateur National d'Ions Lourds (GANIL), CEA/DRF–CNRS/IN2P3, Boulevard Henri Becquerel, F-14076 Caen, France}
\author{S.~Valdré}
\affiliation{INFN - Sezione di Firenze, 50019 Sesto Fiorentino, Italy}
\author{G.~Verde}
\affiliation{INFN - Sezione di Catania, 95123 Catania, Italy}
\affiliation{Laboratoire des 2 Infinis - Toulouse (L2IT-IN2P3), Universit\'e de Toulouse, CNRS, UPS, F-31062 Toulouse Cedex 9 (France)}
\author{E.~Vient}
\affiliation{Université de Caen Normandie, ENSICAEN, CNRS/IN2P3, LPC Caen UMR6534, F-14000 Caen, France}

\collaboration{INDRA-FAZIA collaboration}
\noaffiliation
\begin{abstract}
In this work we provide a model-independent experimental evaluation of the degree of isospin equilibration taking place in $^{58,64}$Ni+$^{58,64}$Ni collisions at 32~MeV/nucleon across varying reaction centralities. This result has been obtained by combining the complementary information provided by two different datasets, sharing common characteristics. The first dataset has been acquired with the INDRA setup and has been used to implement a model-independent reconstruction of the impact parameter. The second dataset has been acquired in the first experimental campaign of the coupled INDRA-FAZIA apparatus at GANIL. The neutron-to-proton content of the quasiprojectile remnant measured by FAZIA has been employed as isospin observable. The effect of isospin diffusion has been evidenced by means of the isospin transport ratio, reported as a function of the impact parameter of the collision. The evolution towards isospin equilibration from semiperipheral to more central collisions is clearly extracted.
This experimental result, expanding our previous works (\href{https://doi.org/10.1103/PhysRevC.106.024603}{Phys. Rev. C 106, 024603 (2022)} and \href{https://doi.org/10.1103/PhysRevC.108.054611}{Phys. Rev. C 108, 054611 (2023)}), can be compared with the predictions of any transport model, and can thus be used to set constraints on the behavior of the symmetry energy term of the nuclear Equation of State at sub- to saturation densities.
\end{abstract}

\maketitle

\section{Introduction}
Since decades, many efforts in both theoretical and experimental nuclear physics converge on the study of the nuclear Equation of State (nEoS) \cite{Sorensen2024}.
This fundamental property of nuclear matter, describing its thermodynamic behavior, has direct consequences on the structure of nuclei \cite{RocaMaza2018}, as well as on the dynamics of heavy ion collisions \cite{Colonna2020}.
Further interest stems from the astrophysics context, where the nEoS represents a key ingredient for modeling core collapse supernovae \cite{Yasin2020}, the formation and features of neutrons stars \cite{Margueron2018II, Burgio2021}, and the evolution of compact binary objects mergers \cite{Lattimer2000}, especially in the era of gravitational waves detection.
The bridge between information achieved with terrestrial experiments and astrophysical observations has been deeply investigated recently, demonstrating the close connection between previously separated fields \cite{Huth2022, Lynch2022, Pang2023, Tsang2024}. 
At present, most of the work is focused on better constraining the nuclear density dependence of the symmetry energy term $E_\text{sym}$ of the nEoS, necessary to describe the behavior of asymmetric nuclear matter.

The study of the nEoS in laboratories is efficiently made with heavy ion reactions, that produce transient nuclear systems far from normal conditions and offer a plethora of phenomena to observe \cite{Colonna2020}. As an example, at Fermi energies one can probe subsaturation densities, possibly reached in semiperipheral collisions within elongated neck-like structures formed between the outgoing quasiprojectile (QP) and quasitarget (QT). The isotopic composition of the particles emerging from the reaction can be studied and compared to model predictions to collect information on the microscopic mechanisms driving the system evolution. Among these, the isospin transport phenomena have been the subject of many experimental and theoretical investigations \cite{Colonna2020, Baran2005, McIntosh2019} aiming at studying the density behavior of the symmetry energy. Within a hydrodynamical framework, two contributions to the proton/neutron currents are generally distinguished: the \textit{isospin drift}, activated by any density gradient in the system and ruled by the first derivative of $E_\text{sym}$ with respect to the nuclear density $\rho$, and the \textit{isospin diffusion}, driven by isospin gradients and dependent on the $E_\text{sym}$ value. Experimental indications of these two contributions have been found in the neutron enrichment of midvelocity emissions \cite{Lionti2005, Lombardo2010, Barlini2013, Piantelli2021, Fable2023} and on the isospin equilibration in asymmetric reactions \cite{Johnston1996, Lombardo2010, Barlini2013, Piantelli2021, Camaiani2021, Fable2023, Ciampi2022}, respectively.

At the nuclear densities explored at Fermi energies, the uncertainties on the $E_\text{sym}$ behavior have been narrowed down thanks to the results from different approaches, so that the differences in the model predictions on experimental observables assuming different functionals are not much pronounced and are easily washed out either by the variability of the transport model predictions on other ingredients still not fully under control \cite{Wolter2022} or by the effects of statistical evaporation and experimental limitations. It is therefore of primary importance identifying the most suitable experimental strategy to access isotopic observables less affected by such issues \cite{Napolitani2010}. A viable method relies on the isospin transport ratio, first proposed by Rami \textit{et al.} \cite{Rami2000} and extensively studied in the past \cite{Tsang2004, Chen2005, Liu2007, Tsang2009, Napolitani2010, Coupland2011, Camaiani2020, Camaiani2021, Ciampi2022, Ciampi2023, Fable2023,Fable2024} to highlight the isospin equilibration enhancing possible differences associated with different $E_\text{sym}$ parametrizations. Given a neutron-rich nuclide $A$ and a neutron-deficient nuclide $B$, considering the reactions between their four possible combinations $i=AA,AB,BA,BB$, the isospin transport ratio is defined as:
\begin{equation}
 R(x_{i}) = \frac{2x_{i}-x_{AA}-x_{BB}}{x_{AA}-x_{BB}}
\end{equation}
where $x$ is an isospin observable sensitive to the isospin diffusion effect, such as the isospin content of the QP remnant, which in the literature is regarded as one of the best probes to gain information on $E_\text{sym}$ \cite{Napolitani2010, Baran2005, Mallik2022}. By construction, $R(x_{AA,BB})=\pm1$: these values represent the non-equilibrated limit in the mixed reactions, while in the case of complete equilibration $R(x_{AB})=R(x_{BA})$. If $x$ is evaluated in identical experimental conditions in the four cases, by considering the differences in the outcomes of symmetric and asymmetric systems, the ratio is regarded as a method to eliminate common effects such as the systematic uncertainties introduced by the apparatus \cite{Rami2000}. $R(x)$ is also largely unaffected by dynamical emissions and statistical de-excitation \cite{Camaiani2020, Mallik2022}: more specifically, Ref.~\cite{Camaiani2020} points out that statistical evaporation can slightly perturb $R(x)$ only for the lowest excitation energies of the primary fragments ($E^*/A<2$~MeV/nucleon), associated with the most peripheral reactions, while no evident dependence of the ratio from the excitation energy is found in Ref.~\cite{Mallik2022}.
This property of $R(x)$ allows to better bring out dynamical effects including the nEoS consequences on isospin diffusion, and it can also be of use when defining the comparison protocol between experiment and model predictions. In fact, transport models are generally stopped at freeze-out and coupled to an afterburner simulating the statistical evaporation phase, but this procedure may introduce multiple sources of error \cite{Piantelli2019, Mallik2022} which could be avoided by just bypassing the secondary phase in the calculations.

In previous experimental works exploiting the isospin transport ratio \cite{Camaiani2021, Ciampi2022, Ciampi2023, Fable2023}, the evolution towards isospin equilibration has been followed as a function of a purposely selected centrality related observable: the characterization of the equilibration process as a function of centrality is indeed crucial to highlight effects associated with different conditions such as density and interaction times or with different output channels \cite{Tsang2004, Ciampi2023}. 
However, a model-independent transformation of such secondary order parameter into the impact parameter $b$ is of interest, both to disentangle the result from the model prediction of the order parameter itself and to allow for the comparison with the primary $R(x)$ calculated with different transport models.
Due to the statistical nature of the processes involved, in this transformation it is of key importance taking into account also the intrinsic fluctuations of the centrality observable associated with any given set of initial conditions. If not correctly treated, the consequent mixing over intervals of impact parameters for any selection of the employed observable can strongly affect the final observations and even hinder the comparison with model predictions, particularly for increasingly central collisions, associated with gradually decreasing cross sections and stronger fluctuations \cite{Li2018, Frankland2021}. In the literature multiple procedures to extract the impact parameter distribution associated with selections on different observables have been proposed, starting with the widely used sharp cut-off approximation \cite{Cavata1990}. A quite recent model-independent method, first proposed for high energy heavy ion collisions \cite{Das2018,Rogly2018} and then adapted for intermediate energies \cite{Frankland2021,Chen2023}, allows to extract the relationship between the impact parameter and the centrality observable, together with the corresponding fluctuations, by means of a fit of the experimental inclusive distribution of the same observable in a ``minimum-bias'' dataset.

This paper further extends the analyses detailed in Refs.~\cite{Ciampi2022,Ciampi2023}, presenting a fully model-independent isospin transport ratio calculated from the isotopic characteristics of the detected QP remnant as a function of the reaction centrality evaluated with the method of Ref.~\cite{Frankland2021}. Section~\ref{sec:experimental_setup} summarizes the characteristics of the setup and of the experimental datasets used for this work. The application of the impact parameter reconstruction method is described in Sec.~\ref{sec:impact_parameter}, and in Sec.~\ref{sec:isospin_transport_ratio} the isospin transport ratio as a function of centrality is presented; moreover, its consistency with the result of our previous analyses is presented in Appendix~\ref{app:comparison}. This model-independent experimental result can in principle be compared with any theoretical prediction, both for primary and secondary QP fragments.

\section{Experimental data}\label{sec:experimental_setup}
For the present work, two datasets of Ni-Ni collisions at 32~MeV/nucleon bearing complementary information have been combined in order to build the final result. The details of the experimental setups and reactions for each dataset are given in the following paragraphs.

 \subsection{INDRA dataset}
 INDRA is a multidetector apparatus operating at GANIL (Caen, France) since the 1990s, optimized for the detection and identification of charged products of heavy ion collisions at Fermi energies. Its original configuration, described in detail in Refs.~\cite{Pouthas1995,Pouthas1996}, was composed of 336 independent telescope modules arranged in 17 rings with cylindrical symmetry around the beam axis, providing a large angular coverage (90\% of the solid angle), and low detection thresholds thanks to the presence of ionization chambers (ChIo) as first stage.
 The number of modules and their telescope configuration depend on the ring \cite{Pouthas1995}: phoswich detectors for the most forward ring (2\textdegree$<\theta_\text{lab}<$3\textdegree), ChIo-Si-CsI(Tl) for rings 2 to 9 (3\textdegree$<\theta_\text{lab}<$45\textdegree) and ChIo-CsI(Tl) for rings 10 to 17 ($\theta_\text{lab}>$45\textdegree). Charge identification was achieved up to uranium via the $\Delta$E-E method, or Pulse Shape Analysis (PSA) in CsI(Tl) detectors for light particles. Isotopic identification was achieved up to $Z\sim4-5$.

 In this work, a dataset acquired with the original configuration in 1994 including the reaction $^{58}$Ni+$^{58}$Ni at 32~MeV/nucleon \cite{Cussol2002} has been employed for centrality evaluation purposes. A total statistics of $7\times10^6$ events is available. In that experiment, the online trigger setting was a total number of fired telescopes over the whole apparatus $M_\text{tot}\geq4$: as it will be explained in the following, such ``minimum-bias'' trigger condition is essential in order to fully apply the method of Ref.~\cite{Frankland2021}.
 
 \subsection{INDRA-FAZIA dataset}
 FAZIA is a last-generation telescope array able to achieve state-of-the-art performances of charge and mass identification of nuclear fragments produced in intermediate-energy collisions, up to the heaviest QP remnants. The basic unit is the \textit{block}, composed of 16 Si(300~$\mu$m)-Si(500~$\mu$m)-CsI(Tl)(10~cm) telescopes able to provide identification via the $\Delta$E-E method and via PSA both in Si and in CsI(Tl). Depending on the method, isotopic identification can be achieved up to $Z\approx25$. More details on the detector characteristics and the associated custom electronics can be found in Refs.~\cite{Bougault2014, Valdre2019}.

 In view of exploiting their complementary characteristics, INDRA and FAZIA have been coupled at GANIL in 2019 \cite{Ciampi2022, Ciampi2023, Casini2022}. The polar angles greater than 14\textdegree~are covered by INDRA rings 6 to 17 corresponding to 80\% of the solid angle; in this experiment, INDRA's ionization chambers have not been used. The first five rings have been replaced by 12 FAZIA blocks, covering the most forward polar angles (1.4\textdegree$<\theta_\text{lab}<$12.6\textdegree~ for a 1~m distance from target) to employ the optimal identification performance for the heavy fragments associated with the QP phase space.
 The two acquisition systems have been coupled through a time-stamp distribution provided by the GANIL VXI CENTRUM module \cite{CENTRUM_ref} allowing for the online merging of coincident events.
 
 The dataset of the first INDRA-FAZIA experiment includes the four reactions $^{58,64}$Ni+$^{58,64}$Ni at 32~MeV/nucleon, aimed at studying the isospin diffusion by comparing the isospin characteristics of the QP remnants produced in symmetric and asymmetric reactions: the isotopic identification capability of FAZIA is indeed necessary to assess the isospin content of such relatively heavy fragment. Due to the limitations posed by the dead time of the former INDRA acquisition, later renewed for the second experiment \cite{Frankland2022}, the trigger logic has been essentially set on the multiplicity of fired telescopes in FAZIA, $M_\text{FAZIA}$. Events satisfying two different conditions have been acquired: the main one was $M_\text{FAZIA}>1$, while the less biasing condition was $M_\text{FAZIA}\geq1$ downscaled by a factor of 100. The two triggers can be distinguished in the offline analysis through the trigger pattern recorded in each event. A total statistics of $3\times10^7$ events has been acquired for each system. We note that this dataset is the same analyzed in Refs.~\cite{Ciampi2022,Ciampi2023}: a comparison of the analyses, including differences and similarities as well as the overall consistency with the findings of the present one, is provided in Appendix~\ref{app:comparison}.
 
The following analysis will be subdivided into two main steps. First, the reconstruction of the impact parameter distributions will be presented: in this phase we make use of both datasets. The result of this procedure will be then applied in the second phase, where the INDRA-FAZIA dataset, that includes the information on the isospin content of the QP remnant, will be employed to show the isospin diffusion as a function of the reconstructed impact parameter. 

\section{Impact parameter reconstruction}\label{sec:impact_parameter}
 The method proposed in Ref.~\cite{Frankland2021} allows to extract in a model-independent way the correspondence between any sample distribution of an experimentally-measured centrality-related observable $X$ and the associated impact parameter distribution. For a complete description of the method, we refer the reader to Ref.~\cite{Frankland2021}. This method was also recently applied in Ref.~\cite{Fable2024} for the investigation of isospin diffusion.
 The procedure is essentially based on modeling the conditional probability distribution of $X$ for a given centrality, $P(X|c_b)$, where the centrality $c_b$ is defined as:
 \begin{equation}
     c_b\equiv\int_0^b P(b')\,db'
 \end{equation}
 In this work, $P(X|c_b)$ is parametrized using a gamma distribution as the fluctuation kernel. 
 The parameters defining its behavior are extracted by fitting the experimental inclusive distribution $P(X)$. Once this conditional probability distribution is obtained, Bayes' theorem allows to deduce the centrality distribution associated with any experimentally measured distribution of $X$. 
 
 However, in order to retrieve absolute $b$ distributions from the deduced centrality distributions, the full impact parameter distribution $P(b)$ corresponding to the experimentally measured inclusive $P(X)$ distribution used in the fit must be known (or estimated) \cite{Frankland2021}.
 Without any bias introduced by the experimental setup, the simplest assumption would be the geometrical cross section leading to a triangular distribution $P(b) \propto 2\pi b$. The effect of the apparatus acceptance, especially through the trigger condition, will be to discard part of the events thus modifying this initial distribution: given the experimental conditions for the two datasets here inspected, we can expect a more prominent role of their minimum-multiplicity triggers on semi- to  peripheral collisions.
 
 In particular, for the INDRA dataset, the effect of the $M_\text{tot}\geq4$ trigger is a smearing of the $P(b)$ for large impact parameters \cite{Vient2018}, which can be well approximated as \cite{Frankland2021}:
 \begin{equation}
 \label{eq:bdist_total}
     P(b) = \frac{2\pi b}{1+\exp\Bigl(\frac{b-b_0}{\Delta b}\Bigr)}
 \end{equation}
 The distribution of eq.~\eqref{eq:bdist_total} has been verified on this and similar ``minimum-bias'' INDRA datasets for various reactions using different models, where in general it has been found $\Delta b\approx0.4$~fm \cite{Frankland2021}. The other parameter $b_0$ can be calculated model-independently for a given $\Delta b$ from the measured total reaction cross section $\sigma_R$ by inverting the relation \cite{Frankland2021}:
 \begin{equation}
     \sigma_R = -2\pi(\Delta b)^2\text{Li}_2\biggl[ -\exp\Bigl( \frac{b_0}{\Delta b}\Bigr)\biggr]
 \end{equation}
 where $\text{Li}_2(z)$ is the dilogarithm function.\\
 The function of eq.~\eqref{eq:bdist_total}, however, cannot be assumed for the reactions in the INDRA-FAZIA dataset: due to the trigger condition based on a multiplicity defined over a limited forward angle, even for the less biasing setting $M_\text{FAZIA}\geq1$, some of the events are discarded in a non-trivial way. We have checked with different filtered model predictions that the corresponding $P(b)$ features some evident depletions on the semiperipheral side that cannot be correctly described by eq.~\eqref{eq:bdist_total}.

 \subsection{Application of the method}
 \label{ssec:impact_parameter_application}
 Considering the previous observations, taking advantage of the presence in both datasets of the same reaction $^{58}$Ni+$^{58}$Ni at 32~MeV/nucleon, we apply the impact parameter reconstruction method on the minimum-bias INDRA dataset, in view of exploiting the deduced relationship between $X$ and $b$ for the centrality estimation in the INDRA-FAZIA dataset analysis. To use this procedure, the selected centrality observable $X$ should behave reasonably the same in the two experiments. We employ the multiplicity of both identified and unidentified charged particles collected in INDRA rings 6 to 17, simply denoted $M$ in the following: this global observable is indeed defined over the part of the apparatus which is in common for the two datasets, thus ensuring that the same solid angle is considered (the number of malfunctioning detectors for each ring in the two experiments has also been taken into account). Moreover, by considering all charged particles, regardless of their identification result, we mitigate the possible effect of the identification thresholds, which were sensibly different in the two cases, due to the absence of INDRA ionization chambers in the first INDRA-FAZIA experiment.  
 
 \begin{figure*}
     \includegraphics[width=0.4\textwidth]{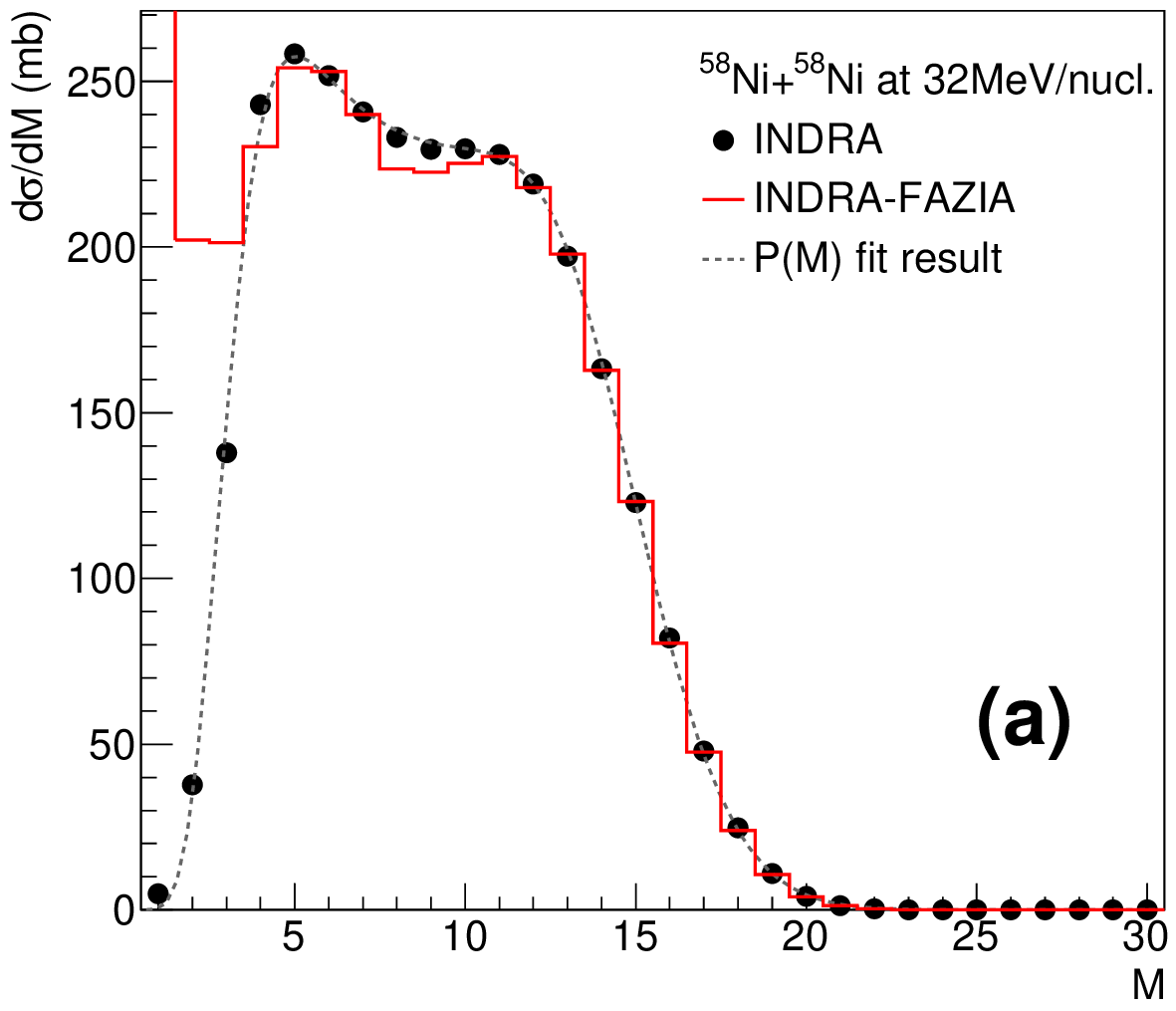}\qquad\qquad
     \includegraphics[width=0.4\textwidth]{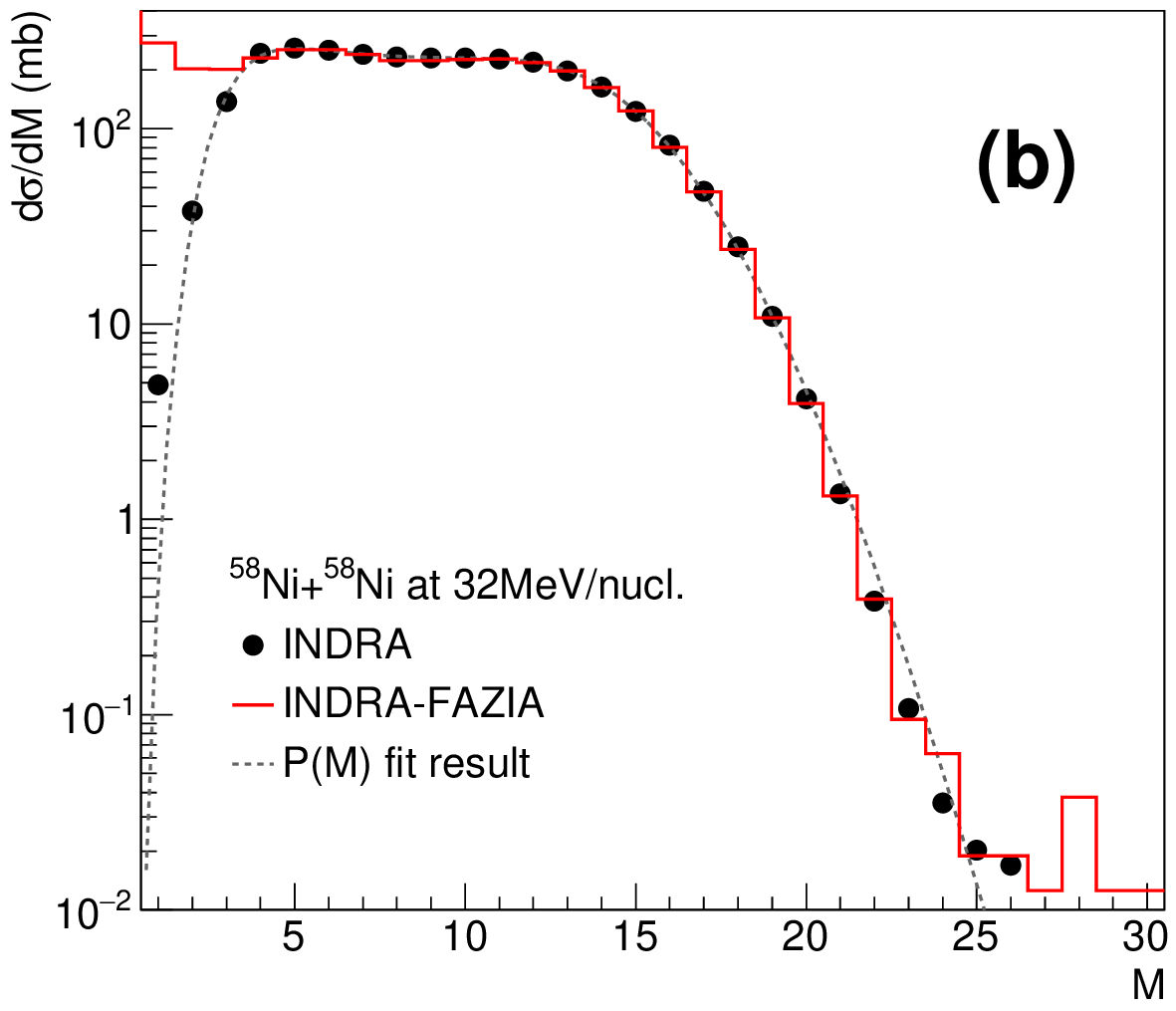}
     \caption{Inclusive distributions of the multiplicity $M$ of identified and unidentified charged particles collected in INDRA rings 6 to 17 for the reaction $^{58}$Ni+$^{58}$Ni at 32~MeV/nucleon in the INDRA dataset (black markers) and result of the fit (gray dashed line), in (a) linear and (b) logarithmic $y$-axis scale, presented as differential cross sections. The red histogram shows the $M$ distribution obtained for the same reaction in the INDRA-FAZIA dataset for the $M_\text{FAZIA}\geq1$ trigger condition, rescaled to take into account the correction due to the different trigger effects (see text) to better show the good agreement between the shapes of the $M>10$ tails.}
     \label{fig:P_M_fit}
 \end{figure*}
The $P(M)$ distributions for the reaction $^{58}$Ni+$^{58}$Ni at 32~MeV/nucleon are shown in Fig.~\ref{fig:P_M_fit} in both linear and logarithmic scale, with black markers for the INDRA dataset and with a red histogram for the INDRA-FAZIA one considering the events acquired with the less biasing $M_\text{FAZIA}\geq1$ trigger condition. The normalization of these plots is described later in this section. First of all, in support of our centrality observable choice, we note that the shape of the high-multiplicity tails of the two distributions matches remarkably well: this part of the distribution, associated with more central events, is indeed less affected by the bias introduced by the two multiplicity-based triggers. On the other hand, the differences that start to develop for $M\lesssim10$ can be ascribed to the role of the trigger conditions acting differently in the two cases, affecting the acquisition of events in different centrality intervals on the more peripheral side.

As introduced, in order to fully apply the method on the $P(M)$ for the INDRA dataset and extract the information on $b$, we adopt an overall impact parameter distribution as in eq.~\eqref{eq:bdist_total}; here, following the prescription of Ref.~\cite{Frankland2021}, we assume $\Delta b=0.4$~fm, but to set the $b_0$ parameter model-independently a measurement of the total reaction cross section $\sigma_R$ is needed. Due to the difficulties encountered in retrieving a measured $\sigma_R$ for the old INDRA experiment, in order to obtain an estimate we combine the information enclosed in the two datasets. A reference is provided by the $M_\text{FAZIA}\geq1$ events acquired in the INDRA-FAZIA dataset, which include the elastic scattering collected with a relatively high statistics by the FAZIA telescopes positioned at the most forward polar angles. The counts associated with the elastic events are clearly identified as a peak in the Si1-Si2 $\Delta$E-E correlations, and by exploiting the geometrical information of the setup they can be related to the corresponding Rutherford cross section integrated over the active area of the telescope. This provides an absolute normalization allowing to transform any number of counts into the corresponding cross section for the reaction in the INDRA-FAZIA dataset. 
The normalization from number of counts to cross sections must be however transferred to the same reaction in the INDRA dataset, whose $\sigma_R$ is needed to set the parameter $b_0$. To this end, we use the $M>10$ tails of the $P(M)$ distributions of Fig.~\ref{fig:P_M_fit} as a reference to relate the number of counts of the two experiments since, as said, these central events are less affected by deformations introduced by the triggers. However, we do not exclude \textit{a priori} a minor effect of the different triggers also in this region. More precisely, taking into account the high geometrical coverage and the trigger condition, for the high-multiplicity central events, the INDRA dataset can be considered as the unbiased one \cite{Frankland2021,Vient2018}, while on the other hand for the same events, there may be a small probability to be discarded by the INDRA-FAZIA trigger $M_\text{FAZIA}\geq1$ in case no particle hits the forward angle covered by FAZIA: in this respect, the INDRA-FAZIA trigger would select a subset of the events acquired by the INDRA trigger for the most central class. To give a rough estimate of the impact of this condition we have applied an INDRA-FAZIA-like trigger on the INDRA data, finding approximately a $\sim$90\% relative detection efficiency in the region $M>10$.
The $P(M)$ distribution shown in Fig.~\ref{fig:P_M_fit} for the INDRA-FAZIA dataset is already multiplied by the corresponding correction factor\footnote{It should be noted that such approximate correction has been evaluated only for the multiplicity interval of interest for our purpose and is likely not valid for lower multiplicities.}: the interval for $M>10$ of this distribution is used as a reference in order to normalize the $P(M)$ from the INDRA dataset, as shown in the same figure. From the integral of the latter, we obtain $\sigma_R=(3.0\pm0.5)$~b corresponding to $b_0=(9.8\pm0.7)$~fm; as a reference, the total reaction cross section according to the Kox-Shen parametrization \cite{KoxShenSihver2014} is $\sim3.7$~b.
%https://bioapp.gsi.de/cross-section-db/
For the uncertainties on these values, we took into account a maximum estimate of the error introduced in the rescaling procedure between the two experiments, combined with that related to the evaluation of the absolute Rutherford cross section associated with the FAZIA telescopes used for the normalization, which depends on the knowledge of their position relative to the beam axis: the latter contribution, which is the dominant one, has been estimated by considering a maximum displacement both in $x$ and in $y$ with respect to the nominal position of the detector equal to $\pm2$~mm, also taking care of the rigid structures in each FAZIA block (i.e., quartets, see Refs.~\cite{Bougault2014,Valdre2019}), fixing the relative positions of some pairs of neighboring telescopes.

 \begin{figure}
     \centering
     \includegraphics[width=0.9\columnwidth]{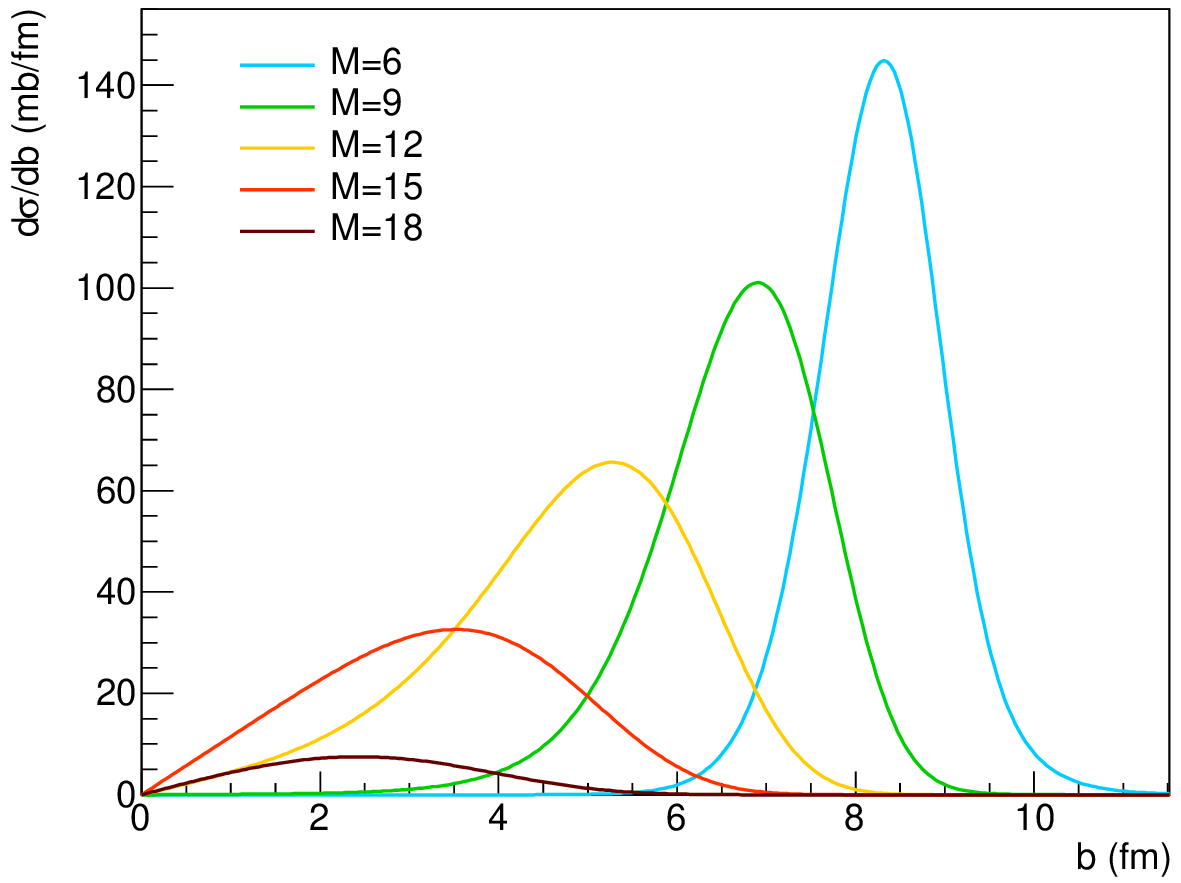}
     \caption{Model-independent impact parameter distributions for some selected values of the multiplicity $M$ of identified and unidentified particles collected in INDRA rings 6 to 17 for the reaction $^{58}$Ni+$^{58}$Ni at 32~MeV/nucleon.}
     \label{fig:b_distributions}
 \end{figure}
Finally, the result of the fit to the inclusive $P(M)$ obtained from the INDRA dataset is plotted in Fig.~\ref{fig:P_M_fit} with a gray dashed line, that well follows the experimental distribution plotted with black markers. The reduced $\chi^2$ is 20.1, in line with what obtained in Ref.~\cite{Frankland2021} for a similarly defined global observable. Figure~\ref{fig:b_distributions} presents some $b$ distributions associated with five different multiplicities $M$, obtained assuming eq.~\eqref{eq:bdist_total} with the aforementioned parameter settings. This plot shows the important role of the intrinsic fluctuations, as relatively different selections of our centrality observable can populate partly (or entirely) superimposed $b$ intervals, particularly for less peripheral collisions.

 \begin{figure*}
     \includegraphics[width=0.45\textwidth]{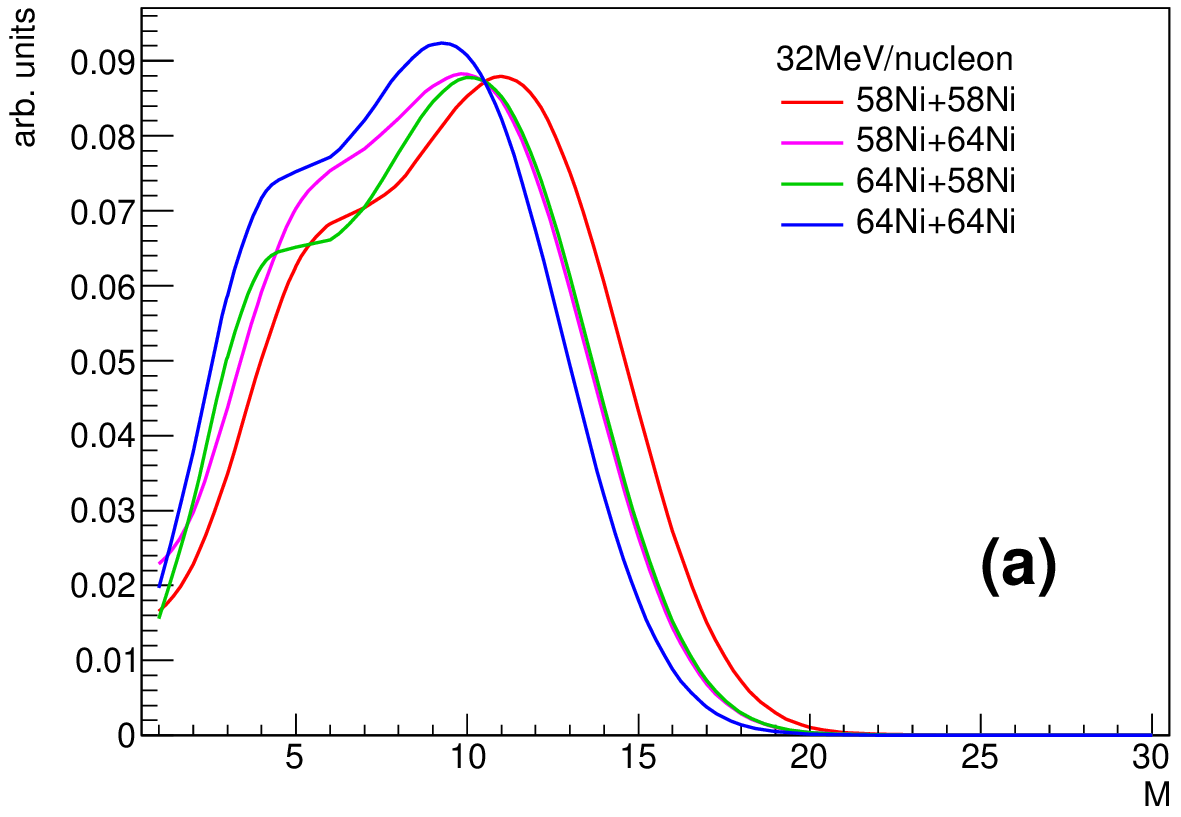}\qquad\qquad
     \includegraphics[width=0.45\textwidth]{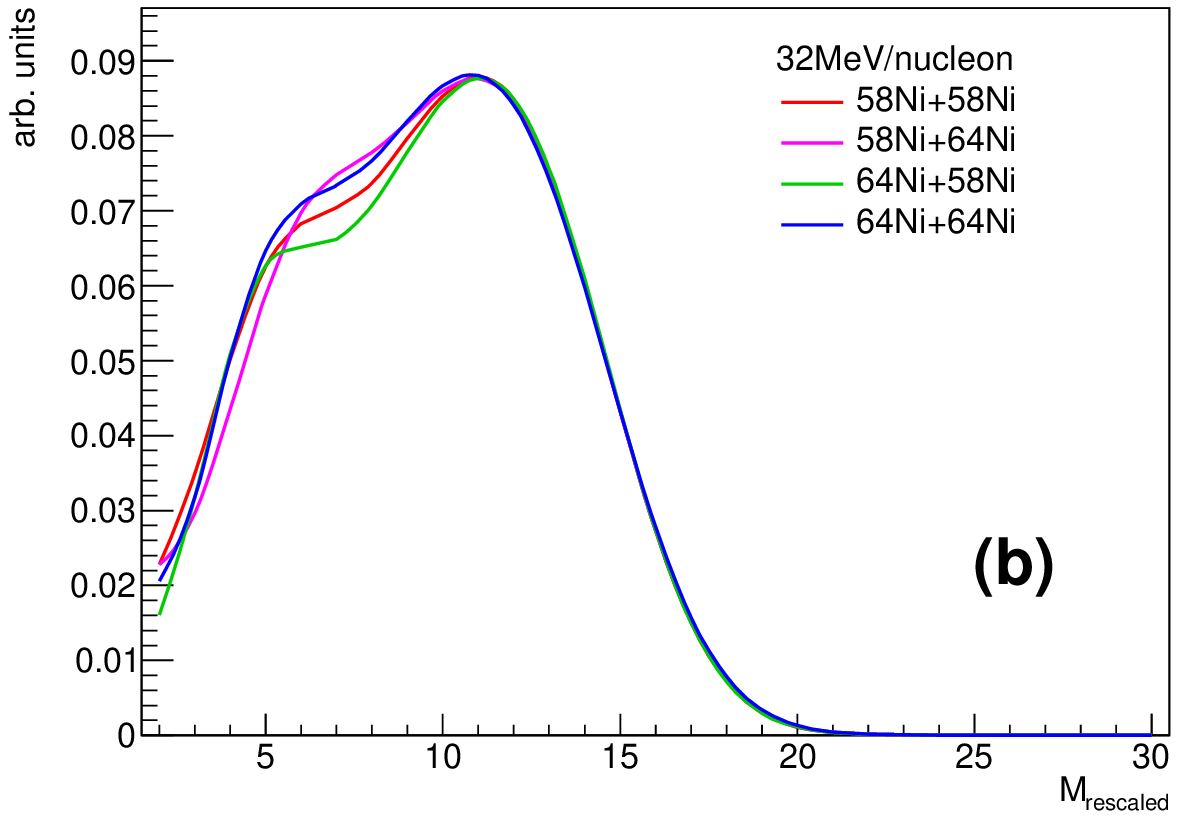}
     \caption{Distributions of the multiplicity $M$ of identified and unidentified particles collected in INDRA rings 6 to 17 for the four reactions $^{58,64}$Ni+$^{58,64}$Ni at 32~MeV/nucleon (a) before and (b) after rescaling (see text), obtained for the main trigger $M_\text{FAZIA}>1$ in the INDRA-FAZIA dataset.}
     \label{fig:Mrescaling}
 \end{figure*}
\begin{figure}
     \centering
     \includegraphics[width=0.9\columnwidth]{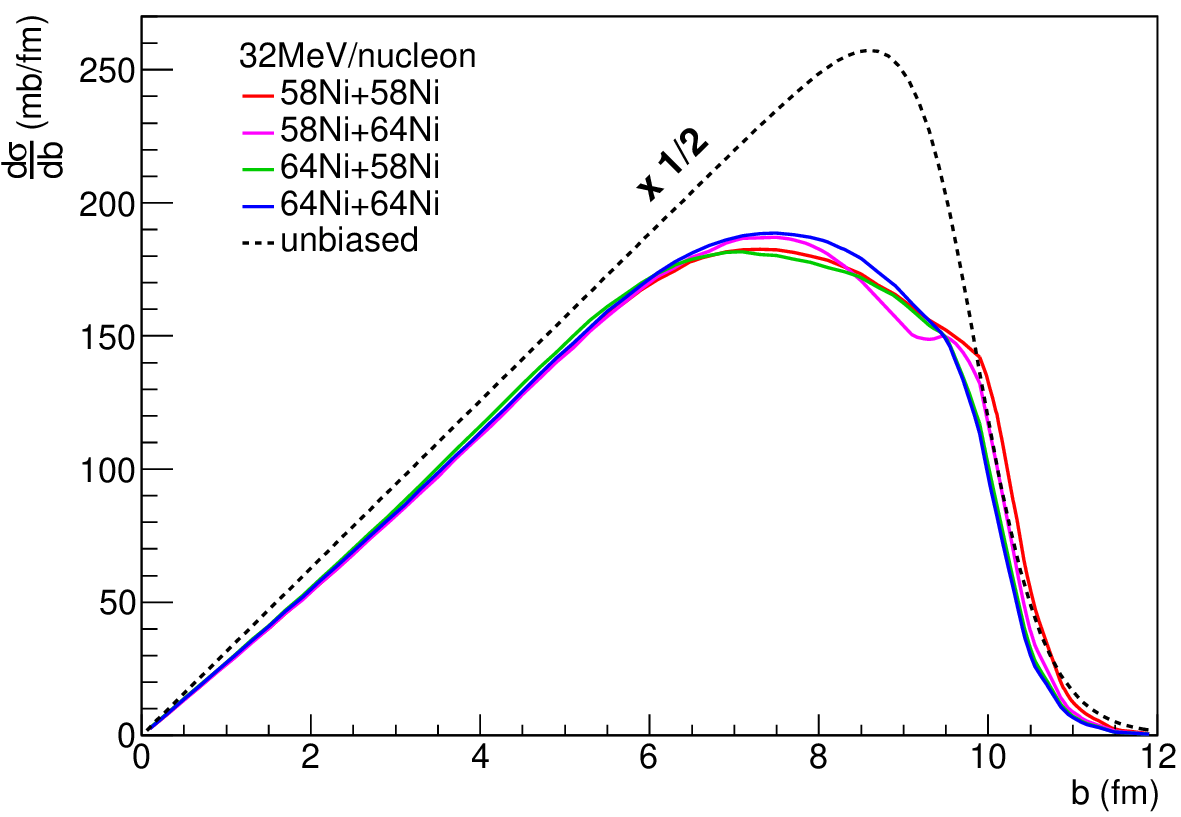}
     \caption{Model-independent global impact parameter distribution for the four reactions $^{58,64}$Ni+$^{58,64}$Ni at 32~MeV/nucleon, obtained for the main trigger $M_\text{FAZIA}>1$ in the INDRA-FAZIA dataset. The black dotted line represents the inclusive $P(b)$, rescaled for easier shape comparison, based on the $b_0$ and $\Delta b$ parameters employed for the INDRA dataset. The effect of the $b_0$ uncertainty is omitted to allow for a clearer comparison of the right-side tails of the distributions.}
     \label{fig:b_dist_4sys}
 \end{figure}

 These results are valid for the reaction $^{58}$Ni+$^{58}$Ni at 32~MeV/nucleon. However, the INDRA-FAZIA dataset includes other three Ni-Ni systems at the same energy for which there is no corresponding measurement in the INDRA dataset to proceed as explained above. The multiplicity $M$ of identified and unidentified charged particles collected in INDRA rings 6 to 17 that we use for the centrality estimation depends on the reaction systems, as suggested by the different endpoints of the $M$ distributions shown for the four systems in Fig.~\ref{fig:Mrescaling}(a). Such difference could already be expected considering that a more neutron-rich primary fragment can partly de-excite by emitting free neutrons which are not detected by the apparatus, therefore, for a given degree of dissipation, we can expect to detect globally higher multiplicities for neutron-poorer systems.
 In order to treat the three systems other than $^{58}$Ni+$^{58}$Ni we find a suitable transformation to convert the detected multiplicity $M_\text{sys}$ into a multiplicity $M_\text{resc}$ rescaled as a corresponding $M$ value for $^{58}$Ni+$^{58}$Ni. The simplest assumption is a linear transformation, that due to the discrete nature of $M$ is implemented as following:
 \begin{equation}
 \label{eq:M_rescaling}
     M_\text{resc} = \bigl\lfloor \alpha\cdot (M_\text{sys}+r) +\beta \bigr\rfloor
 \end{equation}
 where $r\sim U([0,1])$ is a uniformly distributed random variable taking values in $[0,1]$. % and the parentheses indicate the truncation.
 To determine the parameters $\alpha$ and $\beta$ of the transformation we use as reference the multiplicities associated with the most central collisions: the high multiplicity part of the $M_\text{sys}$ distributions shown in Fig.~\ref{fig:Mrescaling}(a) is matched with the one for $M>10$ for the system $^{58}$Ni+$^{58}$Ni. The result is shown in Fig.~\ref{fig:Mrescaling}(b), where it can be noticed how applying the transformation \eqref{eq:M_rescaling} allows in all three cases to reproduce remarkably well the shape of the right tail of the $^{58}$Ni+$^{58}$Ni $M$ distribution.
 These rescaled multiplicities are then treated as the $M$ for the $^{58}$Ni+$^{58}$Ni reaction, and the information on the relationship between $M$ and the impact parameter extracted for the latter is applied to all systems.
 Figure~\ref{fig:b_dist_4sys} presents the corresponding global impact parameter distribution for the four reactions in the INDRA-FAZIA dataset, obtained for the events acquired with the main trigger and considered in the analysis of the following section. The inclusive $P(b)$ assumed for the INDRA dataset is also plotted (black dotted line) as a reference for shape comparison. The bias introduced in the INDRA-FAZIA dataset by the combined influence of the experimental apparatus and, in particular, by the trigger settings is evidenced by the deformations on the right side of the distribution, which become apparent from semiperipheral to peripheral collisions.
 
\section{Isospin analysis}\label{sec:isospin_transport_ratio}
The isospin analysis described in the following is carried out exploiting the four systems in the INDRA-FAZIA dataset, focusing on the events acquired with the main trigger $M_\text{FAZIA}>1$; only a few ($\sim1\%$) events associated with some spurious coincidences, violating charge and momentum conservation laws, have been discarded.  

\subsection{QP remnant selection}
\begin{figure}
     \includegraphics[width=0.49\columnwidth]{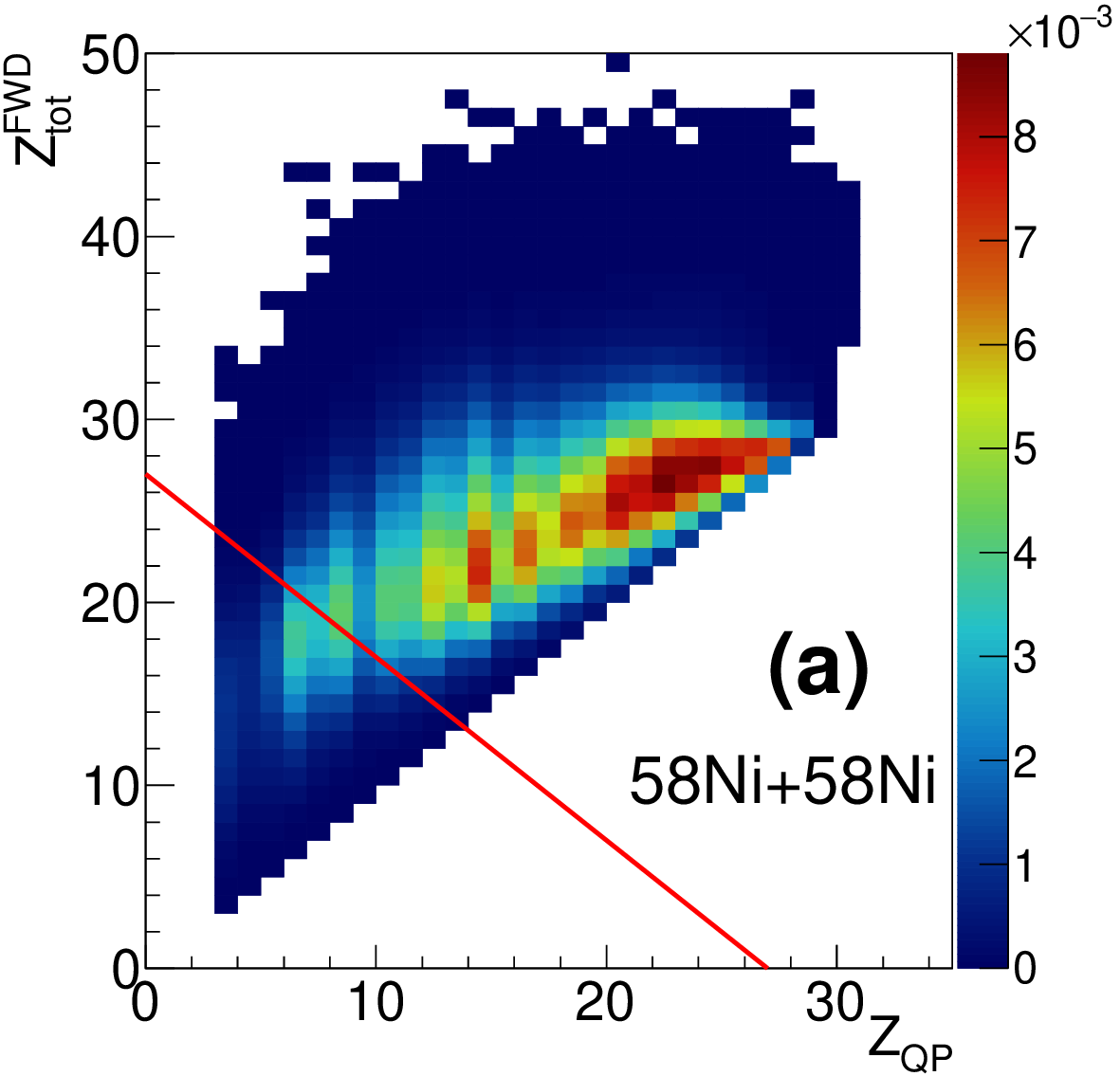}
     \includegraphics[width=0.49\columnwidth]{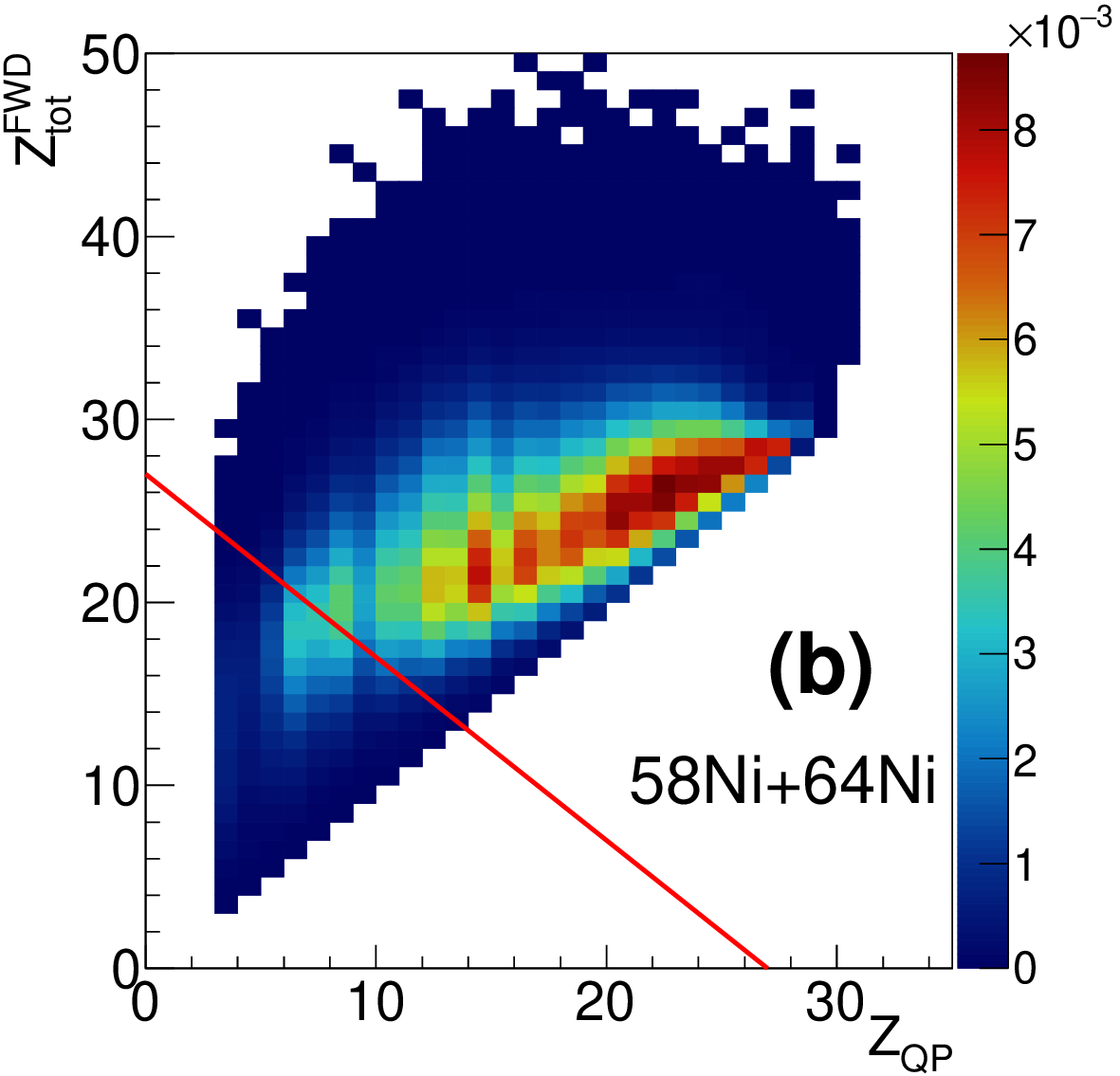}\\
     \includegraphics[width=0.49\columnwidth]{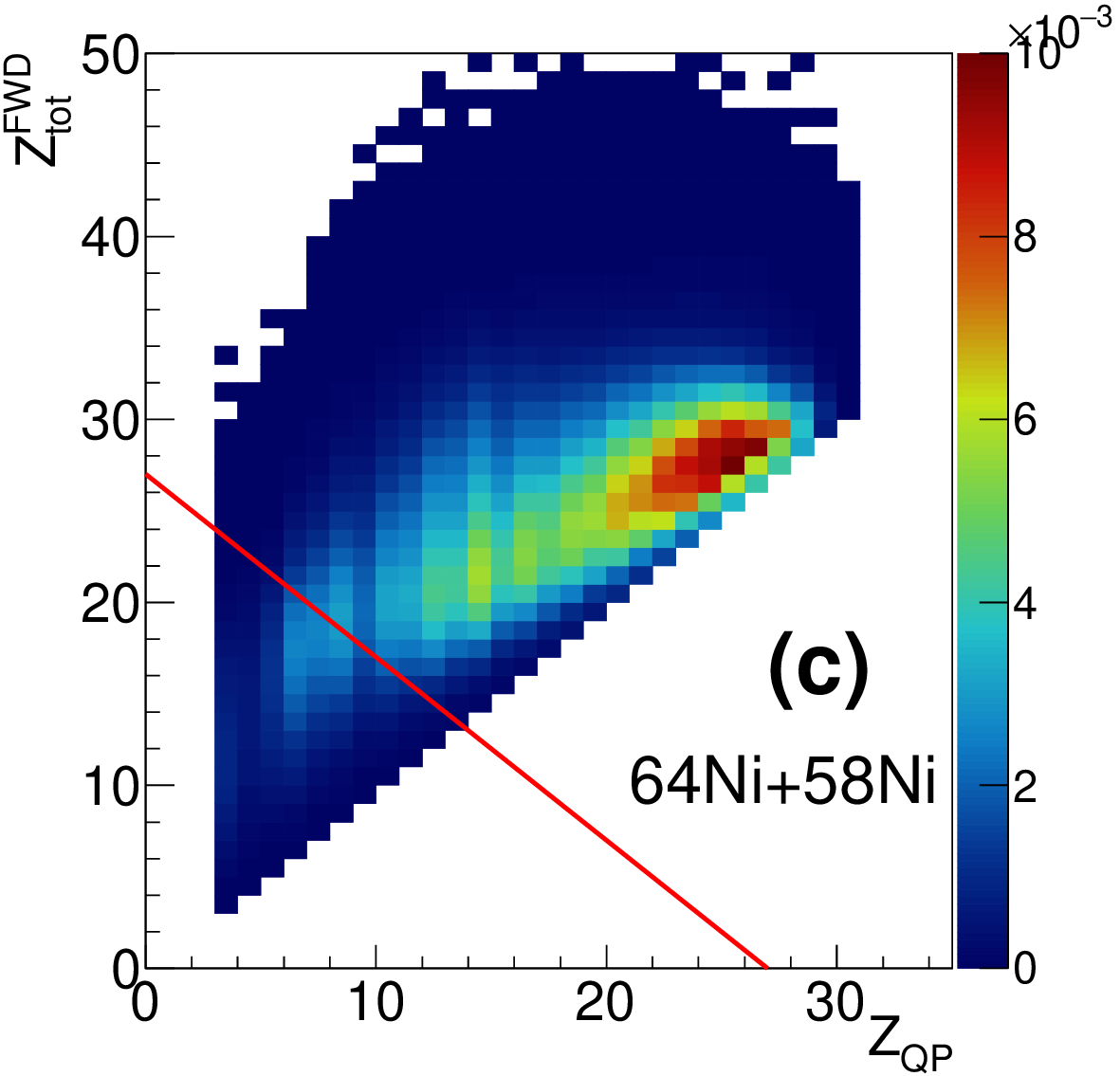}
     \includegraphics[width=0.49\columnwidth]{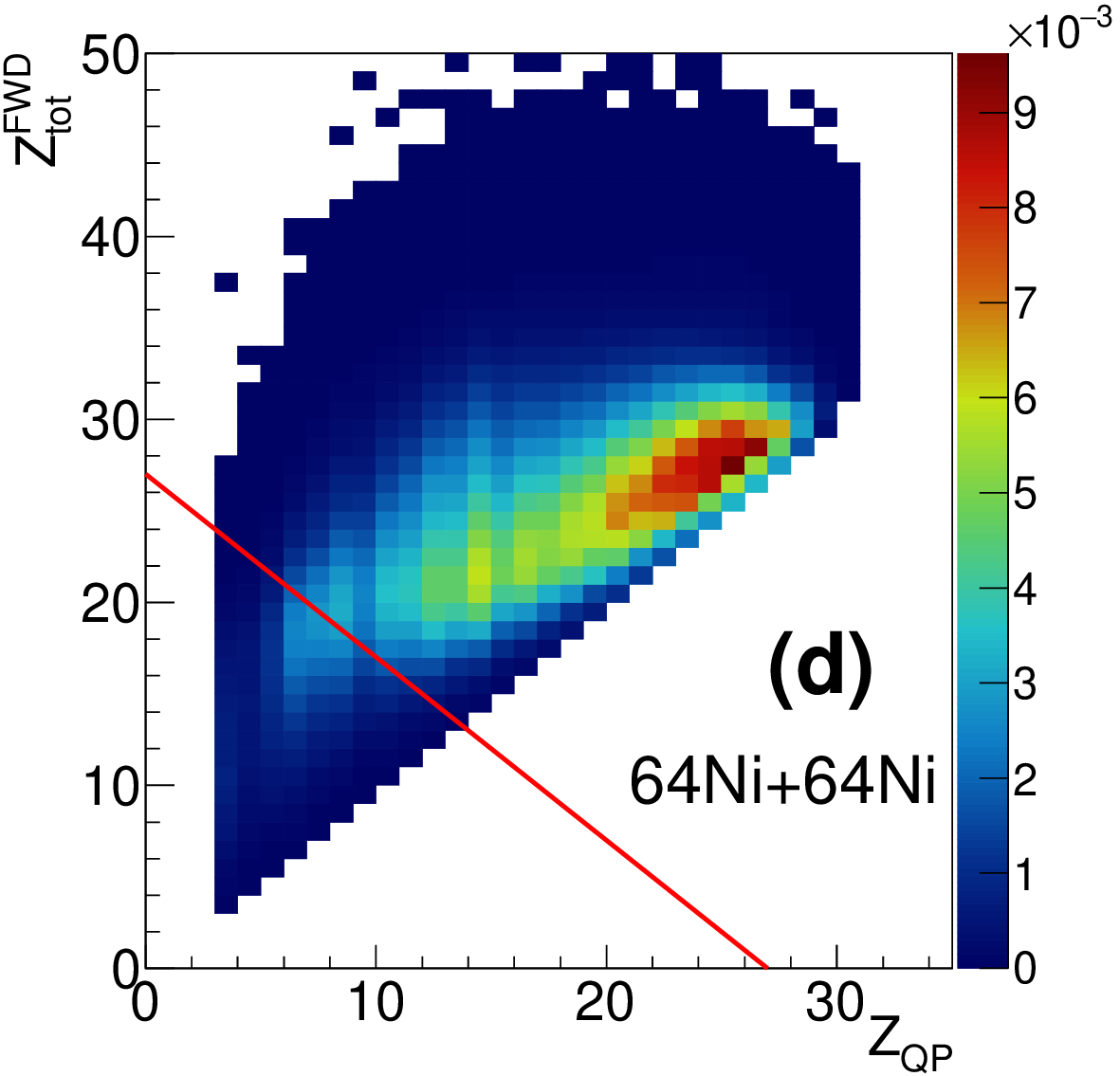}
     \caption{Correlation between the total charge $Z_\text{tot}^\text{FWD}$ detected in the forward hemisphere in the c.m. reference frame and the charge $Z_\text{QP}$ of the heaviest fragment in the same hemisphere, considered as the remnant of the QP, for the four systems in the INDRA-FAZIA dataset.
     The red line represents the lower limit associated with the condition $Z_\text{QP}\geq Z_\text{Ni}-Z_\text{tot}^\text{FWD}$.
     A clear staggering effect is visible on the $Z_\text{QP}$ distribution, more evident for the reactions induced by the neutron-poorer projectile, in agreement with previous observations \cite{Lombardo2011}.}
     \label{fig:completeness}
 \end{figure}

In order to present a model-independent result as general as possible, easily comparable with both primary and secondary model predictions, we aim at setting a quite inclusive analysis, reducing to a reasonable minimum the amount of conditions and selections on the experimental data. Therefore, at variance with previous works having different goals \cite{Ciampi2022,Ciampi2023}, we avoid multiplicity-based distinctions among different output channels. For each event, we designate as candidate for the QP remnant the fragment with the largest charge $Z_\text{QP}$ identified in the forward hemisphere in the center-of-mass (c.m.)~reference frame\footnote{This basic selection, chosen for its straightforward applicability to theoretical simulations, incorporates in the analysis heavy fragments produced in various dynamical mechanisms. While it includes true binary events (which dominate the statistics), it also includes smaller contributions from ternary or quaternary events. For further discussions on this point see Appendix~\ref{app:comparison}.}; if more than one forward-emitted fragment with the same largest charge $Z_\text{QP}$ is found, the one with the largest velocity component along the beam axis $v_z^\text{c.m.}$ is selected. 
In order to be considered a QP remnant, the fragment is required to have a minimum charge $Z_\text{QP}^\text{min} = 5$. In principle, such minimum charge threshold is set very low in order to prevent the exclusion of some very light QP remnants that may result from the most dissipative events, which is necessary to correctly take into account also more central collisions. However, on the other hand, setting a very low $Z_\text{QP}^\text{min}$ increases the probability of including in the analysis some incomplete events where the heaviest fragment has not been detected, leading to the erroneous identification of the second heaviest as QP remnant. To avoid this scenario, it is necessary to carefully check the ``completeness'' of the event information in relation to our purpose: more specifically, the events where the total undetected charge in the forward hemisphere does not exceed the charge of the QP remnant $Z_\text{QP}$ can be safely accepted. Assuming that on average we can expect a total charge in the forward hemisphere around the atomic number of the original projectile\footnote{As a consequence of the same effect of isospin diffusion, proton exchanges during the projectile-target interaction can shift such average total charge value expected in the forward and backward hemispheres for the two asymmetric systems. Therefore, for these two reactions, we have tested some variations of the completeness condition, by properly shifting the total expected forward charge by a few units, up to $\pm3$, observing that the final results hereby presented are negligibly modified and are thus stable against this possible effect.} ($Z_\text{Ni}=28$), the total undetected forward-emitted charge can be roughly estimated as the difference between such value and the total charge $Z_\text{tot}^\text{FWD}$ of all identified particles with $v_z^\text{c.m.}>0$: therefore, we impose the completeness condition $Z_\text{QP}\geq Z_\text{Ni}-Z_\text{tot}^\text{FWD}$, represented by the red line in the correlation plots of $Z_\text{tot}^\text{FWD}$ vs $Z_\text{QP}$ shown in Fig.~\ref{fig:completeness} for the four systems in the INDRA-FAZIA dataset. 
We have verified that when this condition is imposed, by discarding less than 13\% of the events for each system, the final result hereby presented becomes stable against reasonable variations of the $Z_\text{QP}^\text{min}$ threshold, which has been tested within the interval $3<Z_\text{QP}^\text{min}<10$. Such independence from this specific setting is of particular interest for comparing this experimental result with theoretical predictions for primary QP fragments, on which imposing a similar condition on $Z_\text{QP}$ would not act the same due to the unaccounted role played by the statistical de-excitation.

\subsection{Model-independent isospin transport ratio}
\label{ssec:final_isospin_transport}
In this section we present the isospin analysis of the QP remnants defined as above,  clearly requiring their isotopic identification. In order to follow the isospin evolution as a function of the reaction centrality, we now employ the relationship between the rescaled multiplicity $M_\text{resc}$ of eq.~\eqref{eq:M_rescaling} in the four systems and the impact parameter $b$, extracted model-independently as explained in Sec.~\ref{sec:impact_parameter}. 

\begin{figure}[t!]
     \centering
     \includegraphics[width=\columnwidth]{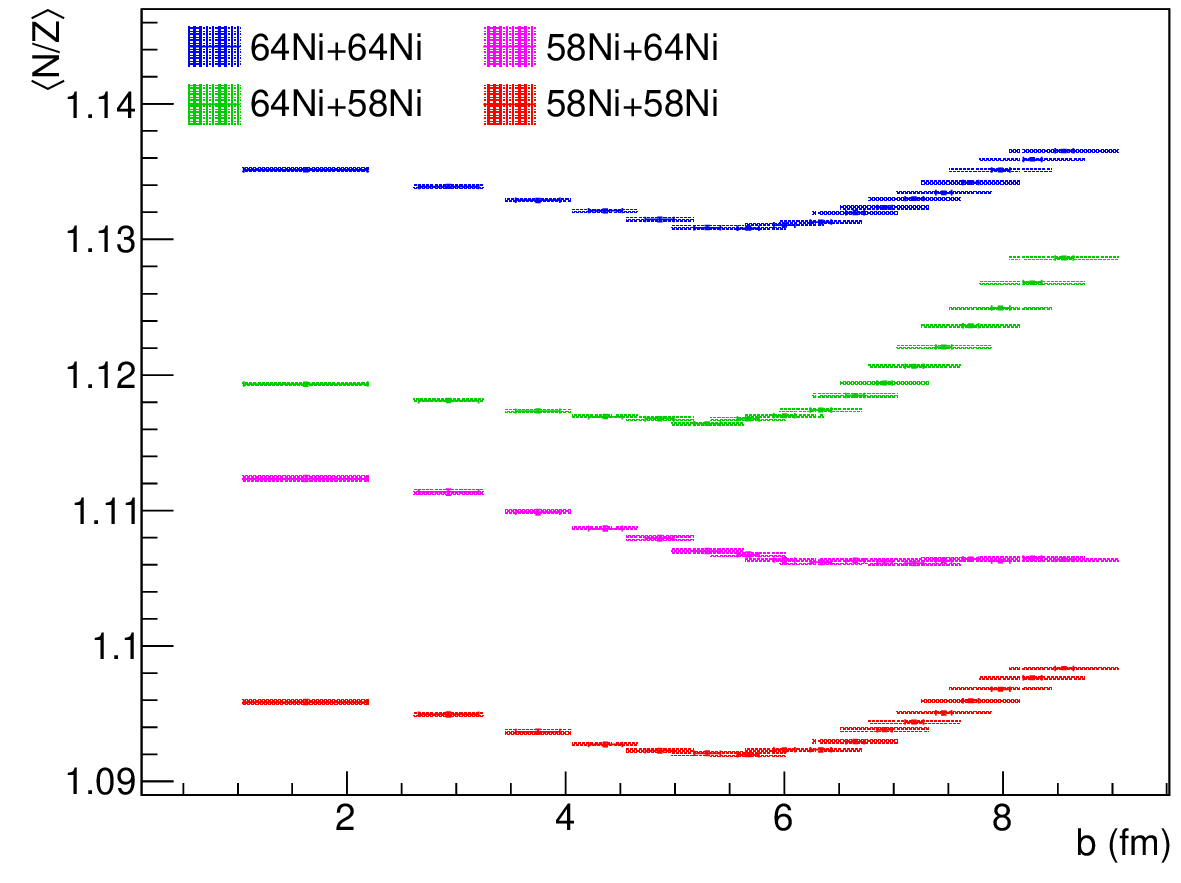}
     \caption{Neutron-to-proton ratio $\langle N/Z\rangle$ of the heaviest forward fragment (QP remnant) as a function of the impact parameter $b$ for the four systems in the INDRA-FAZIA dataset. Statistical errors are plotted on the $y$-axis. The error associated with the extraction of the $b_0$ parameter is reported on the $x$-axis.}
     \label{fig:NsuZ_QP}
 \end{figure}

More specifically, in our analysis, each event is assigned an impact parameter value randomly drawn from the $b$ distribution associated with its corresponding $M_\text{resc}$ value (see Fig.~\ref{fig:b_distributions}).
This method allows us to correctly take into account the full information on both the average $b$ for any given $M_\text{resc}$ and the intrinsic fluctuations around this value. 
Custom impact parameter bins are defined consistently in the four systems, such that each one contains approximately the same statistics.
The resulting average neutron-to-proton ratio $\langle N/Z\rangle$ for the QP remnant as a function of $b$ obtained for the four systems in the INDRA-FAZIA dataset is shown in Fig.~\ref{fig:NsuZ_QP}. 
The $x$-axis value of each data point is the barycenter of the $b$ distribution for the corresponding bin, and the horizontal error bars represent the standard deviation. The vertical error bars, often smaller than the marker size, represent statistical errors. For each data point, we also show a shaded rectangle representing our global uncertainty, where along the $x$-axis the standard deviation of $b$ is combined with the error associated with the $b_0$ uncertainty  (Sec.~\ref{ssec:impact_parameter_application}). The latter contribution affects less central collisions to a greater extent. 
Moreover, in Fig.~\ref{fig:NsuZ_QP}, the points corresponding to the most peripheral collisions have been removed. In fact, large $b$ values are primarily related to low $M_\text{resc}$ and to heavy QP remnants for which the required isotopic identification may not be achieved. Consequently, in this isospin analysis the bins at large $b$ are populated essentially only by the tails of higher-multiplicity events and are therefore considered unreliable.
However, the effect of isospin equilibration clearly emerges from the relative behavior of the four plots: in the two mixed systems, the QP remnant tends to assume a $\langle N/Z\rangle$ close to the one obtained for the symmetric reaction with the same projectile for the more peripheral reactions, while moving towards more central collisions the results for the two asymmetric reactions tend towards each other, signaling the evolution towards equilibration.
We note that the path towards equilibrium is system dependent.

 \begin{figure}[]
     \centering
     \includegraphics[width=\columnwidth]{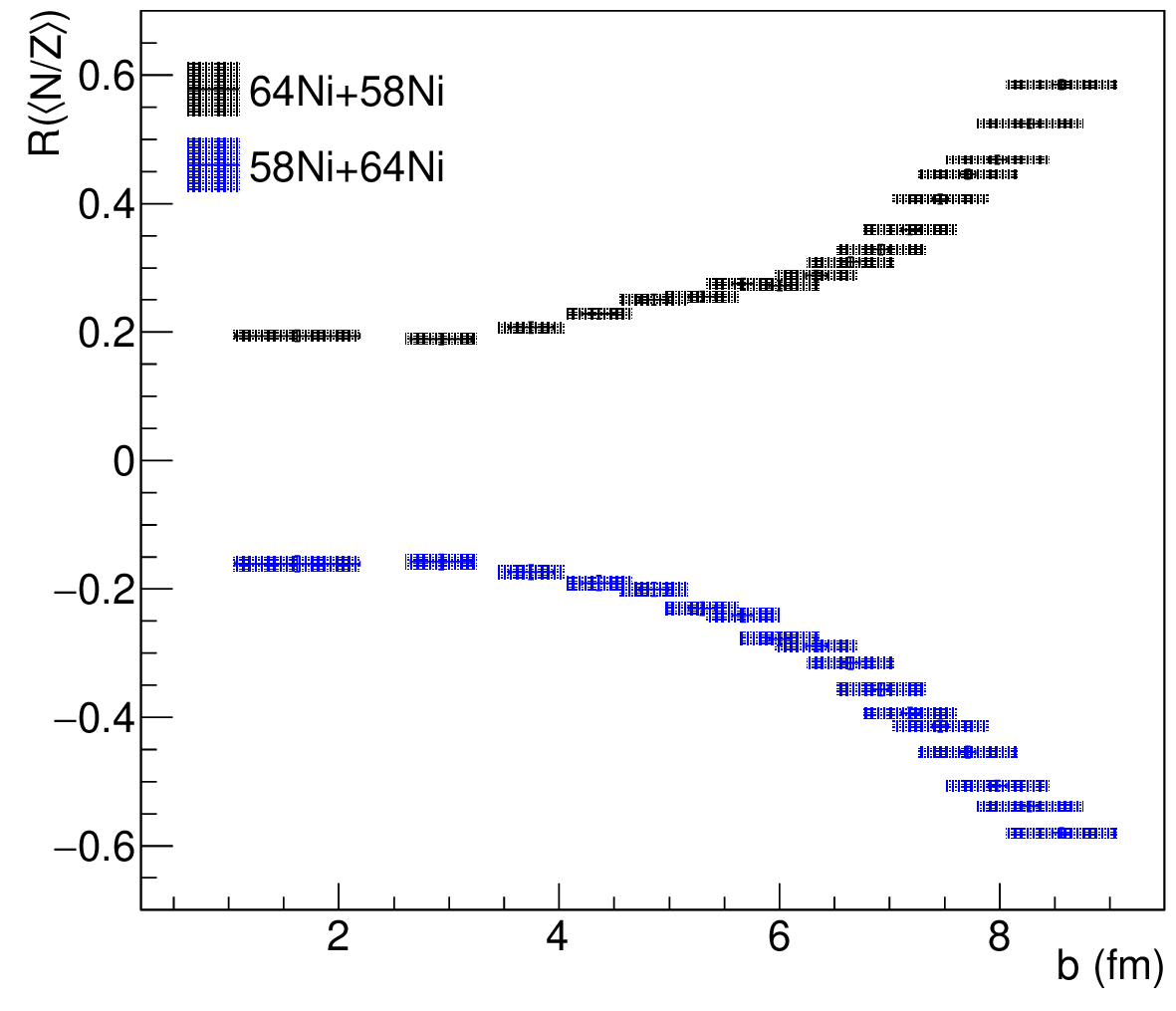}
     \caption{Isospin transport ratio calculated with the $\langle N/Z\rangle$ of the heaviest forward fragment (QP remnant) as a function of the impact parameter $b$. Statistical errors are plotted on the $y$-axis. The error associated with the extraction of the $b_0$ parameter is reported on the $x$-axis.}
     \label{fig:ratioNsuZ_QP}
 \end{figure}
 Finally, the $\langle N/Z \rangle$ plots shown in Fig.~\ref{fig:NsuZ_QP} can be used to compute the isospin transport ratio $R(\langle N/Z \rangle)$ as a function of the impact parameter. The obtained $R(\langle N/Z \rangle)$ vs $b$ for the two asymmetric systems are in Fig.~\ref{fig:ratioNsuZ_QP}, where vertical error bars denote statistical errors, and horizontal error bars should be interpreted in the same way as in Fig.~\ref{fig:NsuZ_QP}.
 Here the regular trend towards equilibration for increasing centrality is further highlighted, and is in general agreement with other observations available in the literature \cite{Sun2010,Camaiani2021}, including our previous results of Refs.~\cite{Ciampi2022,Ciampi2023}, as detailed in Appendix~\ref{app:comparison}. 
 The system dependent path towards equilibrium visible in Fig.~\ref{fig:NsuZ_QP} has now disappeared thanks to the use of the ratio.
 As anticipated, due to our limitations in terms of isotopic identification, with this analysis we are not able to efficiently inspect less dissipative collisions, which justifies the fact that the values $R(\langle N/Z \rangle)= \pm1$ are not reached by the two branches of the ratio. On the other hand, we note that even for the most central collisions considered here, the two mixed systems do not asymptotically tend toward the same $R(\langle N/Z \rangle)$ value, and the condition of full equilibration is thus never obtained: this may signal that the projectile-target interaction timescale between medium-sized nuclei as those here inspected is never sufficient for full $N/Z$ equilibration to occur. Of course the degree of equilibration is also expected to be determined by the density dependence of the nuclear symmetry energy. A similar behavior of $R(\langle N/Z \rangle)$ with $b$ has been found in transport model predictions with different nEoS for the same reactions \cite{Mallik2022}.  Moreover, in  Ref.~\cite{Mallik2022}, calculations using the BUU@VECC-McGill transport code employing the nEoS metamodeling technique of Ref.~\cite{Margueron2018I} have shown that the isospin transport ratio for a given impact parameter is particularly sensitive to different nEoS parametrizations. A thorough comparison of our data with that model's predictions will be the focus of a separate article.
 
 To summarize, the model-independent isospin transport ratio hereby presented has been evaluated based on the direct measurement of the isospin content of a fragment unambiguously generated from the QP, generally regarded as the best probe for similar investigations \cite{Napolitani2010,Coupland2011,Mallik2022}, combined with a detailed global characterization of the event: to our knowledge, this result represents the most comprehensive experimental assessment to date of the isospin diffusion effect across varying reaction centralities.

\section{Conclusions}
In this work, we have reported an experimental measurement of the isospin equilibration phenomenon in $^{58,64}$Ni+$^{58,64}$Ni collisions at 32~MeV/nucleon, obtained by means of a completely model-independent analysis. In order to achieve this goal, we have made use of two different datasets sharing many common characteristics, but providing complementary information.

The first one, acquired with the INDRA apparatus, has been essentially employed to assess the relationship between the selected centrality-related global observable and the impact parameter: the model-independent method proposed in Ref.~\cite{Frankland2021} has been applied, thus obtaining both the average relative behavior the two quantities and the associated intrinsic fluctuations. The centrality observable, namely the multiplicity $M$ of identified and unidentified particles in INDRA rings 6 to 17, has been carefully selected to ensure a comparable behavior in the two datasets.

The isospin-related information on the fragment originated as the remnant of the QP that emerges from the reaction has been retrieved from a full dataset including the four systems $^{58,64}$Ni+$^{58,64}$Ni at 32~MeV/nucleon, recently acquired with the coupled INDRA-FAZIA apparatus. This complete set of reactions allows to exploit the isospin transport ratio, a powerful method that helps to effectively bring out the degree of isospin equilibration, minimizing the impact of other mechanisms, such as statistical de-excitation, that can easily wash out the underlying dependence from the details of the symmetry energy of the nEoS.

In view of producing a general result and facilitate the comparison with a variety of model predictions, we have avoided a strictly exclusive analysis, simply selecting the largest forward-oriented fragment as QP remnant.
A basic ``completeness'' condition is imposed in order to exclude the events in which the heaviest fragment may have not been detected.
The $\langle N/Z\rangle$ of the QP remnant is employed as isospin observable, and its evolution with the reaction centrality is investigated. 
In this phase, we have implemented a random impact parameter assignment to each event based on the full $b$ distribution associated with its measured multiplicity, so to take into account the intrinsic fluctuations within the $b$ vs $M_\text{resc}$ correlation previously extracted. It should be noted that a similar treatment of the impact parameter is important for comparing data with the results of transport models, particularly when the centrality-related observables are not well reproduced.
The resulting $\langle N/Z\rangle$ vs $b$ for the four reactions have been finally employed to compute the isospin transport ratio $R(\langle N/Z \rangle)$ for the two mixed systems as a function of the impact parameter, that reveals a very clear evolution towards isospin equilibration for increasing centrality: this result, as shown in Fig.~\ref{fig:literature_comparison} in Appendix~\ref{app:comparison} is fully consistent with the results of our previous analyses \cite{Ciampi2022,Ciampi2023}.

We stress that, thanks to the advantageous properties of the isospin transport ratio, this model-independent experimental measurement of isospin diffusion can represent a robust benchmark to test the performance of transport models and to gain some insight on the nEoS behavior for sub- to saturation densities. Some ongoing comparisons with calculations for primary fragments carried out with the BUU@VECC-McGill transport code \cite{Mallik2022}, assuming different parametrizations of the nEoS \cite{Margueron2018I}, indicate that this approach is effective in preserving the original transport model sensitivity to the details of the $E_\text{sym}$.
In this respect, it is of significant interest providing similar measurements for different reactions, varying, e.g., the initial isospin imbalance between the colliding partners and/or the beam energy: this would allow to check and validate the model predictions for different conditions explored by the interacting system, also simultaneously testing different features of the transport models which are currently under investigation \cite{Wolter2022}.

 \begin{acknowledgments} 
 This work was supported by the National Research Foundation of Korea (NRF; Grant No. 2018R1A5A1025563), and by the Institute of Basic Science, Republic of Korea (IBS; Grant No.~IBS-R031-D1).
 We acknowledge support from Région Normandie under Réseau d’Intérêt Normand FIDNEOS (RIN/FIDNEOS).
 Many thanks are due to the accelerator staff of GANIL for delivering a very good quality beam and to the technical staff for the continuous support.
\end{acknowledgments}

\appendix
\section{Comparison with previous results}
\label{app:comparison}
As introduced, the effect of isospin diffusion has been already evidenced in the INDRA-FAZIA dataset here employed: such analyses are the subject of Refs.~\cite{Ciampi2022,Ciampi2023}.

Our previous works focused on different goals and hence adopted a different methodology.
In Ref.~\cite{Ciampi2022}, we carried out a comparison of the degree of isospin equilibration reached in semiperipheral to peripheral collisions at two different beam energies in the Fermi regime, carefully selecting the QP evaporation channel: by employing two different isospin-sensitive observables, one associated with the QP remnant and the other with its evaporative emissions, we coherently showed that more equilibration is reached for the lower beam energy.
In Ref.~\cite{Ciampi2023}, the results obtained for the main QP evaporation channel have been compared to what obtained for the less populated QP breakup channel, finding a stronger equilibration associated with the latter outcome.
In both cases, a more exclusive analysis has been carried out, in order to isolate the isospin behavior associated with specific output channels. The reaction centrality has been evaluated exploiting a QP-related characteristics, namely the reduced QP momentum $p_\text{red}=p^\text{QP}_z/p_\text{beam}$ \cite{Camaiani2021}, finally converted into the reduced impact parameter $b_\text{red}=b/b_\text{grazing}$ by using AMD+\textsc{Gemini++} simulations \cite{Ono1992,Charity2010} of the reactions.

On the other hand, the aim of the analysis hereby presented is to provide a general result without imposing conditions that might bias further comparisons with theoretical calculations. For the same reason, special care has been dedicated to the reconstruction of the impact parameter which, although not entirely free of assumptions, does not depend on any specific transport model prediction. Moreover, the less exclusive selections, together with the different choice of centrality observable, allowed to extend the analysis to more central collisions.

\begin{figure}[]
    \centering
    \includegraphics[width=\columnwidth]{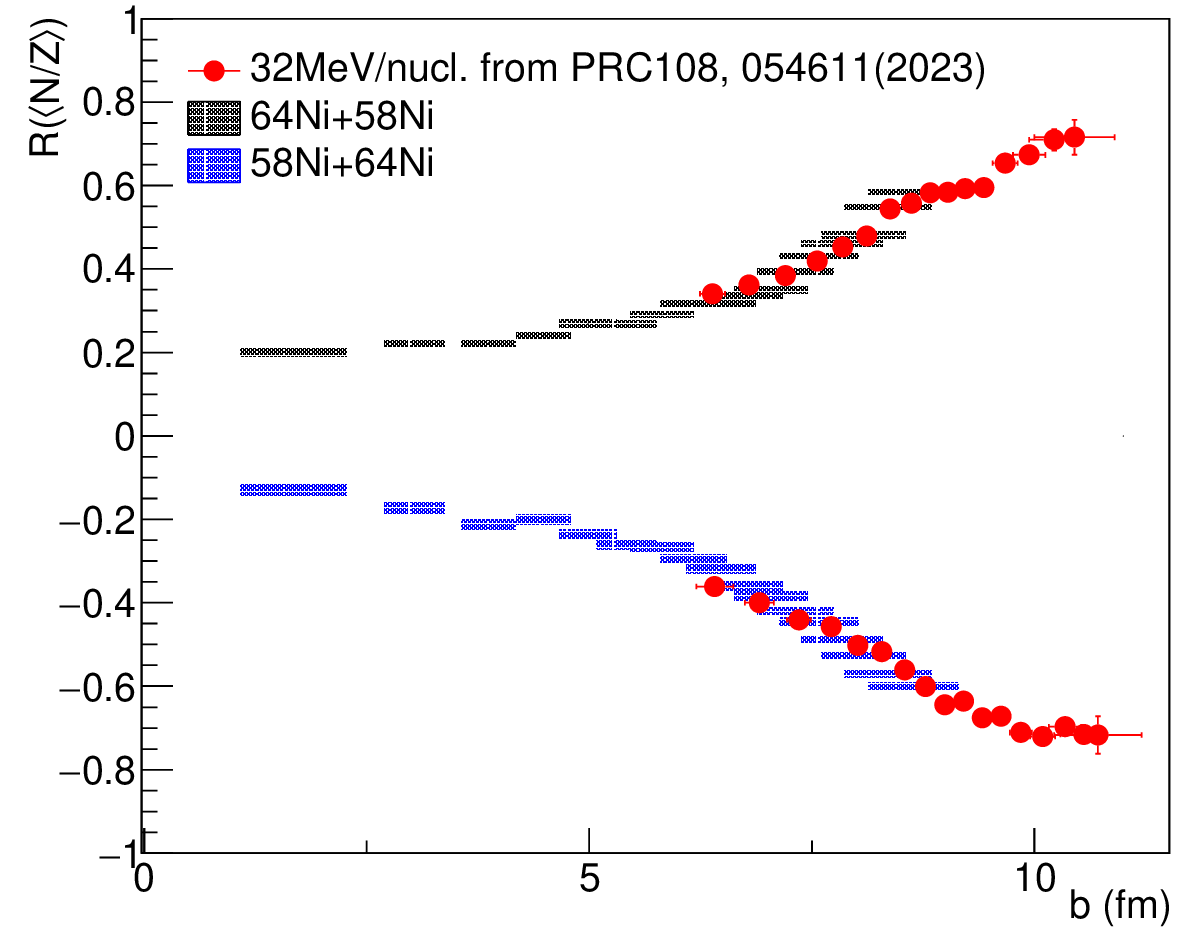}
    \caption{Comparison between the isospin transport ratio calculated from the $\langle N/Z\rangle$ of the QP remnant in the evaporation channel, reported as a function of the impact parameter as obtained with the centrality treatment of the present work and as in the already published results of Ref.~\cite{Ciampi2023}. The results of Ref.~\cite{Ciampi2023} are reported with red markers as a function of $b$, obtained by rescaling $b_\text{red}$ using the average $b_\text{grazing}$ value of 10.8~fm. The results from the new centrality treatment are reported with shaded rectangles representing the global uncertainties as explained in Sec.~\ref{ssec:final_isospin_transport}. }
    \label{fig:literature_comparison}
\end{figure}

It is nevertheless interesting comparing the results of this work with the previously published ones: this evaluation is done focusing on the most populated channel, i.e. the binary one associated with QP evaporation. The analysis and the centrality treatment described in Sec.~\ref{sec:isospin_transport_ratio} are now carried out with the only additional requirement to have only one heavy fragment (i.e., with $Z>5$) in the forward hemisphere, in order to exclude events associated with QP breakups. At variance with Refs.~\cite{Ciampi2022,Ciampi2023}, here we do not require a minimum $Z_\text{QP}=15$, in order to still consider very damped events and show the global result extended to the most central collisions. The corresponding isospin transport ratio is shown with shaded rectangles in Fig.~\ref{fig:literature_comparison}: with respect to the outcome of the more inclusive analysis of Fig.~\ref{fig:ratioNsuZ_QP}, we generally observe a very small shift towards a less equilibrated condition, which is coherent with our result of Ref.~\cite{Ciampi2023} since the QP breakups are now strongly suppressed.
The red markers in Fig.~\ref{fig:literature_comparison} show the results published in Refs.~\cite{Ciampi2022,Ciampi2023} for the QP evaporation channel in semiperipheral to peripheral collisions, where we note that the impact parameter has been extracted from AMD+\textsc{Gemini++} simulations. In the $b$ interval in common between the two analyses, we observe a very good agreement within error bars, supporting the validity of the model-dependent centrality transformation used in the previous works. In conclusion, we highlight the overall consistency of the findings obtained from this first INDRA-FAZIA dataset, even when different analytical approaches are applied.

 %\bibliography{E789_BUUcomparison}

\begin{thebibliography}{0}%
\makeatletter
\providecommand \@ifxundefined [1]{%
 \@ifx{#1\undefined}
}%
\providecommand \@ifnum [1]{%
 \ifnum #1\expandafter \@firstoftwo
 \else \expandafter \@secondoftwo
 \fi
}%
\providecommand \@ifx [1]{%
 \ifx #1\expandafter \@firstoftwo
 \else \expandafter \@secondoftwo
 \fi
}%
\providecommand \natexlab [1]{#1}%
\providecommand \enquote  [1]{``#1''}%
\providecommand \bibnamefont  [1]{#1}%
\providecommand \bibfnamefont [1]{#1}%
\providecommand \citenamefont [1]{#1}%
\providecommand \href@noop [0]{\@secondoftwo}%
\providecommand \href [0]{\begingroup \@sanitize@url \@href}%
\providecommand \@href[1]{\@@startlink{#1}\@@href}%
\providecommand \@@href[1]{\endgroup#1\@@endlink}%
\providecommand \@sanitize@url [0]{\catcode `\\12\catcode `\$12\catcode
  `\&12\catcode `\#12\catcode `\^12\catcode `\_12\catcode `\%12\relax}%
\providecommand \@@startlink[1]{}%
\providecommand \@@endlink[0]{}%
\providecommand \url  [0]{\begingroup\@sanitize@url \@url }%
\providecommand \@url [1]{\endgroup\@href {#1}{\urlprefix }}%
\providecommand \urlprefix  [0]{URL }%
\providecommand \Eprint [0]{\href }%
\providecommand \doibase [0]{https://doi.org/}%
\providecommand \selectlanguage [0]{\@gobble}%
\providecommand \bibinfo  [0]{\@secondoftwo}%
\providecommand \bibfield  [0]{\@secondoftwo}%
\providecommand \translation [1]{[#1]}%
\providecommand \BibitemOpen [0]{}%
\providecommand \bibitemStop [0]{}%
\providecommand \bibitemNoStop [0]{.\EOS\space}%
\providecommand \EOS [0]{\spacefactor3000\relax}%
\providecommand \BibitemShut  [1]{\csname bibitem#1\endcsname}%
\let\auto@bib@innerbib\@empty
%</preamble>
\end{thebibliography}%


\begin{thebibliography}{55}%
\makeatletter
\providecommand \@ifxundefined [1]{%
 \@ifx{#1\undefined}
}%
\providecommand \@ifnum [1]{%
 \ifnum #1\expandafter \@firstoftwo
 \else \expandafter \@secondoftwo
 \fi
}%
\providecommand \@ifx [1]{%
 \ifx #1\expandafter \@firstoftwo
 \else \expandafter \@secondoftwo
 \fi
}%
\providecommand \natexlab [1]{#1}%
\providecommand \enquote  [1]{``#1''}%
\providecommand \bibnamefont  [1]{#1}%
\providecommand \bibfnamefont [1]{#1}%
\providecommand \citenamefont [1]{#1}%
\providecommand \href@noop [0]{\@secondoftwo}%
\providecommand \href [0]{\begingroup \@sanitize@url \@href}%
\providecommand \@href[1]{\@@startlink{#1}\@@href}%
\providecommand \@@href[1]{\endgroup#1\@@endlink}%
\providecommand \@sanitize@url [0]{\catcode `\\12\catcode `\$12\catcode
  `\&12\catcode `\#12\catcode `\^12\catcode `\_12\catcode `\%12\relax}%
\providecommand \@@startlink[1]{}%
\providecommand \@@endlink[0]{}%
\providecommand \url  [0]{\begingroup\@sanitize@url \@url }%
\providecommand \@url [1]{\endgroup\@href {#1}{\urlprefix }}%
\providecommand \urlprefix  [0]{URL }%
\providecommand \Eprint [0]{\href }%
\providecommand \doibase [0]{https://doi.org/}%
\providecommand \selectlanguage [0]{\@gobble}%
\providecommand \bibinfo  [0]{\@secondoftwo}%
\providecommand \bibfield  [0]{\@secondoftwo}%
\providecommand \translation [1]{[#1]}%
\providecommand \BibitemOpen [0]{}%
\providecommand \bibitemStop [0]{}%
\providecommand \bibitemNoStop [0]{.\EOS\space}%
\providecommand \EOS [0]{\spacefactor3000\relax}%
\providecommand \BibitemShut  [1]{\csname bibitem#1\endcsname}%
\let\auto@bib@innerbib\@empty
%</preamble>
\bibitem [{\citenamefont {Sorensen}\ \emph {et~al.}(2024)\citenamefont
  {Sorensen}, \citenamefont {Agarwal}, \citenamefont {Brown}, \citenamefont
  {Chajecki}, \citenamefont {Danielewicz}, \citenamefont {Drischler},
  \citenamefont {Gandolfi}, \citenamefont {Holt}, \citenamefont {Kaminski},
  \citenamefont {Ko}, \citenamefont {Kumar}, \citenamefont {Li}, \citenamefont
  {Lynch}, \citenamefont {McIntosh}, \citenamefont {Newton}, \citenamefont
  {Pratt}, \citenamefont {Savchuk}, \citenamefont {Stefaniak}, \citenamefont
  {Tews}, \citenamefont {Tsang}, \citenamefont {Vogt}, \citenamefont {Wolter},
  \citenamefont {Zbroszczyk}, \citenamefont {Abbasi}, \citenamefont {Aichelin},
  \citenamefont {Andronic}, \citenamefont {Bass}, \citenamefont {Becattini},
  \citenamefont {Blaschke}, \citenamefont {Bleicher}, \citenamefont {Blume},
  \citenamefont {Bratkovskaya}, \citenamefont {Brown}, \citenamefont {Brown},
  \citenamefont {Camaiani}, \citenamefont {Casini}, \citenamefont
  {Chatziioannou}, \citenamefont {Chbihi}, \citenamefont {Colonna},
  \citenamefont {Cozma}, \citenamefont {Dexheimer}, \citenamefont {Dong},
  \citenamefont {Dore}, \citenamefont {Du}, \citenamefont {Dueñas},
  \citenamefont {Elfner}, \citenamefont {Florkowski}, \citenamefont {Fujimoto},
  \citenamefont {Furnstahl}, \citenamefont {Gade}, \citenamefont {Galatyuk},
  \citenamefont {Gale}, \citenamefont {Geurts}, \citenamefont {Gramegna},
  \citenamefont {Grozdanov}, \citenamefont {Hagel}, \citenamefont {Harris},
  \citenamefont {Haxton}, \citenamefont {Heinz}, \citenamefont {Heller},
  \citenamefont {Hen}, \citenamefont {Hergert}, \citenamefont {Herrmann},
  \citenamefont {Huang}, \citenamefont {Huang}, \citenamefont {Ikeno},
  \citenamefont {Inghirami}, \citenamefont {Jankowski}, \citenamefont {Jia},
  \citenamefont {Jiménez}, \citenamefont {Kapusta}, \citenamefont {Kardan},
  \citenamefont {Karpenko}, \citenamefont {Keane}, \citenamefont {Kharzeev},
  \citenamefont {Kugler}, \citenamefont {{Le Fèvre}}, \citenamefont {Lee},
  \citenamefont {Liu}, \citenamefont {Lisa}, \citenamefont {Llope},
  \citenamefont {Lombardo}, \citenamefont {Lorenz}, \citenamefont {Marchi},
  \citenamefont {McLerran}, \citenamefont {Mosel}, \citenamefont {Motornenko},
  \citenamefont {Müller}, \citenamefont {Napolitani}, \citenamefont
  {Natowitz}, \citenamefont {Nazarewicz}, \citenamefont {Noronha},
  \citenamefont {Noronha-Hostler}, \citenamefont {Odyniec}, \citenamefont
  {Papakonstantinou}, \citenamefont {Paulínyová}, \citenamefont
  {Piekarewicz}, \citenamefont {Pisarski}, \citenamefont {Plumberg},
  \citenamefont {Prakash}, \citenamefont {Randrup}, \citenamefont {Ratti},
  \citenamefont {Rau}, \citenamefont {Reddy}, \citenamefont {Schmidt},
  \citenamefont {Russotto}, \citenamefont {Ryblewski}, \citenamefont
  {Schäfer}, \citenamefont {Schenke}, \citenamefont {Sen}, \citenamefont
  {Senger}, \citenamefont {Seto}, \citenamefont {Shen}, \citenamefont
  {Sherrill}, \citenamefont {Singh}, \citenamefont {Skokov}, \citenamefont
  {Spaliński}, \citenamefont {Steinheimer}, \citenamefont {Stephanov},
  \citenamefont {Stroth}, \citenamefont {Sturm}, \citenamefont {Sun},
  \citenamefont {Tang}, \citenamefont {Torrieri}, \citenamefont {Trautmann},
  \citenamefont {Verde}, \citenamefont {Vovchenko}, \citenamefont {Wada},
  \citenamefont {Wang}, \citenamefont {Wang}, \citenamefont {Werner},
  \citenamefont {Xu}, \citenamefont {Xu}, \citenamefont {Yee}, \citenamefont
  {Yennello},\ and\ \citenamefont {Yin}}]{Sorensen2024}%
  \BibitemOpen
  \bibfield  {author} {\bibinfo {author} {\bibfnamefont {A.}~\bibnamefont
  {Sorensen}}, \bibinfo {author} {\bibfnamefont {K.}~\bibnamefont {Agarwal}},
  \bibinfo {author} {\bibfnamefont {K.~W.}\ \bibnamefont {Brown}}, \bibinfo
  {author} {\bibfnamefont {Z.}~\bibnamefont {Chajecki}}, \bibinfo {author}
  {\bibfnamefont {P.}~\bibnamefont {Danielewicz}}, \bibinfo {author}
  {\bibfnamefont {C.}~\bibnamefont {Drischler}}, \bibinfo {author}
  {\bibfnamefont {S.}~\bibnamefont {Gandolfi}}, \bibinfo {author}
  {\bibfnamefont {J.~W.}\ \bibnamefont {Holt}}, \bibinfo {author}
  {\bibfnamefont {M.}~\bibnamefont {Kaminski}}, \bibinfo {author}
  {\bibfnamefont {C.-M.}\ \bibnamefont {Ko}}, \bibinfo {author} {\bibfnamefont
  {R.}~\bibnamefont {Kumar}}, \bibinfo {author} {\bibfnamefont {B.-A.}\
  \bibnamefont {Li}}, \bibinfo {author} {\bibfnamefont {W.~G.}\ \bibnamefont
  {Lynch}}, \bibinfo {author} {\bibfnamefont {A.~B.}\ \bibnamefont {McIntosh}},
  \bibinfo {author} {\bibfnamefont {W.~G.}\ \bibnamefont {Newton}}, \bibinfo
  {author} {\bibfnamefont {S.}~\bibnamefont {Pratt}}, \bibinfo {author}
  {\bibfnamefont {O.}~\bibnamefont {Savchuk}}, \bibinfo {author} {\bibfnamefont
  {M.}~\bibnamefont {Stefaniak}}, \bibinfo {author} {\bibfnamefont
  {I.}~\bibnamefont {Tews}}, \bibinfo {author} {\bibfnamefont {M.~B.}\
  \bibnamefont {Tsang}}, \bibinfo {author} {\bibfnamefont {R.}~\bibnamefont
  {Vogt}}, \bibinfo {author} {\bibfnamefont {H.}~\bibnamefont {Wolter}},
  \bibinfo {author} {\bibfnamefont {H.}~\bibnamefont {Zbroszczyk}}, \bibinfo
  {author} {\bibfnamefont {N.}~\bibnamefont {Abbasi}}, \bibinfo {author}
  {\bibfnamefont {J.}~\bibnamefont {Aichelin}}, \bibinfo {author}
  {\bibfnamefont {A.}~\bibnamefont {Andronic}}, \bibinfo {author}
  {\bibfnamefont {S.~A.}\ \bibnamefont {Bass}}, \bibinfo {author}
  {\bibfnamefont {F.}~\bibnamefont {Becattini}}, \bibinfo {author}
  {\bibfnamefont {D.}~\bibnamefont {Blaschke}}, \bibinfo {author}
  {\bibfnamefont {M.}~\bibnamefont {Bleicher}}, \bibinfo {author}
  {\bibfnamefont {C.}~\bibnamefont {Blume}}, \bibinfo {author} {\bibfnamefont
  {E.}~\bibnamefont {Bratkovskaya}}, \bibinfo {author} {\bibfnamefont {B.~A.}\
  \bibnamefont {Brown}}, \bibinfo {author} {\bibfnamefont {D.~A.}\ \bibnamefont
  {Brown}}, \bibinfo {author} {\bibfnamefont {A.}~\bibnamefont {Camaiani}},
  \bibinfo {author} {\bibfnamefont {G.}~\bibnamefont {Casini}}, \bibinfo
  {author} {\bibfnamefont {K.}~\bibnamefont {Chatziioannou}}, \bibinfo {author}
  {\bibfnamefont {A.}~\bibnamefont {Chbihi}}, \bibinfo {author} {\bibfnamefont
  {M.}~\bibnamefont {Colonna}}, \bibinfo {author} {\bibfnamefont {M.~D.}\
  \bibnamefont {Cozma}}, \bibinfo {author} {\bibfnamefont {V.}~\bibnamefont
  {Dexheimer}}, \bibinfo {author} {\bibfnamefont {X.}~\bibnamefont {Dong}},
  \bibinfo {author} {\bibfnamefont {T.}~\bibnamefont {Dore}}, \bibinfo {author}
  {\bibfnamefont {L.}~\bibnamefont {Du}}, \bibinfo {author} {\bibfnamefont
  {J.~A.}\ \bibnamefont {Dueñas}}, \bibinfo {author} {\bibfnamefont
  {H.}~\bibnamefont {Elfner}}, \bibinfo {author} {\bibfnamefont
  {W.}~\bibnamefont {Florkowski}}, \bibinfo {author} {\bibfnamefont
  {Y.}~\bibnamefont {Fujimoto}}, \bibinfo {author} {\bibfnamefont {R.~J.}\
  \bibnamefont {Furnstahl}}, \bibinfo {author} {\bibfnamefont {A.}~\bibnamefont
  {Gade}}, \bibinfo {author} {\bibfnamefont {T.}~\bibnamefont {Galatyuk}},
  \bibinfo {author} {\bibfnamefont {C.}~\bibnamefont {Gale}}, \bibinfo {author}
  {\bibfnamefont {F.}~\bibnamefont {Geurts}}, \bibinfo {author} {\bibfnamefont
  {F.}~\bibnamefont {Gramegna}}, \bibinfo {author} {\bibfnamefont
  {S.}~\bibnamefont {Grozdanov}}, \bibinfo {author} {\bibfnamefont
  {K.}~\bibnamefont {Hagel}}, \bibinfo {author} {\bibfnamefont {S.~P.}\
  \bibnamefont {Harris}}, \bibinfo {author} {\bibfnamefont {W.}~\bibnamefont
  {Haxton}}, \bibinfo {author} {\bibfnamefont {U.}~\bibnamefont {Heinz}},
  \bibinfo {author} {\bibfnamefont {M.~P.}\ \bibnamefont {Heller}}, \bibinfo
  {author} {\bibfnamefont {O.}~\bibnamefont {Hen}}, \bibinfo {author}
  {\bibfnamefont {H.}~\bibnamefont {Hergert}}, \bibinfo {author} {\bibfnamefont
  {N.}~\bibnamefont {Herrmann}}, \bibinfo {author} {\bibfnamefont {H.~Z.}\
  \bibnamefont {Huang}}, \bibinfo {author} {\bibfnamefont {X.-G.}\ \bibnamefont
  {Huang}}, \bibinfo {author} {\bibfnamefont {N.}~\bibnamefont {Ikeno}},
  \bibinfo {author} {\bibfnamefont {G.}~\bibnamefont {Inghirami}}, \bibinfo
  {author} {\bibfnamefont {J.}~\bibnamefont {Jankowski}}, \bibinfo {author}
  {\bibfnamefont {J.}~\bibnamefont {Jia}}, \bibinfo {author} {\bibfnamefont
  {J.~C.}\ \bibnamefont {Jiménez}}, \bibinfo {author} {\bibfnamefont
  {J.}~\bibnamefont {Kapusta}}, \bibinfo {author} {\bibfnamefont
  {B.}~\bibnamefont {Kardan}}, \bibinfo {author} {\bibfnamefont
  {I.}~\bibnamefont {Karpenko}}, \bibinfo {author} {\bibfnamefont
  {D.}~\bibnamefont {Keane}}, \bibinfo {author} {\bibfnamefont
  {D.}~\bibnamefont {Kharzeev}}, \bibinfo {author} {\bibfnamefont
  {A.}~\bibnamefont {Kugler}}, \bibinfo {author} {\bibfnamefont
  {A.}~\bibnamefont {{Le Fèvre}}}, \bibinfo {author} {\bibfnamefont
  {D.}~\bibnamefont {Lee}}, \bibinfo {author} {\bibfnamefont {H.}~\bibnamefont
  {Liu}}, \bibinfo {author} {\bibfnamefont {M.~A.}\ \bibnamefont {Lisa}},
  \bibinfo {author} {\bibfnamefont {W.~J.}\ \bibnamefont {Llope}}, \bibinfo
  {author} {\bibfnamefont {I.}~\bibnamefont {Lombardo}}, \bibinfo {author}
  {\bibfnamefont {M.}~\bibnamefont {Lorenz}}, \bibinfo {author} {\bibfnamefont
  {T.}~\bibnamefont {Marchi}}, \bibinfo {author} {\bibfnamefont
  {L.}~\bibnamefont {McLerran}}, \bibinfo {author} {\bibfnamefont
  {U.}~\bibnamefont {Mosel}}, \bibinfo {author} {\bibfnamefont
  {A.}~\bibnamefont {Motornenko}}, \bibinfo {author} {\bibfnamefont
  {B.}~\bibnamefont {Müller}}, \bibinfo {author} {\bibfnamefont
  {P.}~\bibnamefont {Napolitani}}, \bibinfo {author} {\bibfnamefont {J.~B.}\
  \bibnamefont {Natowitz}}, \bibinfo {author} {\bibfnamefont {W.}~\bibnamefont
  {Nazarewicz}}, \bibinfo {author} {\bibfnamefont {J.}~\bibnamefont {Noronha}},
  \bibinfo {author} {\bibfnamefont {J.}~\bibnamefont {Noronha-Hostler}},
  \bibinfo {author} {\bibfnamefont {G.}~\bibnamefont {Odyniec}}, \bibinfo
  {author} {\bibfnamefont {P.}~\bibnamefont {Papakonstantinou}}, \bibinfo
  {author} {\bibfnamefont {Z.}~\bibnamefont {Paulínyová}}, \bibinfo {author}
  {\bibfnamefont {J.}~\bibnamefont {Piekarewicz}}, \bibinfo {author}
  {\bibfnamefont {R.~D.}\ \bibnamefont {Pisarski}}, \bibinfo {author}
  {\bibfnamefont {C.}~\bibnamefont {Plumberg}}, \bibinfo {author}
  {\bibfnamefont {M.}~\bibnamefont {Prakash}}, \bibinfo {author} {\bibfnamefont
  {J.}~\bibnamefont {Randrup}}, \bibinfo {author} {\bibfnamefont
  {C.}~\bibnamefont {Ratti}}, \bibinfo {author} {\bibfnamefont
  {P.}~\bibnamefont {Rau}}, \bibinfo {author} {\bibfnamefont {S.}~\bibnamefont
  {Reddy}}, \bibinfo {author} {\bibfnamefont {H.-R.}\ \bibnamefont {Schmidt}},
  \bibinfo {author} {\bibfnamefont {P.}~\bibnamefont {Russotto}}, \bibinfo
  {author} {\bibfnamefont {R.}~\bibnamefont {Ryblewski}}, \bibinfo {author}
  {\bibfnamefont {A.}~\bibnamefont {Schäfer}}, \bibinfo {author}
  {\bibfnamefont {B.}~\bibnamefont {Schenke}}, \bibinfo {author} {\bibfnamefont
  {S.}~\bibnamefont {Sen}}, \bibinfo {author} {\bibfnamefont {P.}~\bibnamefont
  {Senger}}, \bibinfo {author} {\bibfnamefont {R.}~\bibnamefont {Seto}},
  \bibinfo {author} {\bibfnamefont {C.}~\bibnamefont {Shen}}, \bibinfo {author}
  {\bibfnamefont {B.}~\bibnamefont {Sherrill}}, \bibinfo {author}
  {\bibfnamefont {M.}~\bibnamefont {Singh}}, \bibinfo {author} {\bibfnamefont
  {V.}~\bibnamefont {Skokov}}, \bibinfo {author} {\bibfnamefont
  {M.}~\bibnamefont {Spaliński}}, \bibinfo {author} {\bibfnamefont
  {J.}~\bibnamefont {Steinheimer}}, \bibinfo {author} {\bibfnamefont
  {M.}~\bibnamefont {Stephanov}}, \bibinfo {author} {\bibfnamefont
  {J.}~\bibnamefont {Stroth}}, \bibinfo {author} {\bibfnamefont
  {C.}~\bibnamefont {Sturm}}, \bibinfo {author} {\bibfnamefont {K.-J.}\
  \bibnamefont {Sun}}, \bibinfo {author} {\bibfnamefont {A.}~\bibnamefont
  {Tang}}, \bibinfo {author} {\bibfnamefont {G.}~\bibnamefont {Torrieri}},
  \bibinfo {author} {\bibfnamefont {W.}~\bibnamefont {Trautmann}}, \bibinfo
  {author} {\bibfnamefont {G.}~\bibnamefont {Verde}}, \bibinfo {author}
  {\bibfnamefont {V.}~\bibnamefont {Vovchenko}}, \bibinfo {author}
  {\bibfnamefont {R.}~\bibnamefont {Wada}}, \bibinfo {author} {\bibfnamefont
  {F.}~\bibnamefont {Wang}}, \bibinfo {author} {\bibfnamefont {G.}~\bibnamefont
  {Wang}}, \bibinfo {author} {\bibfnamefont {K.}~\bibnamefont {Werner}},
  \bibinfo {author} {\bibfnamefont {N.}~\bibnamefont {Xu}}, \bibinfo {author}
  {\bibfnamefont {Z.}~\bibnamefont {Xu}}, \bibinfo {author} {\bibfnamefont
  {H.-U.}\ \bibnamefont {Yee}}, \bibinfo {author} {\bibfnamefont
  {S.}~\bibnamefont {Yennello}},\ and\ \bibinfo {author} {\bibfnamefont
  {Y.}~\bibnamefont {Yin}},\ }\href
  {https://doi.org/https://doi.org/10.1016/j.ppnp.2023.104080} {\bibfield
  {journal} {\bibinfo  {journal} {Progress in Particle and Nuclear Physics}\
  }\textbf {\bibinfo {volume} {134}},\ \bibinfo {pages} {104080} (\bibinfo
  {year} {2024})}\BibitemShut {NoStop}%
\bibitem [{\citenamefont {Roca-Maza}\ and\ \citenamefont
  {Paar}(2018)}]{RocaMaza2018}%
  \BibitemOpen
  \bibfield  {author} {\bibinfo {author} {\bibfnamefont {X.}~\bibnamefont
  {Roca-Maza}}\ and\ \bibinfo {author} {\bibfnamefont {N.}~\bibnamefont
  {Paar}},\ }\href {https://doi.org/https://doi.org/10.1016/j.ppnp.2018.04.001}
  {\bibfield  {journal} {\bibinfo  {journal} {Progress in Particle and Nuclear
  Physics}\ }\textbf {\bibinfo {volume} {101}},\ \bibinfo {pages} {96}
  (\bibinfo {year} {2018})}\BibitemShut {NoStop}%
\bibitem [{\citenamefont {Colonna}(2020)}]{Colonna2020}%
  \BibitemOpen
  \bibfield  {author} {\bibinfo {author} {\bibfnamefont {M.}~\bibnamefont
  {Colonna}},\ }\href
  {https://doi.org/https://doi.org/10.1016/j.ppnp.2020.103775} {\bibfield
  {journal} {\bibinfo  {journal} {Progress in Particle and Nuclear Physics}\
  }\textbf {\bibinfo {volume} {113}},\ \bibinfo {pages} {103775} (\bibinfo
  {year} {2020})}\BibitemShut {NoStop}%
\bibitem [{\citenamefont {Yasin}\ \emph {et~al.}(2020)\citenamefont {Yasin},
  \citenamefont {Sch\"afer}, \citenamefont {Arcones},\ and\ \citenamefont
  {Schwenk}}]{Yasin2020}%
  \BibitemOpen
  \bibfield  {author} {\bibinfo {author} {\bibfnamefont {H.}~\bibnamefont
  {Yasin}}, \bibinfo {author} {\bibfnamefont {S.}~\bibnamefont {Sch\"afer}},
  \bibinfo {author} {\bibfnamefont {A.}~\bibnamefont {Arcones}},\ and\ \bibinfo
  {author} {\bibfnamefont {A.}~\bibnamefont {Schwenk}},\ }\href
  {https://doi.org/10.1103/PhysRevLett.124.092701} {\bibfield  {journal}
  {\bibinfo  {journal} {Phys. Rev. Lett.}\ }\textbf {\bibinfo {volume} {124}},\
  \bibinfo {pages} {092701} (\bibinfo {year} {2020})}\BibitemShut {NoStop}%
\bibitem [{\citenamefont {Margueron}\ \emph
  {et~al.}(2018{\natexlab{a}})\citenamefont {Margueron}, \citenamefont
  {Hoffmann~Casali},\ and\ \citenamefont {Gulminelli}}]{Margueron2018II}%
  \BibitemOpen
  \bibfield  {author} {\bibinfo {author} {\bibfnamefont {J.}~\bibnamefont
  {Margueron}}, \bibinfo {author} {\bibfnamefont {R.}~\bibnamefont
  {Hoffmann~Casali}},\ and\ \bibinfo {author} {\bibfnamefont {F.}~\bibnamefont
  {Gulminelli}},\ }\href {https://doi.org/10.1103/PhysRevC.97.025806}
  {\bibfield  {journal} {\bibinfo  {journal} {Phys. Rev. C}\ }\textbf {\bibinfo
  {volume} {97}},\ \bibinfo {pages} {025806} (\bibinfo {year}
  {2018}{\natexlab{a}})}\BibitemShut {NoStop}%
\bibitem [{\citenamefont {Burgio}\ \emph {et~al.}(2021)\citenamefont {Burgio},
  \citenamefont {Schulze}, \citenamefont {Vidaña},\ and\ \citenamefont
  {Wei}}]{Burgio2021}%
  \BibitemOpen
  \bibfield  {author} {\bibinfo {author} {\bibfnamefont {G.}~\bibnamefont
  {Burgio}}, \bibinfo {author} {\bibfnamefont {H.-J.}\ \bibnamefont {Schulze}},
  \bibinfo {author} {\bibfnamefont {I.}~\bibnamefont {Vidaña}},\ and\ \bibinfo
  {author} {\bibfnamefont {J.-B.}\ \bibnamefont {Wei}},\ }\href
  {https://doi.org/https://doi.org/10.1016/j.ppnp.2021.103879} {\bibfield
  {journal} {\bibinfo  {journal} {Progress in Particle and Nuclear Physics}\
  }\textbf {\bibinfo {volume} {120}},\ \bibinfo {pages} {103879} (\bibinfo
  {year} {2021})}\BibitemShut {NoStop}%
\bibitem [{\citenamefont {Lattimer}\ and\ \citenamefont
  {Prakash}(2000)}]{Lattimer2000}%
  \BibitemOpen
  \bibfield  {author} {\bibinfo {author} {\bibfnamefont {J.~M.}\ \bibnamefont
  {Lattimer}}\ and\ \bibinfo {author} {\bibfnamefont {M.}~\bibnamefont
  {Prakash}},\ }\href
  {https://doi.org/https://doi.org/10.1016/S0370-1573(00)00019-3} {\bibfield
  {journal} {\bibinfo  {journal} {Physics Reports}\ }\textbf {\bibinfo {volume}
  {333-334}},\ \bibinfo {pages} {121} (\bibinfo {year} {2000})}\BibitemShut
  {NoStop}%
\bibitem [{\citenamefont {Huth}\ \emph {et~al.}(2022)\citenamefont {Huth},
  \citenamefont {Pang}, \citenamefont {Tews}, \citenamefont {Dietrich},
  \citenamefont {{Le Fèvre}}, \citenamefont {Schwenk}, \citenamefont
  {Trautmann}, \citenamefont {Agarwal}, \citenamefont {Bulla}, \citenamefont
  {Coughlin},\ and\ \citenamefont {{Van Den Broeck}}}]{Huth2022}%
  \BibitemOpen
  \bibfield  {author} {\bibinfo {author} {\bibfnamefont {S.}~\bibnamefont
  {Huth}}, \bibinfo {author} {\bibfnamefont {P.~T.~H.}\ \bibnamefont {Pang}},
  \bibinfo {author} {\bibfnamefont {I.}~\bibnamefont {Tews}}, \bibinfo {author}
  {\bibfnamefont {T.}~\bibnamefont {Dietrich}}, \bibinfo {author}
  {\bibfnamefont {A.}~\bibnamefont {{Le Fèvre}}}, \bibinfo {author}
  {\bibfnamefont {A.}~\bibnamefont {Schwenk}}, \bibinfo {author} {\bibfnamefont
  {W.}~\bibnamefont {Trautmann}}, \bibinfo {author} {\bibfnamefont
  {K.}~\bibnamefont {Agarwal}}, \bibinfo {author} {\bibfnamefont
  {M.}~\bibnamefont {Bulla}}, \bibinfo {author} {\bibfnamefont {M.~W.}\
  \bibnamefont {Coughlin}},\ and\ \bibinfo {author} {\bibfnamefont
  {C.}~\bibnamefont {{Van Den Broeck}}},\ }\href
  {https://doi.org//10.1038/s41586-022-04750-w} {\bibfield  {journal} {\bibinfo
   {journal} {Nature}\ }\textbf {\bibinfo {volume} {606}},\ \bibinfo {pages}
  {276} (\bibinfo {year} {2022})}\BibitemShut {NoStop}%
\bibitem [{\citenamefont {Lynch}\ and\ \citenamefont
  {Tsang}(2022)}]{Lynch2022}%
  \BibitemOpen
  \bibfield  {author} {\bibinfo {author} {\bibfnamefont {W.~G.}\ \bibnamefont
  {Lynch}}\ and\ \bibinfo {author} {\bibfnamefont {M.~B.}\ \bibnamefont
  {Tsang}},\ }\href
  {https://doi.org/https://doi.org/10.1016/j.physletb.2022.137098} {\bibfield
  {journal} {\bibinfo  {journal} {Physics Letters B}\ }\textbf {\bibinfo
  {volume} {830}},\ \bibinfo {pages} {137098} (\bibinfo {year}
  {2022})}\BibitemShut {NoStop}%
\bibitem [{\citenamefont {Pang}\ \emph {et~al.}(2023)\citenamefont {Pang},
  \citenamefont {Dietrich}, \citenamefont {Coughlin}, \citenamefont {Bulla},
  \citenamefont {Tews}, \citenamefont {Almualla}, \citenamefont {Barna},
  \citenamefont {Kiendrebeogo}, \citenamefont {Kunert}, \citenamefont
  {Mansingh}, \citenamefont {Reed}, \citenamefont {Sravan}, \citenamefont
  {Toivonen}, \citenamefont {Antier}, \citenamefont {VandenBerg}, \citenamefont
  {Heinzel}, \citenamefont {Nedora}, \citenamefont {Salehi}, \citenamefont
  {Sharma}, \citenamefont {Somasundaram},\ and\ \citenamefont {Van
  Den~Broeck}}]{Pang2023}%
  \BibitemOpen
  \bibfield  {author} {\bibinfo {author} {\bibfnamefont {P.~T.~H.}\
  \bibnamefont {Pang}}, \bibinfo {author} {\bibfnamefont {T.}~\bibnamefont
  {Dietrich}}, \bibinfo {author} {\bibfnamefont {M.~W.}\ \bibnamefont
  {Coughlin}}, \bibinfo {author} {\bibfnamefont {M.}~\bibnamefont {Bulla}},
  \bibinfo {author} {\bibfnamefont {I.}~\bibnamefont {Tews}}, \bibinfo {author}
  {\bibfnamefont {M.}~\bibnamefont {Almualla}}, \bibinfo {author}
  {\bibfnamefont {T.}~\bibnamefont {Barna}}, \bibinfo {author} {\bibfnamefont
  {R.~W.}\ \bibnamefont {Kiendrebeogo}}, \bibinfo {author} {\bibfnamefont
  {N.}~\bibnamefont {Kunert}}, \bibinfo {author} {\bibfnamefont
  {G.}~\bibnamefont {Mansingh}}, \bibinfo {author} {\bibfnamefont
  {B.}~\bibnamefont {Reed}}, \bibinfo {author} {\bibfnamefont {N.}~\bibnamefont
  {Sravan}}, \bibinfo {author} {\bibfnamefont {A.}~\bibnamefont {Toivonen}},
  \bibinfo {author} {\bibfnamefont {S.}~\bibnamefont {Antier}}, \bibinfo
  {author} {\bibfnamefont {R.~O.}\ \bibnamefont {VandenBerg}}, \bibinfo
  {author} {\bibfnamefont {J.}~\bibnamefont {Heinzel}}, \bibinfo {author}
  {\bibfnamefont {V.}~\bibnamefont {Nedora}}, \bibinfo {author} {\bibfnamefont
  {P.}~\bibnamefont {Salehi}}, \bibinfo {author} {\bibfnamefont
  {R.}~\bibnamefont {Sharma}}, \bibinfo {author} {\bibfnamefont
  {R.}~\bibnamefont {Somasundaram}},\ and\ \bibinfo {author} {\bibfnamefont
  {C.}~\bibnamefont {Van Den~Broeck}},\ }\href
  {https://doi.org/10.1038/s41467-023-43932-6} {\bibfield  {journal} {\bibinfo
  {journal} {Nature Communications}\ }\textbf {\bibinfo {volume} {14}},\
  \bibinfo {pages} {8352} (\bibinfo {year} {2023})}\BibitemShut {NoStop}%
\bibitem [{\citenamefont {Tsang}\ \emph {et~al.}(2024)\citenamefont {Tsang},
  \citenamefont {Tsang}, \citenamefont {Lynch}, \citenamefont {Kumar},\ and\
  \citenamefont {Horowitz}}]{Tsang2024}%
  \BibitemOpen
  \bibfield  {author} {\bibinfo {author} {\bibfnamefont {C.~Y.}\ \bibnamefont
  {Tsang}}, \bibinfo {author} {\bibfnamefont {M.~B.}\ \bibnamefont {Tsang}},
  \bibinfo {author} {\bibfnamefont {W.~G.}\ \bibnamefont {Lynch}}, \bibinfo
  {author} {\bibfnamefont {R.}~\bibnamefont {Kumar}},\ and\ \bibinfo {author}
  {\bibfnamefont {C.~J.}\ \bibnamefont {Horowitz}},\ }\href
  {https://doi.org/10.1038/s41550-023-02161-z} {\bibfield  {journal} {\bibinfo
  {journal} {Nature Astronomy}\ }\textbf {\bibinfo {volume} {8}},\ \bibinfo
  {pages} {328} (\bibinfo {year} {2024})}\BibitemShut {NoStop}%
\bibitem [{\citenamefont {Baran}\ \emph {et~al.}(2005)\citenamefont {Baran},
  \citenamefont {Colonna}, \citenamefont {Di~Toro}, \citenamefont
  {Zielinska-Pfab\'e},\ and\ \citenamefont {Wolter}}]{Baran2005}%
  \BibitemOpen
  \bibfield  {author} {\bibinfo {author} {\bibfnamefont {V.}~\bibnamefont
  {Baran}}, \bibinfo {author} {\bibfnamefont {M.}~\bibnamefont {Colonna}},
  \bibinfo {author} {\bibfnamefont {M.}~\bibnamefont {Di~Toro}}, \bibinfo
  {author} {\bibfnamefont {M.}~\bibnamefont {Zielinska-Pfab\'e}},\ and\
  \bibinfo {author} {\bibfnamefont {H.~H.}\ \bibnamefont {Wolter}},\ }\href
  {https://doi.org/10.1103/PhysRevC.72.064620} {\bibfield  {journal} {\bibinfo
  {journal} {Phys. Rev. C}\ }\textbf {\bibinfo {volume} {72}},\ \bibinfo
  {pages} {064620} (\bibinfo {year} {2005})}\BibitemShut {NoStop}%
\bibitem [{\citenamefont {McIntosh}\ and\ \citenamefont
  {Yennello}(2019)}]{McIntosh2019}%
  \BibitemOpen
  \bibfield  {author} {\bibinfo {author} {\bibfnamefont {A.~B.}\ \bibnamefont
  {McIntosh}}\ and\ \bibinfo {author} {\bibfnamefont {S.~J.}\ \bibnamefont
  {Yennello}},\ }\href
  {https://doi.org/https://doi.org/10.1016/j.ppnp.2019.06.001} {\bibfield
  {journal} {\bibinfo  {journal} {Progress in Particle and Nuclear Physics}\
  }\textbf {\bibinfo {volume} {108}},\ \bibinfo {pages} {103707} (\bibinfo
  {year} {2019})}\BibitemShut {NoStop}%
\bibitem [{\citenamefont {Lionti}\ \emph {et~al.}(2005)\citenamefont {Lionti},
  \citenamefont {Baran}, \citenamefont {Colonna},\ and\ \citenamefont {{Di
  Toro}}}]{Lionti2005}%
  \BibitemOpen
  \bibfield  {author} {\bibinfo {author} {\bibfnamefont {R.}~\bibnamefont
  {Lionti}}, \bibinfo {author} {\bibfnamefont {V.}~\bibnamefont {Baran}},
  \bibinfo {author} {\bibfnamefont {M.}~\bibnamefont {Colonna}},\ and\ \bibinfo
  {author} {\bibfnamefont {M.}~\bibnamefont {{Di Toro}}},\ }\href
  {https://doi.org/https://doi.org/10.1016/j.physletb.2005.08.044} {\bibfield
  {journal} {\bibinfo  {journal} {Phys. Lett. B}\ }\textbf {\bibinfo {volume}
  {625}},\ \bibinfo {pages} {33} (\bibinfo {year} {2005})}\BibitemShut
  {NoStop}%
\bibitem [{\citenamefont {Lombardo}\ \emph {et~al.}(2010)\citenamefont
  {Lombardo}, \citenamefont {Agodi}, \citenamefont {Alba}, \citenamefont
  {Amorini}, \citenamefont {Anzalone}, \citenamefont {Berceanu}, \citenamefont
  {Cardella}, \citenamefont {Cavallaro}, \citenamefont {Chatterjee},
  \citenamefont {De~Filippo}, \citenamefont {Di~Pietro}, \citenamefont
  {Figuera}, \citenamefont {Geraci}, \citenamefont {Giuliani}, \citenamefont
  {Grassi}, \citenamefont {Grzeszczuk}, \citenamefont {Han}, \citenamefont
  {La~Guidara}, \citenamefont {Lanzalone}, \citenamefont {{Le Neindre}},
  \citenamefont {Maiolino}, \citenamefont {Pagano}, \citenamefont {Papa},
  \citenamefont {Pirrone}, \citenamefont {Politi}, \citenamefont {Pop},
  \citenamefont {Porto}, \citenamefont {Rizzo}, \citenamefont {Russotto},
  \citenamefont {Santonocito},\ and\ \citenamefont {Verde}}]{Lombardo2010}%
  \BibitemOpen
  \bibfield  {author} {\bibinfo {author} {\bibfnamefont {I.}~\bibnamefont
  {Lombardo}}, \bibinfo {author} {\bibfnamefont {C.}~\bibnamefont {Agodi}},
  \bibinfo {author} {\bibfnamefont {R.}~\bibnamefont {Alba}}, \bibinfo {author}
  {\bibfnamefont {F.}~\bibnamefont {Amorini}}, \bibinfo {author} {\bibfnamefont
  {A.}~\bibnamefont {Anzalone}}, \bibinfo {author} {\bibfnamefont
  {I.}~\bibnamefont {Berceanu}}, \bibinfo {author} {\bibfnamefont
  {G.}~\bibnamefont {Cardella}}, \bibinfo {author} {\bibfnamefont
  {S.}~\bibnamefont {Cavallaro}}, \bibinfo {author} {\bibfnamefont {M.~B.}\
  \bibnamefont {Chatterjee}}, \bibinfo {author} {\bibfnamefont
  {E.}~\bibnamefont {De~Filippo}}, \bibinfo {author} {\bibfnamefont
  {A.}~\bibnamefont {Di~Pietro}}, \bibinfo {author} {\bibfnamefont
  {P.}~\bibnamefont {Figuera}}, \bibinfo {author} {\bibfnamefont
  {E.}~\bibnamefont {Geraci}}, \bibinfo {author} {\bibfnamefont
  {G.}~\bibnamefont {Giuliani}}, \bibinfo {author} {\bibfnamefont
  {L.}~\bibnamefont {Grassi}}, \bibinfo {author} {\bibfnamefont
  {A.}~\bibnamefont {Grzeszczuk}}, \bibinfo {author} {\bibfnamefont
  {J.}~\bibnamefont {Han}}, \bibinfo {author} {\bibfnamefont {E.}~\bibnamefont
  {La~Guidara}}, \bibinfo {author} {\bibfnamefont {G.}~\bibnamefont
  {Lanzalone}}, \bibinfo {author} {\bibfnamefont {N.}~\bibnamefont {{Le
  Neindre}}}, \bibinfo {author} {\bibfnamefont {C.}~\bibnamefont {Maiolino}},
  \bibinfo {author} {\bibfnamefont {A.}~\bibnamefont {Pagano}}, \bibinfo
  {author} {\bibfnamefont {M.}~\bibnamefont {Papa}}, \bibinfo {author}
  {\bibfnamefont {S.}~\bibnamefont {Pirrone}}, \bibinfo {author} {\bibfnamefont
  {G.}~\bibnamefont {Politi}}, \bibinfo {author} {\bibfnamefont
  {A.}~\bibnamefont {Pop}}, \bibinfo {author} {\bibfnamefont {F.}~\bibnamefont
  {Porto}}, \bibinfo {author} {\bibfnamefont {F.}~\bibnamefont {Rizzo}},
  \bibinfo {author} {\bibfnamefont {P.}~\bibnamefont {Russotto}}, \bibinfo
  {author} {\bibfnamefont {D.}~\bibnamefont {Santonocito}},\ and\ \bibinfo
  {author} {\bibfnamefont {G.}~\bibnamefont {Verde}},\ }\href
  {https://doi.org/10.1103/PhysRevC.82.014608} {\bibfield  {journal} {\bibinfo
  {journal} {Phys. Rev. C}\ }\textbf {\bibinfo {volume} {82}},\ \bibinfo
  {pages} {014608} (\bibinfo {year} {2010})}\BibitemShut {NoStop}%
\bibitem [{\citenamefont {Barlini}\ \emph {et~al.}(2013)\citenamefont
  {Barlini}, \citenamefont {Piantelli}, \citenamefont {Casini}, \citenamefont
  {Maurenzig}, \citenamefont {Olmi}, \citenamefont {Bini}, \citenamefont
  {Carboni}, \citenamefont {Pasquali}, \citenamefont {Poggi}, \citenamefont
  {Stefanini}, \citenamefont {Bougault}, \citenamefont {Bonnet}, \citenamefont
  {Borderie}, \citenamefont {Chbihi}, \citenamefont {Frankland}, \citenamefont
  {Gruyer}, \citenamefont {Lopez}, \citenamefont {{Le Neindre}}, \citenamefont
  {P\^arlog}, \citenamefont {Rivet}, \citenamefont {Vient}, \citenamefont
  {Rosato}, \citenamefont {Spadaccini}, \citenamefont {Vigilante},
  \citenamefont {Bruno}, \citenamefont {Marchi}, \citenamefont {Morelli},
  \citenamefont {Cinausero}, \citenamefont {Degerlier}, \citenamefont
  {Gramegna}, \citenamefont {Kozik}, \citenamefont {Twar\'og}, \citenamefont
  {Alba}, \citenamefont {Maiolino},\ and\ \citenamefont
  {Santonocito}}]{Barlini2013}%
  \BibitemOpen
  \bibfield  {author} {\bibinfo {author} {\bibfnamefont {S.}~\bibnamefont
  {Barlini}}, \bibinfo {author} {\bibfnamefont {S.}~\bibnamefont {Piantelli}},
  \bibinfo {author} {\bibfnamefont {G.}~\bibnamefont {Casini}}, \bibinfo
  {author} {\bibfnamefont {P.~R.}\ \bibnamefont {Maurenzig}}, \bibinfo {author}
  {\bibfnamefont {A.}~\bibnamefont {Olmi}}, \bibinfo {author} {\bibfnamefont
  {M.}~\bibnamefont {Bini}}, \bibinfo {author} {\bibfnamefont {S.}~\bibnamefont
  {Carboni}}, \bibinfo {author} {\bibfnamefont {G.}~\bibnamefont {Pasquali}},
  \bibinfo {author} {\bibfnamefont {G.}~\bibnamefont {Poggi}}, \bibinfo
  {author} {\bibfnamefont {A.~A.}\ \bibnamefont {Stefanini}}, \bibinfo {author}
  {\bibfnamefont {R.}~\bibnamefont {Bougault}}, \bibinfo {author}
  {\bibfnamefont {E.}~\bibnamefont {Bonnet}}, \bibinfo {author} {\bibfnamefont
  {B.}~\bibnamefont {Borderie}}, \bibinfo {author} {\bibfnamefont
  {A.}~\bibnamefont {Chbihi}}, \bibinfo {author} {\bibfnamefont {J.~D.}\
  \bibnamefont {Frankland}}, \bibinfo {author} {\bibfnamefont {D.}~\bibnamefont
  {Gruyer}}, \bibinfo {author} {\bibfnamefont {O.}~\bibnamefont {Lopez}},
  \bibinfo {author} {\bibfnamefont {N.}~\bibnamefont {{Le Neindre}}}, \bibinfo
  {author} {\bibfnamefont {M.}~\bibnamefont {P\^arlog}}, \bibinfo {author}
  {\bibfnamefont {M.~F.}\ \bibnamefont {Rivet}}, \bibinfo {author}
  {\bibfnamefont {E.}~\bibnamefont {Vient}}, \bibinfo {author} {\bibfnamefont
  {E.}~\bibnamefont {Rosato}}, \bibinfo {author} {\bibfnamefont
  {G.}~\bibnamefont {Spadaccini}}, \bibinfo {author} {\bibfnamefont
  {M.}~\bibnamefont {Vigilante}}, \bibinfo {author} {\bibfnamefont
  {M.}~\bibnamefont {Bruno}}, \bibinfo {author} {\bibfnamefont
  {T.}~\bibnamefont {Marchi}}, \bibinfo {author} {\bibfnamefont
  {L.}~\bibnamefont {Morelli}}, \bibinfo {author} {\bibfnamefont
  {M.}~\bibnamefont {Cinausero}}, \bibinfo {author} {\bibfnamefont
  {M.}~\bibnamefont {Degerlier}}, \bibinfo {author} {\bibfnamefont
  {F.}~\bibnamefont {Gramegna}}, \bibinfo {author} {\bibfnamefont
  {T.}~\bibnamefont {Kozik}}, \bibinfo {author} {\bibfnamefont
  {T.}~\bibnamefont {Twar\'og}}, \bibinfo {author} {\bibfnamefont
  {R.}~\bibnamefont {Alba}}, \bibinfo {author} {\bibfnamefont {C.}~\bibnamefont
  {Maiolino}},\ and\ \bibinfo {author} {\bibfnamefont {D.}~\bibnamefont
  {Santonocito}} (\bibinfo {collaboration} {FAZIA Collaboration}),\ }\href
  {https://doi.org/10.1103/PhysRevC.87.054607} {\bibfield  {journal} {\bibinfo
  {journal} {Phys. Rev. C}\ }\textbf {\bibinfo {volume} {87}},\ \bibinfo
  {pages} {054607} (\bibinfo {year} {2013})}\BibitemShut {NoStop}%
\bibitem [{\citenamefont {Piantelli}\ \emph {et~al.}(2021)\citenamefont
  {Piantelli}, \citenamefont {Casini}, \citenamefont {Ono}, \citenamefont
  {Poggi}, \citenamefont {Pastore}, \citenamefont {Barlini}, \citenamefont
  {Bini}, \citenamefont {Boiano}, \citenamefont {Bonnet}, \citenamefont
  {Borderie}, \citenamefont {Bougault}, \citenamefont {Bruno}, \citenamefont
  {Buccola}, \citenamefont {Camaiani}, \citenamefont {Chbihi}, \citenamefont
  {Ciampi}, \citenamefont {Cicerchia}, \citenamefont {Cinausero}, \citenamefont
  {Degerlier}, \citenamefont {Due\~nas}, \citenamefont {Fable}, \citenamefont
  {Fabris}, \citenamefont {Frankland}, \citenamefont {Frosin}, \citenamefont
  {Gramegna}, \citenamefont {Gruyer}, \citenamefont {Kordyasz}, \citenamefont
  {Kozik}, \citenamefont {Lemari\'e}, \citenamefont {Le~Neindre}, \citenamefont
  {Lombardo}, \citenamefont {Lopez}, \citenamefont {Mantovani}, \citenamefont
  {Marchi}, \citenamefont {Henri}, \citenamefont {Morelli}, \citenamefont
  {Olmi}, \citenamefont {Ottanelli}, \citenamefont {P\^arlog}, \citenamefont
  {Pasquali}, \citenamefont {Quicray}, \citenamefont {Stefanini}, \citenamefont
  {Tortone}, \citenamefont {Upadhyaya}, \citenamefont {Valdr\'e}, \citenamefont
  {Verde}, \citenamefont {Vient},\ and\ \citenamefont
  {Vigilante}}]{Piantelli2021}%
  \BibitemOpen
  \bibfield  {author} {\bibinfo {author} {\bibfnamefont {S.}~\bibnamefont
  {Piantelli}}, \bibinfo {author} {\bibfnamefont {G.}~\bibnamefont {Casini}},
  \bibinfo {author} {\bibfnamefont {A.}~\bibnamefont {Ono}}, \bibinfo {author}
  {\bibfnamefont {G.}~\bibnamefont {Poggi}}, \bibinfo {author} {\bibfnamefont
  {G.}~\bibnamefont {Pastore}}, \bibinfo {author} {\bibfnamefont
  {S.}~\bibnamefont {Barlini}}, \bibinfo {author} {\bibfnamefont
  {M.}~\bibnamefont {Bini}}, \bibinfo {author} {\bibfnamefont {A.}~\bibnamefont
  {Boiano}}, \bibinfo {author} {\bibfnamefont {E.}~\bibnamefont {Bonnet}},
  \bibinfo {author} {\bibfnamefont {B.}~\bibnamefont {Borderie}}, \bibinfo
  {author} {\bibfnamefont {R.}~\bibnamefont {Bougault}}, \bibinfo {author}
  {\bibfnamefont {M.}~\bibnamefont {Bruno}}, \bibinfo {author} {\bibfnamefont
  {A.}~\bibnamefont {Buccola}}, \bibinfo {author} {\bibfnamefont
  {A.}~\bibnamefont {Camaiani}}, \bibinfo {author} {\bibfnamefont
  {A.}~\bibnamefont {Chbihi}}, \bibinfo {author} {\bibfnamefont
  {C.}~\bibnamefont {Ciampi}}, \bibinfo {author} {\bibfnamefont
  {M.}~\bibnamefont {Cicerchia}}, \bibinfo {author} {\bibfnamefont
  {M.}~\bibnamefont {Cinausero}}, \bibinfo {author} {\bibfnamefont
  {M.}~\bibnamefont {Degerlier}}, \bibinfo {author} {\bibfnamefont {J.~A.}\
  \bibnamefont {Due\~nas}}, \bibinfo {author} {\bibfnamefont {Q.}~\bibnamefont
  {Fable}}, \bibinfo {author} {\bibfnamefont {D.}~\bibnamefont {Fabris}},
  \bibinfo {author} {\bibfnamefont {J.~D.}\ \bibnamefont {Frankland}}, \bibinfo
  {author} {\bibfnamefont {C.}~\bibnamefont {Frosin}}, \bibinfo {author}
  {\bibfnamefont {F.}~\bibnamefont {Gramegna}}, \bibinfo {author}
  {\bibfnamefont {D.}~\bibnamefont {Gruyer}}, \bibinfo {author} {\bibfnamefont
  {A.}~\bibnamefont {Kordyasz}}, \bibinfo {author} {\bibfnamefont
  {T.}~\bibnamefont {Kozik}}, \bibinfo {author} {\bibfnamefont
  {J.}~\bibnamefont {Lemari\'e}}, \bibinfo {author} {\bibfnamefont
  {N.}~\bibnamefont {Le~Neindre}}, \bibinfo {author} {\bibfnamefont
  {I.}~\bibnamefont {Lombardo}}, \bibinfo {author} {\bibfnamefont
  {O.}~\bibnamefont {Lopez}}, \bibinfo {author} {\bibfnamefont
  {G.}~\bibnamefont {Mantovani}}, \bibinfo {author} {\bibfnamefont
  {T.}~\bibnamefont {Marchi}}, \bibinfo {author} {\bibfnamefont
  {M.}~\bibnamefont {Henri}}, \bibinfo {author} {\bibfnamefont
  {L.}~\bibnamefont {Morelli}}, \bibinfo {author} {\bibfnamefont
  {A.}~\bibnamefont {Olmi}}, \bibinfo {author} {\bibfnamefont {P.}~\bibnamefont
  {Ottanelli}}, \bibinfo {author} {\bibfnamefont {M.}~\bibnamefont {P\^arlog}},
  \bibinfo {author} {\bibfnamefont {G.}~\bibnamefont {Pasquali}}, \bibinfo
  {author} {\bibfnamefont {J.}~\bibnamefont {Quicray}}, \bibinfo {author}
  {\bibfnamefont {A.~A.}\ \bibnamefont {Stefanini}}, \bibinfo {author}
  {\bibfnamefont {G.}~\bibnamefont {Tortone}}, \bibinfo {author} {\bibfnamefont
  {S.}~\bibnamefont {Upadhyaya}}, \bibinfo {author} {\bibfnamefont
  {S.}~\bibnamefont {Valdr\'e}}, \bibinfo {author} {\bibfnamefont
  {G.}~\bibnamefont {Verde}}, \bibinfo {author} {\bibfnamefont
  {E.}~\bibnamefont {Vient}},\ and\ \bibinfo {author} {\bibfnamefont
  {M.}~\bibnamefont {Vigilante}} (\bibinfo {collaboration} {FAZIA
  Collaboration}),\ }\href {https://doi.org/10.1103/PhysRevC.103.014603}
  {\bibfield  {journal} {\bibinfo  {journal} {Phys. Rev. C}\ }\textbf {\bibinfo
  {volume} {103}},\ \bibinfo {pages} {014603} (\bibinfo {year}
  {2021})}\BibitemShut {NoStop}%
\bibitem [{\citenamefont {Fable}\ \emph {et~al.}(2023)\citenamefont {Fable},
  \citenamefont {Chbihi}, \citenamefont {Frankland}, \citenamefont
  {Napolitani}, \citenamefont {Verde}, \citenamefont {Bonnet}, \citenamefont
  {Borderie}, \citenamefont {Bougault}, \citenamefont {Galichet}, \citenamefont
  {G\'enard}, \citenamefont {Gruyer}, \citenamefont {Henri}, \citenamefont
  {La~Commara}, \citenamefont {Le~F\`evre}, \citenamefont {Lemari\'e},
  \citenamefont {Le~Neindre}, \citenamefont {Lopez}, \citenamefont {Marini},
  \citenamefont {P\^arlog}, \citenamefont {Rebillard-Souli\'e}, \citenamefont
  {Trautmann}, \citenamefont {Vient},\ and\ \citenamefont
  {Vigilante}}]{Fable2023}%
  \BibitemOpen
  \bibfield  {author} {\bibinfo {author} {\bibfnamefont {Q.}~\bibnamefont
  {Fable}}, \bibinfo {author} {\bibfnamefont {A.}~\bibnamefont {Chbihi}},
  \bibinfo {author} {\bibfnamefont {J.~D.}\ \bibnamefont {Frankland}}, \bibinfo
  {author} {\bibfnamefont {P.}~\bibnamefont {Napolitani}}, \bibinfo {author}
  {\bibfnamefont {G.}~\bibnamefont {Verde}}, \bibinfo {author} {\bibfnamefont
  {E.}~\bibnamefont {Bonnet}}, \bibinfo {author} {\bibfnamefont
  {B.}~\bibnamefont {Borderie}}, \bibinfo {author} {\bibfnamefont
  {R.}~\bibnamefont {Bougault}}, \bibinfo {author} {\bibfnamefont
  {E.}~\bibnamefont {Galichet}}, \bibinfo {author} {\bibfnamefont
  {T.}~\bibnamefont {G\'enard}}, \bibinfo {author} {\bibfnamefont
  {D.}~\bibnamefont {Gruyer}}, \bibinfo {author} {\bibfnamefont
  {M.}~\bibnamefont {Henri}}, \bibinfo {author} {\bibfnamefont
  {M.}~\bibnamefont {La~Commara}}, \bibinfo {author} {\bibfnamefont
  {A.}~\bibnamefont {Le~F\`evre}}, \bibinfo {author} {\bibfnamefont
  {J.}~\bibnamefont {Lemari\'e}}, \bibinfo {author} {\bibfnamefont
  {N.}~\bibnamefont {Le~Neindre}}, \bibinfo {author} {\bibfnamefont
  {O.}~\bibnamefont {Lopez}}, \bibinfo {author} {\bibfnamefont
  {P.}~\bibnamefont {Marini}}, \bibinfo {author} {\bibfnamefont
  {M.}~\bibnamefont {P\^arlog}}, \bibinfo {author} {\bibfnamefont
  {A.}~\bibnamefont {Rebillard-Souli\'e}}, \bibinfo {author} {\bibfnamefont
  {W.}~\bibnamefont {Trautmann}}, \bibinfo {author} {\bibfnamefont
  {E.}~\bibnamefont {Vient}},\ and\ \bibinfo {author} {\bibfnamefont
  {M.}~\bibnamefont {Vigilante}} (\bibinfo {collaboration} {INDRA
  Collaboration}),\ }\href {https://doi.org/10.1103/PhysRevC.107.014604}
  {\bibfield  {journal} {\bibinfo  {journal} {Phys. Rev. C}\ }\textbf {\bibinfo
  {volume} {107}},\ \bibinfo {pages} {014604} (\bibinfo {year}
  {2023})}\BibitemShut {NoStop}%
\bibitem [{\citenamefont {Johnston}\ \emph {et~al.}(1996)\citenamefont
  {Johnston}, \citenamefont {White}, \citenamefont {Winger}, \citenamefont
  {Rowland}, \citenamefont {Hurst}, \citenamefont {Gimeno-Nogues},
  \citenamefont {O'Kelly},\ and\ \citenamefont {Yennello}}]{Johnston1996}%
  \BibitemOpen
  \bibfield  {author} {\bibinfo {author} {\bibfnamefont {H.}~\bibnamefont
  {Johnston}}, \bibinfo {author} {\bibfnamefont {T.}~\bibnamefont {White}},
  \bibinfo {author} {\bibfnamefont {J.}~\bibnamefont {Winger}}, \bibinfo
  {author} {\bibfnamefont {D.}~\bibnamefont {Rowland}}, \bibinfo {author}
  {\bibfnamefont {B.}~\bibnamefont {Hurst}}, \bibinfo {author} {\bibfnamefont
  {F.}~\bibnamefont {Gimeno-Nogues}}, \bibinfo {author} {\bibfnamefont
  {D.}~\bibnamefont {O'Kelly}},\ and\ \bibinfo {author} {\bibfnamefont
  {S.}~\bibnamefont {Yennello}},\ }\href
  {https://doi.org/https://doi.org/10.1016/0370-2693(96)00019-6} {\bibfield
  {journal} {\bibinfo  {journal} {Physics Letters B}\ }\textbf {\bibinfo
  {volume} {371}},\ \bibinfo {pages} {186} (\bibinfo {year}
  {1996})}\BibitemShut {NoStop}%
\bibitem [{\citenamefont {Camaiani}\ \emph {et~al.}(2021)\citenamefont
  {Camaiani}, \citenamefont {Casini}, \citenamefont {Piantelli}, \citenamefont
  {Ono}, \citenamefont {Bonnet}, \citenamefont {Alba}, \citenamefont {Barlini},
  \citenamefont {Borderie}, \citenamefont {Bougault}, \citenamefont {Ciampi},
  \citenamefont {Chbihi}, \citenamefont {Cicerchia}, \citenamefont {Cinausero},
  \citenamefont {Due\~nas}, \citenamefont {Dell'Aquila}, \citenamefont {Fable},
  \citenamefont {Fabris}, \citenamefont {Frosin}, \citenamefont {Frankland},
  \citenamefont {Gramegna}, \citenamefont {Gruyer}, \citenamefont {Hahn},
  \citenamefont {Henri}, \citenamefont {Hong}, \citenamefont {Kim},
  \citenamefont {Kordyasz}, \citenamefont {Kweon}, \citenamefont {Lee},
  \citenamefont {Lemari\'e}, \citenamefont {LeNeindre}, \citenamefont
  {Lombardo}, \citenamefont {Lopez}, \citenamefont {Marchi}, \citenamefont
  {Nam}, \citenamefont {Ottanelli}, \citenamefont {Parlog}, \citenamefont
  {Pasquali}, \citenamefont {Poggi}, \citenamefont {Quicray}, \citenamefont
  {Stefanini}, \citenamefont {Upadhyaya}, \citenamefont {Valdr\'e},\ and\
  \citenamefont {Vient}}]{Camaiani2021}%
  \BibitemOpen
  \bibfield  {author} {\bibinfo {author} {\bibfnamefont {A.}~\bibnamefont
  {Camaiani}}, \bibinfo {author} {\bibfnamefont {G.}~\bibnamefont {Casini}},
  \bibinfo {author} {\bibfnamefont {S.}~\bibnamefont {Piantelli}}, \bibinfo
  {author} {\bibfnamefont {A.}~\bibnamefont {Ono}}, \bibinfo {author}
  {\bibfnamefont {E.}~\bibnamefont {Bonnet}}, \bibinfo {author} {\bibfnamefont
  {R.}~\bibnamefont {Alba}}, \bibinfo {author} {\bibfnamefont {S.}~\bibnamefont
  {Barlini}}, \bibinfo {author} {\bibfnamefont {B.}~\bibnamefont {Borderie}},
  \bibinfo {author} {\bibfnamefont {R.}~\bibnamefont {Bougault}}, \bibinfo
  {author} {\bibfnamefont {C.}~\bibnamefont {Ciampi}}, \bibinfo {author}
  {\bibfnamefont {A.}~\bibnamefont {Chbihi}}, \bibinfo {author} {\bibfnamefont
  {M.}~\bibnamefont {Cicerchia}}, \bibinfo {author} {\bibfnamefont
  {M.}~\bibnamefont {Cinausero}}, \bibinfo {author} {\bibfnamefont {J.~A.}\
  \bibnamefont {Due\~nas}}, \bibinfo {author} {\bibfnamefont {D.}~\bibnamefont
  {Dell'Aquila}}, \bibinfo {author} {\bibfnamefont {Q.}~\bibnamefont {Fable}},
  \bibinfo {author} {\bibfnamefont {D.}~\bibnamefont {Fabris}}, \bibinfo
  {author} {\bibfnamefont {C.}~\bibnamefont {Frosin}}, \bibinfo {author}
  {\bibfnamefont {J.~D.}\ \bibnamefont {Frankland}}, \bibinfo {author}
  {\bibfnamefont {F.}~\bibnamefont {Gramegna}}, \bibinfo {author}
  {\bibfnamefont {D.}~\bibnamefont {Gruyer}}, \bibinfo {author} {\bibfnamefont
  {K.~I.}\ \bibnamefont {Hahn}}, \bibinfo {author} {\bibfnamefont
  {M.}~\bibnamefont {Henri}}, \bibinfo {author} {\bibfnamefont
  {B.}~\bibnamefont {Hong}}, \bibinfo {author} {\bibfnamefont {S.}~\bibnamefont
  {Kim}}, \bibinfo {author} {\bibfnamefont {A.}~\bibnamefont {Kordyasz}},
  \bibinfo {author} {\bibfnamefont {M.~J.}\ \bibnamefont {Kweon}}, \bibinfo
  {author} {\bibfnamefont {H.~J.}\ \bibnamefont {Lee}}, \bibinfo {author}
  {\bibfnamefont {J.}~\bibnamefont {Lemari\'e}}, \bibinfo {author}
  {\bibfnamefont {N.}~\bibnamefont {LeNeindre}}, \bibinfo {author}
  {\bibfnamefont {I.}~\bibnamefont {Lombardo}}, \bibinfo {author}
  {\bibfnamefont {O.}~\bibnamefont {Lopez}}, \bibinfo {author} {\bibfnamefont
  {T.}~\bibnamefont {Marchi}}, \bibinfo {author} {\bibfnamefont {S.~H.}\
  \bibnamefont {Nam}}, \bibinfo {author} {\bibfnamefont {P.}~\bibnamefont
  {Ottanelli}}, \bibinfo {author} {\bibfnamefont {M.}~\bibnamefont {Parlog}},
  \bibinfo {author} {\bibfnamefont {G.}~\bibnamefont {Pasquali}}, \bibinfo
  {author} {\bibfnamefont {G.}~\bibnamefont {Poggi}}, \bibinfo {author}
  {\bibfnamefont {J.}~\bibnamefont {Quicray}}, \bibinfo {author} {\bibfnamefont
  {A.~A.}\ \bibnamefont {Stefanini}}, \bibinfo {author} {\bibfnamefont
  {S.}~\bibnamefont {Upadhyaya}}, \bibinfo {author} {\bibfnamefont
  {S.}~\bibnamefont {Valdr\'e}},\ and\ \bibinfo {author} {\bibfnamefont
  {E.}~\bibnamefont {Vient}},\ }\href
  {https://doi.org/10.1103/PhysRevC.103.014605} {\bibfield  {journal} {\bibinfo
   {journal} {Phys. Rev. C}\ }\textbf {\bibinfo {volume} {103}},\ \bibinfo
  {pages} {014605} (\bibinfo {year} {2021})}\BibitemShut {NoStop}%
\bibitem [{\citenamefont {Ciampi}\ \emph {et~al.}(2022)\citenamefont {Ciampi},
  \citenamefont {Piantelli}, \citenamefont {Casini}, \citenamefont {Pasquali},
  \citenamefont {Quicray}, \citenamefont {Baldesi}, \citenamefont {Barlini},
  \citenamefont {Borderie}, \citenamefont {Bougault}, \citenamefont {Camaiani},
  \citenamefont {Chbihi}, \citenamefont {Dell'Aquila}, \citenamefont
  {Cicerchia}, \citenamefont {Due\~nas}, \citenamefont {Fable}, \citenamefont
  {Fabris}, \citenamefont {Frankland}, \citenamefont {Frosin}, \citenamefont
  {G\'enard}, \citenamefont {Gramegna}, \citenamefont {Gruyer}, \citenamefont
  {Henri}, \citenamefont {Hong}, \citenamefont {Kim}, \citenamefont {Kordyasz},
  \citenamefont {Kozik}, \citenamefont {Kweon}, \citenamefont {Lemari\'e},
  \citenamefont {Le~Neindre}, \citenamefont {Lombardo}, \citenamefont {Lopez},
  \citenamefont {Marchi}, \citenamefont {Nam}, \citenamefont {Ordine},
  \citenamefont {Ottanelli}, \citenamefont {Park}, \citenamefont {Park},
  \citenamefont {P\^arlog}, \citenamefont {Poggi}, \citenamefont
  {Rebillard-Souli\'e}, \citenamefont {Stefanini}, \citenamefont {Upadhyaya},
  \citenamefont {Valdr\'e}, \citenamefont {Verde}, \citenamefont {Vient},\ and\
  \citenamefont {Vigilante}}]{Ciampi2022}%
  \BibitemOpen
  \bibfield  {author} {\bibinfo {author} {\bibfnamefont {C.}~\bibnamefont
  {Ciampi}}, \bibinfo {author} {\bibfnamefont {S.}~\bibnamefont {Piantelli}},
  \bibinfo {author} {\bibfnamefont {G.}~\bibnamefont {Casini}}, \bibinfo
  {author} {\bibfnamefont {G.}~\bibnamefont {Pasquali}}, \bibinfo {author}
  {\bibfnamefont {J.}~\bibnamefont {Quicray}}, \bibinfo {author} {\bibfnamefont
  {L.}~\bibnamefont {Baldesi}}, \bibinfo {author} {\bibfnamefont
  {S.}~\bibnamefont {Barlini}}, \bibinfo {author} {\bibfnamefont
  {B.}~\bibnamefont {Borderie}}, \bibinfo {author} {\bibfnamefont
  {R.}~\bibnamefont {Bougault}}, \bibinfo {author} {\bibfnamefont
  {A.}~\bibnamefont {Camaiani}}, \bibinfo {author} {\bibfnamefont
  {A.}~\bibnamefont {Chbihi}}, \bibinfo {author} {\bibfnamefont
  {D.}~\bibnamefont {Dell'Aquila}}, \bibinfo {author} {\bibfnamefont
  {M.}~\bibnamefont {Cicerchia}}, \bibinfo {author} {\bibfnamefont {J.~A.}\
  \bibnamefont {Due\~nas}}, \bibinfo {author} {\bibfnamefont {Q.}~\bibnamefont
  {Fable}}, \bibinfo {author} {\bibfnamefont {D.}~\bibnamefont {Fabris}},
  \bibinfo {author} {\bibfnamefont {J.~D.}\ \bibnamefont {Frankland}}, \bibinfo
  {author} {\bibfnamefont {C.}~\bibnamefont {Frosin}}, \bibinfo {author}
  {\bibfnamefont {T.}~\bibnamefont {G\'enard}}, \bibinfo {author}
  {\bibfnamefont {F.}~\bibnamefont {Gramegna}}, \bibinfo {author}
  {\bibfnamefont {D.}~\bibnamefont {Gruyer}}, \bibinfo {author} {\bibfnamefont
  {M.}~\bibnamefont {Henri}}, \bibinfo {author} {\bibfnamefont
  {B.}~\bibnamefont {Hong}}, \bibinfo {author} {\bibfnamefont {S.}~\bibnamefont
  {Kim}}, \bibinfo {author} {\bibfnamefont {A.}~\bibnamefont {Kordyasz}},
  \bibinfo {author} {\bibfnamefont {T.}~\bibnamefont {Kozik}}, \bibinfo
  {author} {\bibfnamefont {M.~J.}\ \bibnamefont {Kweon}}, \bibinfo {author}
  {\bibfnamefont {J.}~\bibnamefont {Lemari\'e}}, \bibinfo {author}
  {\bibfnamefont {N.}~\bibnamefont {Le~Neindre}}, \bibinfo {author}
  {\bibfnamefont {I.}~\bibnamefont {Lombardo}}, \bibinfo {author}
  {\bibfnamefont {O.}~\bibnamefont {Lopez}}, \bibinfo {author} {\bibfnamefont
  {T.}~\bibnamefont {Marchi}}, \bibinfo {author} {\bibfnamefont {S.~H.}\
  \bibnamefont {Nam}}, \bibinfo {author} {\bibfnamefont {A.}~\bibnamefont
  {Ordine}}, \bibinfo {author} {\bibfnamefont {P.}~\bibnamefont {Ottanelli}},
  \bibinfo {author} {\bibfnamefont {J.}~\bibnamefont {Park}}, \bibinfo {author}
  {\bibfnamefont {J.~H.}\ \bibnamefont {Park}}, \bibinfo {author}
  {\bibfnamefont {M.}~\bibnamefont {P\^arlog}}, \bibinfo {author}
  {\bibfnamefont {G.}~\bibnamefont {Poggi}}, \bibinfo {author} {\bibfnamefont
  {A.}~\bibnamefont {Rebillard-Souli\'e}}, \bibinfo {author} {\bibfnamefont
  {A.~A.}\ \bibnamefont {Stefanini}}, \bibinfo {author} {\bibfnamefont
  {S.}~\bibnamefont {Upadhyaya}}, \bibinfo {author} {\bibfnamefont
  {S.}~\bibnamefont {Valdr\'e}}, \bibinfo {author} {\bibfnamefont
  {G.}~\bibnamefont {Verde}}, \bibinfo {author} {\bibfnamefont
  {E.}~\bibnamefont {Vient}},\ and\ \bibinfo {author} {\bibfnamefont
  {M.}~\bibnamefont {Vigilante}} (\bibinfo {collaboration} {{INDRA-FAZIA}
  Collaboration}),\ }\href {https://doi.org/10.1103/PhysRevC.106.024603}
  {\bibfield  {journal} {\bibinfo  {journal} {Phys. Rev. C}\ }\textbf {\bibinfo
  {volume} {106}},\ \bibinfo {pages} {024603} (\bibinfo {year}
  {2022})}\BibitemShut {NoStop}%
\bibitem [{\citenamefont {Wolter}\ \emph {et~al.}(2022)\citenamefont {Wolter},
  \citenamefont {Colonna}, \citenamefont {Cozma}, \citenamefont {Danielewicz},
  \citenamefont {Ko}, \citenamefont {Kumar}, \citenamefont {Ono}, \citenamefont
  {Tsang}, \citenamefont {Xu}, \citenamefont {Zhang}, \citenamefont
  {Bratkovskaya}, \citenamefont {Feng}, \citenamefont {Gaitanos}, \citenamefont
  {{Le Fèvre}}, \citenamefont {Ikeno}, \citenamefont {Kim}, \citenamefont
  {Mallik}, \citenamefont {Napolitani}, \citenamefont {Oliinychenko},
  \citenamefont {Ogawa}, \citenamefont {Papa}, \citenamefont {Su},
  \citenamefont {Wang}, \citenamefont {Wang}, \citenamefont {Weil},
  \citenamefont {Zhang}, \citenamefont {Zhang}, \citenamefont {Zhang},
  \citenamefont {Aichelin}, \citenamefont {Cassing}, \citenamefont {Chen},
  \citenamefont {Cheng}, \citenamefont {Elfner}, \citenamefont {Gallmeister},
  \citenamefont {Hartnack}, \citenamefont {Hashimoto}, \citenamefont {Jeon},
  \citenamefont {Kim}, \citenamefont {Kim}, \citenamefont {Li}, \citenamefont
  {Lee}, \citenamefont {Li}, \citenamefont {Li}, \citenamefont {Mosel},
  \citenamefont {Nara}, \citenamefont {Niita}, \citenamefont {Ohnishi},
  \citenamefont {Sato}, \citenamefont {Song}, \citenamefont {Sorensen},
  \citenamefont {Wang},\ and\ \citenamefont {Xie}}]{Wolter2022}%
  \BibitemOpen
  \bibfield  {author} {\bibinfo {author} {\bibfnamefont {H.}~\bibnamefont
  {Wolter}}, \bibinfo {author} {\bibfnamefont {M.}~\bibnamefont {Colonna}},
  \bibinfo {author} {\bibfnamefont {D.}~\bibnamefont {Cozma}}, \bibinfo
  {author} {\bibfnamefont {P.}~\bibnamefont {Danielewicz}}, \bibinfo {author}
  {\bibfnamefont {C.~M.}\ \bibnamefont {Ko}}, \bibinfo {author} {\bibfnamefont
  {R.}~\bibnamefont {Kumar}}, \bibinfo {author} {\bibfnamefont
  {A.}~\bibnamefont {Ono}}, \bibinfo {author} {\bibfnamefont {M.~B.}\
  \bibnamefont {Tsang}}, \bibinfo {author} {\bibfnamefont {J.}~\bibnamefont
  {Xu}}, \bibinfo {author} {\bibfnamefont {Y.-X.}\ \bibnamefont {Zhang}},
  \bibinfo {author} {\bibfnamefont {E.}~\bibnamefont {Bratkovskaya}}, \bibinfo
  {author} {\bibfnamefont {Z.-Q.}\ \bibnamefont {Feng}}, \bibinfo {author}
  {\bibfnamefont {T.}~\bibnamefont {Gaitanos}}, \bibinfo {author}
  {\bibfnamefont {A.}~\bibnamefont {{Le Fèvre}}}, \bibinfo {author}
  {\bibfnamefont {N.}~\bibnamefont {Ikeno}}, \bibinfo {author} {\bibfnamefont
  {Y.}~\bibnamefont {Kim}}, \bibinfo {author} {\bibfnamefont {S.}~\bibnamefont
  {Mallik}}, \bibinfo {author} {\bibfnamefont {P.}~\bibnamefont {Napolitani}},
  \bibinfo {author} {\bibfnamefont {D.}~\bibnamefont {Oliinychenko}}, \bibinfo
  {author} {\bibfnamefont {T.}~\bibnamefont {Ogawa}}, \bibinfo {author}
  {\bibfnamefont {M.}~\bibnamefont {Papa}}, \bibinfo {author} {\bibfnamefont
  {J.}~\bibnamefont {Su}}, \bibinfo {author} {\bibfnamefont {R.}~\bibnamefont
  {Wang}}, \bibinfo {author} {\bibfnamefont {Y.-J.}\ \bibnamefont {Wang}},
  \bibinfo {author} {\bibfnamefont {J.}~\bibnamefont {Weil}}, \bibinfo {author}
  {\bibfnamefont {F.-S.}\ \bibnamefont {Zhang}}, \bibinfo {author}
  {\bibfnamefont {G.-Q.}\ \bibnamefont {Zhang}}, \bibinfo {author}
  {\bibfnamefont {Z.}~\bibnamefont {Zhang}}, \bibinfo {author} {\bibfnamefont
  {J.}~\bibnamefont {Aichelin}}, \bibinfo {author} {\bibfnamefont
  {W.}~\bibnamefont {Cassing}}, \bibinfo {author} {\bibfnamefont {L.-W.}\
  \bibnamefont {Chen}}, \bibinfo {author} {\bibfnamefont {H.-G.}\ \bibnamefont
  {Cheng}}, \bibinfo {author} {\bibfnamefont {H.}~\bibnamefont {Elfner}},
  \bibinfo {author} {\bibfnamefont {K.}~\bibnamefont {Gallmeister}}, \bibinfo
  {author} {\bibfnamefont {C.}~\bibnamefont {Hartnack}}, \bibinfo {author}
  {\bibfnamefont {S.}~\bibnamefont {Hashimoto}}, \bibinfo {author}
  {\bibfnamefont {S.}~\bibnamefont {Jeon}}, \bibinfo {author} {\bibfnamefont
  {K.}~\bibnamefont {Kim}}, \bibinfo {author} {\bibfnamefont {M.}~\bibnamefont
  {Kim}}, \bibinfo {author} {\bibfnamefont {B.-A.}\ \bibnamefont {Li}},
  \bibinfo {author} {\bibfnamefont {C.-H.}\ \bibnamefont {Lee}}, \bibinfo
  {author} {\bibfnamefont {Q.-F.}\ \bibnamefont {Li}}, \bibinfo {author}
  {\bibfnamefont {Z.-X.}\ \bibnamefont {Li}}, \bibinfo {author} {\bibfnamefont
  {U.}~\bibnamefont {Mosel}}, \bibinfo {author} {\bibfnamefont
  {Y.}~\bibnamefont {Nara}}, \bibinfo {author} {\bibfnamefont {K.}~\bibnamefont
  {Niita}}, \bibinfo {author} {\bibfnamefont {A.}~\bibnamefont {Ohnishi}},
  \bibinfo {author} {\bibfnamefont {T.}~\bibnamefont {Sato}}, \bibinfo {author}
  {\bibfnamefont {T.}~\bibnamefont {Song}}, \bibinfo {author} {\bibfnamefont
  {A.}~\bibnamefont {Sorensen}}, \bibinfo {author} {\bibfnamefont
  {N.}~\bibnamefont {Wang}},\ and\ \bibinfo {author} {\bibfnamefont {W.-J.}\
  \bibnamefont {Xie}},\ }\href
  {https://doi.org/https://doi.org/10.1016/j.ppnp.2022.103962} {\bibfield
  {journal} {\bibinfo  {journal} {Progress in Particle and Nuclear Physics}\
  }\textbf {\bibinfo {volume} {125}},\ \bibinfo {pages} {103962} (\bibinfo
  {year} {2022})}\BibitemShut {NoStop}%
\bibitem [{\citenamefont {Napolitani}\ \emph {et~al.}(2010)\citenamefont
  {Napolitani}, \citenamefont {Colonna}, \citenamefont {Gulminelli},
  \citenamefont {Galichet}, \citenamefont {Piantelli}, \citenamefont {Verde},\
  and\ \citenamefont {Vient}}]{Napolitani2010}%
  \BibitemOpen
  \bibfield  {author} {\bibinfo {author} {\bibfnamefont {P.}~\bibnamefont
  {Napolitani}}, \bibinfo {author} {\bibfnamefont {M.}~\bibnamefont {Colonna}},
  \bibinfo {author} {\bibfnamefont {F.}~\bibnamefont {Gulminelli}}, \bibinfo
  {author} {\bibfnamefont {E.}~\bibnamefont {Galichet}}, \bibinfo {author}
  {\bibfnamefont {S.}~\bibnamefont {Piantelli}}, \bibinfo {author}
  {\bibfnamefont {G.}~\bibnamefont {Verde}},\ and\ \bibinfo {author}
  {\bibfnamefont {E.}~\bibnamefont {Vient}},\ }\href
  {https://doi.org/10.1103/PhysRevC.81.044619} {\bibfield  {journal} {\bibinfo
  {journal} {Phys. Rev. C}\ }\textbf {\bibinfo {volume} {81}},\ \bibinfo
  {pages} {044619} (\bibinfo {year} {2010})}\BibitemShut {NoStop}%
\bibitem [{\citenamefont {Rami}\ \emph {et~al.}(2000)\citenamefont {Rami},
  \citenamefont {Leifels}, \citenamefont {de~Schauenburg}, \citenamefont
  {Gobbi}, \citenamefont {Hong}, \citenamefont {Alard}, \citenamefont
  {Andronic}, \citenamefont {Averbeck}, \citenamefont {Barret}, \citenamefont
  {Basrak}, \citenamefont {Bastid}, \citenamefont {Belyaev}, \citenamefont
  {Bendarag}, \citenamefont {Berek}, \citenamefont {\ifmmode~\check{C}\else
  \v{C}\fi{}aplar}, \citenamefont {Cindro}, \citenamefont {Crochet},
  \citenamefont {Devismes}, \citenamefont {Dupieux}, \citenamefont
  {D\ifmmode~\check{z}\else \v{z}\fi{}elalija}, \citenamefont {Eskef},
  \citenamefont {Finck}, \citenamefont {Fodor}, \citenamefont {Folger},
  \citenamefont {Fraysse}, \citenamefont {Genoux-Lubain}, \citenamefont
  {Grigorian}, \citenamefont {Grishkin}, \citenamefont {Herrmann},
  \citenamefont {Hildenbrand}, \citenamefont {Kecskemeti}, \citenamefont {Kim},
  \citenamefont {Koczon}, \citenamefont {Kirejczyk}, \citenamefont {Korolija},
  \citenamefont {Kotte}, \citenamefont {Kowalczyk}, \citenamefont {Kress},
  \citenamefont {Kutsche}, \citenamefont {Lebedev}, \citenamefont {Lee},
  \citenamefont {Manko}, \citenamefont {Merlitz}, \citenamefont {Mohren},
  \citenamefont {Moisa}, \citenamefont {M\"osner}, \citenamefont {Neubert},
  \citenamefont {Nianine}, \citenamefont {Pelte}, \citenamefont {Petrovici},
  \citenamefont {Pinkenburg}, \citenamefont {Plettner}, \citenamefont
  {Reisdorf}, \citenamefont {Ritman}, \citenamefont {Sch\"ull}, \citenamefont
  {Seres}, \citenamefont {Sikora}, \citenamefont {Sim}, \citenamefont {Simion},
  \citenamefont {Siwek-Wilczy\ifmmode~\acute{n}\else \'{n}\fi{}ska},
  \citenamefont {Somov}, \citenamefont {Stockmeier}, \citenamefont {Stoicea},
  \citenamefont {Vasiliev}, \citenamefont {Wagner}, \citenamefont
  {Wi\ifmmode~\acute{s}\else \'{s}\fi{}niewski}, \citenamefont {Wohlfarth},
  \citenamefont {Yang}, \citenamefont {Yushmanov},\ and\ \citenamefont
  {Zhilin}}]{Rami2000}%
  \BibitemOpen
  \bibfield  {author} {\bibinfo {author} {\bibfnamefont {F.}~\bibnamefont
  {Rami}}, \bibinfo {author} {\bibfnamefont {Y.}~\bibnamefont {Leifels}},
  \bibinfo {author} {\bibfnamefont {B.}~\bibnamefont {de~Schauenburg}},
  \bibinfo {author} {\bibfnamefont {A.}~\bibnamefont {Gobbi}}, \bibinfo
  {author} {\bibfnamefont {B.}~\bibnamefont {Hong}}, \bibinfo {author}
  {\bibfnamefont {J.~P.}\ \bibnamefont {Alard}}, \bibinfo {author}
  {\bibfnamefont {A.}~\bibnamefont {Andronic}}, \bibinfo {author}
  {\bibfnamefont {R.}~\bibnamefont {Averbeck}}, \bibinfo {author}
  {\bibfnamefont {V.}~\bibnamefont {Barret}}, \bibinfo {author} {\bibfnamefont
  {Z.}~\bibnamefont {Basrak}}, \bibinfo {author} {\bibfnamefont
  {N.}~\bibnamefont {Bastid}}, \bibinfo {author} {\bibfnamefont
  {I.}~\bibnamefont {Belyaev}}, \bibinfo {author} {\bibfnamefont
  {A.}~\bibnamefont {Bendarag}}, \bibinfo {author} {\bibfnamefont
  {G.}~\bibnamefont {Berek}}, \bibinfo {author} {\bibfnamefont
  {R.}~\bibnamefont {\ifmmode~\check{C}\else \v{C}\fi{}aplar}}, \bibinfo
  {author} {\bibfnamefont {N.}~\bibnamefont {Cindro}}, \bibinfo {author}
  {\bibfnamefont {P.}~\bibnamefont {Crochet}}, \bibinfo {author} {\bibfnamefont
  {A.}~\bibnamefont {Devismes}}, \bibinfo {author} {\bibfnamefont
  {P.}~\bibnamefont {Dupieux}}, \bibinfo {author} {\bibfnamefont
  {M.}~\bibnamefont {D\ifmmode~\check{z}\else \v{z}\fi{}elalija}}, \bibinfo
  {author} {\bibfnamefont {M.}~\bibnamefont {Eskef}}, \bibinfo {author}
  {\bibfnamefont {C.}~\bibnamefont {Finck}}, \bibinfo {author} {\bibfnamefont
  {Z.}~\bibnamefont {Fodor}}, \bibinfo {author} {\bibfnamefont
  {H.}~\bibnamefont {Folger}}, \bibinfo {author} {\bibfnamefont
  {L.}~\bibnamefont {Fraysse}}, \bibinfo {author} {\bibfnamefont
  {A.}~\bibnamefont {Genoux-Lubain}}, \bibinfo {author} {\bibfnamefont
  {Y.}~\bibnamefont {Grigorian}}, \bibinfo {author} {\bibfnamefont
  {Y.}~\bibnamefont {Grishkin}}, \bibinfo {author} {\bibfnamefont
  {N.}~\bibnamefont {Herrmann}}, \bibinfo {author} {\bibfnamefont {K.~D.}\
  \bibnamefont {Hildenbrand}}, \bibinfo {author} {\bibfnamefont
  {J.}~\bibnamefont {Kecskemeti}}, \bibinfo {author} {\bibfnamefont {Y.~J.}\
  \bibnamefont {Kim}}, \bibinfo {author} {\bibfnamefont {P.}~\bibnamefont
  {Koczon}}, \bibinfo {author} {\bibfnamefont {M.}~\bibnamefont {Kirejczyk}},
  \bibinfo {author} {\bibfnamefont {M.}~\bibnamefont {Korolija}}, \bibinfo
  {author} {\bibfnamefont {R.}~\bibnamefont {Kotte}}, \bibinfo {author}
  {\bibfnamefont {M.}~\bibnamefont {Kowalczyk}}, \bibinfo {author}
  {\bibfnamefont {T.}~\bibnamefont {Kress}}, \bibinfo {author} {\bibfnamefont
  {R.}~\bibnamefont {Kutsche}}, \bibinfo {author} {\bibfnamefont
  {A.}~\bibnamefont {Lebedev}}, \bibinfo {author} {\bibfnamefont {K.~S.}\
  \bibnamefont {Lee}}, \bibinfo {author} {\bibfnamefont {V.}~\bibnamefont
  {Manko}}, \bibinfo {author} {\bibfnamefont {H.}~\bibnamefont {Merlitz}},
  \bibinfo {author} {\bibfnamefont {S.}~\bibnamefont {Mohren}}, \bibinfo
  {author} {\bibfnamefont {D.}~\bibnamefont {Moisa}}, \bibinfo {author}
  {\bibfnamefont {J.}~\bibnamefont {M\"osner}}, \bibinfo {author}
  {\bibfnamefont {W.}~\bibnamefont {Neubert}}, \bibinfo {author} {\bibfnamefont
  {A.}~\bibnamefont {Nianine}}, \bibinfo {author} {\bibfnamefont
  {D.}~\bibnamefont {Pelte}}, \bibinfo {author} {\bibfnamefont
  {M.}~\bibnamefont {Petrovici}}, \bibinfo {author} {\bibfnamefont
  {C.}~\bibnamefont {Pinkenburg}}, \bibinfo {author} {\bibfnamefont
  {C.}~\bibnamefont {Plettner}}, \bibinfo {author} {\bibfnamefont
  {W.}~\bibnamefont {Reisdorf}}, \bibinfo {author} {\bibfnamefont
  {J.}~\bibnamefont {Ritman}}, \bibinfo {author} {\bibfnamefont
  {D.}~\bibnamefont {Sch\"ull}}, \bibinfo {author} {\bibfnamefont
  {Z.}~\bibnamefont {Seres}}, \bibinfo {author} {\bibfnamefont
  {B.}~\bibnamefont {Sikora}}, \bibinfo {author} {\bibfnamefont {K.~S.}\
  \bibnamefont {Sim}}, \bibinfo {author} {\bibfnamefont {V.}~\bibnamefont
  {Simion}}, \bibinfo {author} {\bibfnamefont {K.}~\bibnamefont
  {Siwek-Wilczy\ifmmode~\acute{n}\else \'{n}\fi{}ska}}, \bibinfo {author}
  {\bibfnamefont {A.}~\bibnamefont {Somov}}, \bibinfo {author} {\bibfnamefont
  {M.~R.}\ \bibnamefont {Stockmeier}}, \bibinfo {author} {\bibfnamefont
  {G.}~\bibnamefont {Stoicea}}, \bibinfo {author} {\bibfnamefont
  {M.}~\bibnamefont {Vasiliev}}, \bibinfo {author} {\bibfnamefont
  {P.}~\bibnamefont {Wagner}}, \bibinfo {author} {\bibfnamefont
  {K.}~\bibnamefont {Wi\ifmmode~\acute{s}\else \'{s}\fi{}niewski}}, \bibinfo
  {author} {\bibfnamefont {D.}~\bibnamefont {Wohlfarth}}, \bibinfo {author}
  {\bibfnamefont {J.~T.}\ \bibnamefont {Yang}}, \bibinfo {author}
  {\bibfnamefont {I.}~\bibnamefont {Yushmanov}},\ and\ \bibinfo {author}
  {\bibfnamefont {A.}~\bibnamefont {Zhilin}} (\bibinfo {collaboration} {FOPI
  Collaboration}),\ }\href {https://doi.org/10.1103/PhysRevLett.84.1120}
  {\bibfield  {journal} {\bibinfo  {journal} {Phys. Rev. Lett.}\ }\textbf
  {\bibinfo {volume} {84}},\ \bibinfo {pages} {1120} (\bibinfo {year}
  {2000})}\BibitemShut {NoStop}%
\bibitem [{\citenamefont {Tsang}\ \emph {et~al.}(2004)\citenamefont {Tsang},
  \citenamefont {Liu}, \citenamefont {Shi}, \citenamefont {Danielewicz},
  \citenamefont {Gelbke}, \citenamefont {Liu}, \citenamefont {Lynch},
  \citenamefont {Tan}, \citenamefont {Verde}, \citenamefont {Wagner},
  \citenamefont {Xu}, \citenamefont {Friedman}, \citenamefont {Beaulieu},
  \citenamefont {Davin}, \citenamefont {de~Souza}, \citenamefont {Larochelle},
  \citenamefont {Lefort}, \citenamefont {Yanez}, \citenamefont {Viola},
  \citenamefont {Charity},\ and\ \citenamefont {Sobotka}}]{Tsang2004}%
  \BibitemOpen
  \bibfield  {author} {\bibinfo {author} {\bibfnamefont {M.~B.}\ \bibnamefont
  {Tsang}}, \bibinfo {author} {\bibfnamefont {T.~X.}\ \bibnamefont {Liu}},
  \bibinfo {author} {\bibfnamefont {L.}~\bibnamefont {Shi}}, \bibinfo {author}
  {\bibfnamefont {P.}~\bibnamefont {Danielewicz}}, \bibinfo {author}
  {\bibfnamefont {C.~K.}\ \bibnamefont {Gelbke}}, \bibinfo {author}
  {\bibfnamefont {X.~D.}\ \bibnamefont {Liu}}, \bibinfo {author} {\bibfnamefont
  {W.~G.}\ \bibnamefont {Lynch}}, \bibinfo {author} {\bibfnamefont {W.~P.}\
  \bibnamefont {Tan}}, \bibinfo {author} {\bibfnamefont {G.}~\bibnamefont
  {Verde}}, \bibinfo {author} {\bibfnamefont {A.}~\bibnamefont {Wagner}},
  \bibinfo {author} {\bibfnamefont {H.~S.}\ \bibnamefont {Xu}}, \bibinfo
  {author} {\bibfnamefont {W.~A.}\ \bibnamefont {Friedman}}, \bibinfo {author}
  {\bibfnamefont {L.}~\bibnamefont {Beaulieu}}, \bibinfo {author}
  {\bibfnamefont {B.}~\bibnamefont {Davin}}, \bibinfo {author} {\bibfnamefont
  {R.~T.}\ \bibnamefont {de~Souza}}, \bibinfo {author} {\bibfnamefont
  {Y.}~\bibnamefont {Larochelle}}, \bibinfo {author} {\bibfnamefont
  {T.}~\bibnamefont {Lefort}}, \bibinfo {author} {\bibfnamefont
  {R.}~\bibnamefont {Yanez}}, \bibinfo {author} {\bibfnamefont {V.~E.}\
  \bibnamefont {Viola}}, \bibinfo {author} {\bibfnamefont {R.~J.}\ \bibnamefont
  {Charity}},\ and\ \bibinfo {author} {\bibfnamefont {L.~G.}\ \bibnamefont
  {Sobotka}},\ }\href {https://doi.org/10.1103/PhysRevLett.92.062701}
  {\bibfield  {journal} {\bibinfo  {journal} {Phys. Rev. Lett.}\ }\textbf
  {\bibinfo {volume} {92}},\ \bibinfo {pages} {062701} (\bibinfo {year}
  {2004})}\BibitemShut {NoStop}%
\bibitem [{\citenamefont {Chen}\ \emph {et~al.}(2005)\citenamefont {Chen},
  \citenamefont {Ko},\ and\ \citenamefont {Li}}]{Chen2005}%
  \BibitemOpen
  \bibfield  {author} {\bibinfo {author} {\bibfnamefont {L.-W.}\ \bibnamefont
  {Chen}}, \bibinfo {author} {\bibfnamefont {C.~M.}\ \bibnamefont {Ko}},\ and\
  \bibinfo {author} {\bibfnamefont {B.-A.}\ \bibnamefont {Li}},\ }\href
  {https://doi.org/10.1103/PhysRevLett.94.032701} {\bibfield  {journal}
  {\bibinfo  {journal} {Phys. Rev. Lett.}\ }\textbf {\bibinfo {volume} {94}},\
  \bibinfo {pages} {032701} (\bibinfo {year} {2005})}\BibitemShut {NoStop}%
\bibitem [{\citenamefont {Liu}\ \emph {et~al.}(2007)\citenamefont {Liu},
  \citenamefont {Lynch}, \citenamefont {Tsang}, \citenamefont {Liu},
  \citenamefont {Shomin}, \citenamefont {Tan}, \citenamefont {Verde},
  \citenamefont {Wagner}, \citenamefont {Xi}, \citenamefont {Xu}, \citenamefont
  {Davin}, \citenamefont {Larochelle}, \citenamefont {{de Souza}},
  \citenamefont {Charity},\ and\ \citenamefont {Sobotka}}]{Liu2007}%
  \BibitemOpen
  \bibfield  {author} {\bibinfo {author} {\bibfnamefont {T.~X.}\ \bibnamefont
  {Liu}}, \bibinfo {author} {\bibfnamefont {W.~G.}\ \bibnamefont {Lynch}},
  \bibinfo {author} {\bibfnamefont {M.~B.}\ \bibnamefont {Tsang}}, \bibinfo
  {author} {\bibfnamefont {X.~D.}\ \bibnamefont {Liu}}, \bibinfo {author}
  {\bibfnamefont {R.}~\bibnamefont {Shomin}}, \bibinfo {author} {\bibfnamefont
  {W.~P.}\ \bibnamefont {Tan}}, \bibinfo {author} {\bibfnamefont
  {G.}~\bibnamefont {Verde}}, \bibinfo {author} {\bibfnamefont
  {A.}~\bibnamefont {Wagner}}, \bibinfo {author} {\bibfnamefont {H.~F.}\
  \bibnamefont {Xi}}, \bibinfo {author} {\bibfnamefont {H.~S.}\ \bibnamefont
  {Xu}}, \bibinfo {author} {\bibfnamefont {B.}~\bibnamefont {Davin}}, \bibinfo
  {author} {\bibfnamefont {Y.}~\bibnamefont {Larochelle}}, \bibinfo {author}
  {\bibfnamefont {R.~T.}\ \bibnamefont {{de Souza}}}, \bibinfo {author}
  {\bibfnamefont {R.~J.}\ \bibnamefont {Charity}},\ and\ \bibinfo {author}
  {\bibfnamefont {L.~G.}\ \bibnamefont {Sobotka}},\ }\href
  {https://doi.org/10.1103/PhysRevC.76.034603} {\bibfield  {journal} {\bibinfo
  {journal} {Phys. Rev. C}\ }\textbf {\bibinfo {volume} {76}},\ \bibinfo
  {pages} {034603} (\bibinfo {year} {2007})}\BibitemShut {NoStop}%
\bibitem [{\citenamefont {Tsang}\ \emph {et~al.}(2009)\citenamefont {Tsang},
  \citenamefont {Zhang}, \citenamefont {Danielewicz}, \citenamefont {Famiano},
  \citenamefont {Li}, \citenamefont {Lynch},\ and\ \citenamefont
  {Steiner}}]{Tsang2009}%
  \BibitemOpen
  \bibfield  {author} {\bibinfo {author} {\bibfnamefont {M.~B.}\ \bibnamefont
  {Tsang}}, \bibinfo {author} {\bibfnamefont {Y.}~\bibnamefont {Zhang}},
  \bibinfo {author} {\bibfnamefont {P.}~\bibnamefont {Danielewicz}}, \bibinfo
  {author} {\bibfnamefont {M.}~\bibnamefont {Famiano}}, \bibinfo {author}
  {\bibfnamefont {Z.}~\bibnamefont {Li}}, \bibinfo {author} {\bibfnamefont
  {W.~G.}\ \bibnamefont {Lynch}},\ and\ \bibinfo {author} {\bibfnamefont
  {A.~W.}\ \bibnamefont {Steiner}},\ }\href
  {https://doi.org/10.1103/PhysRevLett.102.122701} {\bibfield  {journal}
  {\bibinfo  {journal} {Phys. Rev. Lett.}\ }\textbf {\bibinfo {volume} {102}},\
  \bibinfo {pages} {122701} (\bibinfo {year} {2009})}\BibitemShut {NoStop}%
\bibitem [{\citenamefont {Coupland}\ \emph {et~al.}(2011)\citenamefont
  {Coupland}, \citenamefont {Lynch}, \citenamefont {Tsang}, \citenamefont
  {Danielewicz},\ and\ \citenamefont {Zhang}}]{Coupland2011}%
  \BibitemOpen
  \bibfield  {author} {\bibinfo {author} {\bibfnamefont {D.~D.~S.}\
  \bibnamefont {Coupland}}, \bibinfo {author} {\bibfnamefont {W.~G.}\
  \bibnamefont {Lynch}}, \bibinfo {author} {\bibfnamefont {M.~B.}\ \bibnamefont
  {Tsang}}, \bibinfo {author} {\bibfnamefont {P.}~\bibnamefont {Danielewicz}},\
  and\ \bibinfo {author} {\bibfnamefont {Y.}~\bibnamefont {Zhang}},\ }\href
  {https://doi.org/10.1103/PhysRevC.84.054603} {\bibfield  {journal} {\bibinfo
  {journal} {Phys. Rev. C}\ }\textbf {\bibinfo {volume} {84}},\ \bibinfo
  {pages} {054603} (\bibinfo {year} {2011})}\BibitemShut {NoStop}%
\bibitem [{\citenamefont {Camaiani}\ \emph {et~al.}(2020)\citenamefont
  {Camaiani}, \citenamefont {Piantelli}, \citenamefont {Ono}, \citenamefont
  {Casini}, \citenamefont {Borderie}, \citenamefont {Bougault}, \citenamefont
  {Ciampi}, \citenamefont {Due\~nas}, \citenamefont {Frosin}, \citenamefont
  {Frankland}, \citenamefont {Gruyer}, \citenamefont {LeNeindre}, \citenamefont
  {Lombardo}, \citenamefont {Mantovani}, \citenamefont {Ottanelli},
  \citenamefont {Parlog}, \citenamefont {Pasquali}, \citenamefont {Upadhyaya},
  \citenamefont {Valdr\'e}, \citenamefont {Verde},\ and\ \citenamefont
  {Vient}}]{Camaiani2020}%
  \BibitemOpen
  \bibfield  {author} {\bibinfo {author} {\bibfnamefont {A.}~\bibnamefont
  {Camaiani}}, \bibinfo {author} {\bibfnamefont {S.}~\bibnamefont {Piantelli}},
  \bibinfo {author} {\bibfnamefont {A.}~\bibnamefont {Ono}}, \bibinfo {author}
  {\bibfnamefont {G.}~\bibnamefont {Casini}}, \bibinfo {author} {\bibfnamefont
  {B.}~\bibnamefont {Borderie}}, \bibinfo {author} {\bibfnamefont
  {R.}~\bibnamefont {Bougault}}, \bibinfo {author} {\bibfnamefont
  {C.}~\bibnamefont {Ciampi}}, \bibinfo {author} {\bibfnamefont {J.~A.}\
  \bibnamefont {Due\~nas}}, \bibinfo {author} {\bibfnamefont {C.}~\bibnamefont
  {Frosin}}, \bibinfo {author} {\bibfnamefont {J.~D.}\ \bibnamefont
  {Frankland}}, \bibinfo {author} {\bibfnamefont {D.}~\bibnamefont {Gruyer}},
  \bibinfo {author} {\bibfnamefont {N.}~\bibnamefont {LeNeindre}}, \bibinfo
  {author} {\bibfnamefont {I.}~\bibnamefont {Lombardo}}, \bibinfo {author}
  {\bibfnamefont {G.}~\bibnamefont {Mantovani}}, \bibinfo {author}
  {\bibfnamefont {P.}~\bibnamefont {Ottanelli}}, \bibinfo {author}
  {\bibfnamefont {M.}~\bibnamefont {Parlog}}, \bibinfo {author} {\bibfnamefont
  {G.}~\bibnamefont {Pasquali}}, \bibinfo {author} {\bibfnamefont
  {S.}~\bibnamefont {Upadhyaya}}, \bibinfo {author} {\bibfnamefont
  {S.}~\bibnamefont {Valdr\'e}}, \bibinfo {author} {\bibfnamefont
  {G.}~\bibnamefont {Verde}},\ and\ \bibinfo {author} {\bibfnamefont
  {E.}~\bibnamefont {Vient}},\ }\href
  {https://doi.org/10.1103/PhysRevC.102.044607} {\bibfield  {journal} {\bibinfo
   {journal} {Phys. Rev. C}\ }\textbf {\bibinfo {volume} {102}},\ \bibinfo
  {pages} {044607} (\bibinfo {year} {2020})}\BibitemShut {NoStop}%
\bibitem [{\citenamefont {Ciampi}\ \emph {et~al.}(2023)\citenamefont {Ciampi},
  \citenamefont {Piantelli}, \citenamefont {Casini}, \citenamefont {Ono},
  \citenamefont {Frankland}, \citenamefont {Baldesi}, \citenamefont {Barlini},
  \citenamefont {Borderie}, \citenamefont {Bougault}, \citenamefont {Camaiani},
  \citenamefont {Chbihi}, \citenamefont {Due\~nas}, \citenamefont {Fable},
  \citenamefont {Fabris}, \citenamefont {Frosin}, \citenamefont {G\'enard},
  \citenamefont {Gramegna}, \citenamefont {Gruyer}, \citenamefont {Henri},
  \citenamefont {Hong}, \citenamefont {Kim}, \citenamefont {Kordyasz},
  \citenamefont {Kozik}, \citenamefont {Kweon}, \citenamefont {Le~Neindre},
  \citenamefont {Lombardo}, \citenamefont {Lopez}, \citenamefont {Marchi},
  \citenamefont {Mazurek}, \citenamefont {Nam}, \citenamefont {Park},
  \citenamefont {P\^arlog}, \citenamefont {Pasquali}, \citenamefont {Poggi},
  \citenamefont {Rebillard-Souli\'e}, \citenamefont {Stefanini}, \citenamefont
  {Upadhyaya}, \citenamefont {Valdr\'e}, \citenamefont {Verde}, \citenamefont
  {Vient},\ and\ \citenamefont {Vigilante}}]{Ciampi2023}%
  \BibitemOpen
  \bibfield  {author} {\bibinfo {author} {\bibfnamefont {C.}~\bibnamefont
  {Ciampi}}, \bibinfo {author} {\bibfnamefont {S.}~\bibnamefont {Piantelli}},
  \bibinfo {author} {\bibfnamefont {G.}~\bibnamefont {Casini}}, \bibinfo
  {author} {\bibfnamefont {A.}~\bibnamefont {Ono}}, \bibinfo {author}
  {\bibfnamefont {J.~D.}\ \bibnamefont {Frankland}}, \bibinfo {author}
  {\bibfnamefont {L.}~\bibnamefont {Baldesi}}, \bibinfo {author} {\bibfnamefont
  {S.}~\bibnamefont {Barlini}}, \bibinfo {author} {\bibfnamefont
  {B.}~\bibnamefont {Borderie}}, \bibinfo {author} {\bibfnamefont
  {R.}~\bibnamefont {Bougault}}, \bibinfo {author} {\bibfnamefont
  {A.}~\bibnamefont {Camaiani}}, \bibinfo {author} {\bibfnamefont
  {A.}~\bibnamefont {Chbihi}}, \bibinfo {author} {\bibfnamefont {J.~A.}\
  \bibnamefont {Due\~nas}}, \bibinfo {author} {\bibfnamefont {Q.}~\bibnamefont
  {Fable}}, \bibinfo {author} {\bibfnamefont {D.}~\bibnamefont {Fabris}},
  \bibinfo {author} {\bibfnamefont {C.}~\bibnamefont {Frosin}}, \bibinfo
  {author} {\bibfnamefont {T.}~\bibnamefont {G\'enard}}, \bibinfo {author}
  {\bibfnamefont {F.}~\bibnamefont {Gramegna}}, \bibinfo {author}
  {\bibfnamefont {D.}~\bibnamefont {Gruyer}}, \bibinfo {author} {\bibfnamefont
  {M.}~\bibnamefont {Henri}}, \bibinfo {author} {\bibfnamefont
  {B.}~\bibnamefont {Hong}}, \bibinfo {author} {\bibfnamefont {S.}~\bibnamefont
  {Kim}}, \bibinfo {author} {\bibfnamefont {A.}~\bibnamefont {Kordyasz}},
  \bibinfo {author} {\bibfnamefont {T.}~\bibnamefont {Kozik}}, \bibinfo
  {author} {\bibfnamefont {M.~J.}\ \bibnamefont {Kweon}}, \bibinfo {author}
  {\bibfnamefont {N.}~\bibnamefont {Le~Neindre}}, \bibinfo {author}
  {\bibfnamefont {I.}~\bibnamefont {Lombardo}}, \bibinfo {author}
  {\bibfnamefont {O.}~\bibnamefont {Lopez}}, \bibinfo {author} {\bibfnamefont
  {T.}~\bibnamefont {Marchi}}, \bibinfo {author} {\bibfnamefont
  {K.}~\bibnamefont {Mazurek}}, \bibinfo {author} {\bibfnamefont {S.~H.}\
  \bibnamefont {Nam}}, \bibinfo {author} {\bibfnamefont {J.}~\bibnamefont
  {Park}}, \bibinfo {author} {\bibfnamefont {M.}~\bibnamefont {P\^arlog}},
  \bibinfo {author} {\bibfnamefont {G.}~\bibnamefont {Pasquali}}, \bibinfo
  {author} {\bibfnamefont {G.}~\bibnamefont {Poggi}}, \bibinfo {author}
  {\bibfnamefont {A.}~\bibnamefont {Rebillard-Souli\'e}}, \bibinfo {author}
  {\bibfnamefont {A.~A.}\ \bibnamefont {Stefanini}}, \bibinfo {author}
  {\bibfnamefont {S.}~\bibnamefont {Upadhyaya}}, \bibinfo {author}
  {\bibfnamefont {S.}~\bibnamefont {Valdr\'e}}, \bibinfo {author}
  {\bibfnamefont {G.}~\bibnamefont {Verde}}, \bibinfo {author} {\bibfnamefont
  {E.}~\bibnamefont {Vient}},\ and\ \bibinfo {author} {\bibfnamefont
  {M.}~\bibnamefont {Vigilante}} (\bibinfo {collaboration} {INDRA-FAZIA
  Collaboration}),\ }\href {https://doi.org/10.1103/PhysRevC.108.054611}
  {\bibfield  {journal} {\bibinfo  {journal} {Phys. Rev. C}\ }\textbf {\bibinfo
  {volume} {108}},\ \bibinfo {pages} {054611} (\bibinfo {year}
  {2023})}\BibitemShut {NoStop}%
\bibitem [{\citenamefont {Fable}\ \emph {et~al.}(2024)\citenamefont {Fable},
  \citenamefont {Baldesi}, \citenamefont {Barlini}, \citenamefont {Bonnet},
  \citenamefont {Borderie}, \citenamefont {Bougault}, \citenamefont {Camaiani},
  \citenamefont {Casini}, \citenamefont {Chbihi}, \citenamefont {Ciampi},
  \citenamefont {Due\~nas}, \citenamefont {Frankland}, \citenamefont
  {G\'enard}, \citenamefont {Gruyer}, \citenamefont {Henri}, \citenamefont
  {Hong}, \citenamefont {Kim}, \citenamefont {Kordyasz}, \citenamefont {Kozik},
  \citenamefont {Le~F\`evre}, \citenamefont {Le~Neindre}, \citenamefont
  {Lombardo}, \citenamefont {Lopez}, \citenamefont {Marchi}, \citenamefont
  {Marini}, \citenamefont {Nam}, \citenamefont {Ono}, \citenamefont {Park},
  \citenamefont {P\^arlog}, \citenamefont {Piantelli}, \citenamefont
  {Rebillard-Souli\'e}, \citenamefont {Verde},\ and\ \citenamefont
  {Vient}}]{Fable2024}%
  \BibitemOpen
  \bibfield  {author} {\bibinfo {author} {\bibfnamefont {Q.}~\bibnamefont
  {Fable}}, \bibinfo {author} {\bibfnamefont {L.}~\bibnamefont {Baldesi}},
  \bibinfo {author} {\bibfnamefont {S.}~\bibnamefont {Barlini}}, \bibinfo
  {author} {\bibfnamefont {E.}~\bibnamefont {Bonnet}}, \bibinfo {author}
  {\bibfnamefont {B.}~\bibnamefont {Borderie}}, \bibinfo {author}
  {\bibfnamefont {R.}~\bibnamefont {Bougault}}, \bibinfo {author}
  {\bibfnamefont {A.}~\bibnamefont {Camaiani}}, \bibinfo {author}
  {\bibfnamefont {G.}~\bibnamefont {Casini}}, \bibinfo {author} {\bibfnamefont
  {A.}~\bibnamefont {Chbihi}}, \bibinfo {author} {\bibfnamefont
  {C.}~\bibnamefont {Ciampi}}, \bibinfo {author} {\bibfnamefont {J.~A.}\
  \bibnamefont {Due\~nas}}, \bibinfo {author} {\bibfnamefont {J.~D.}\
  \bibnamefont {Frankland}}, \bibinfo {author} {\bibfnamefont {T.}~\bibnamefont
  {G\'enard}}, \bibinfo {author} {\bibfnamefont {D.}~\bibnamefont {Gruyer}},
  \bibinfo {author} {\bibfnamefont {M.}~\bibnamefont {Henri}}, \bibinfo
  {author} {\bibfnamefont {B.}~\bibnamefont {Hong}}, \bibinfo {author}
  {\bibfnamefont {S.}~\bibnamefont {Kim}}, \bibinfo {author} {\bibfnamefont
  {A.~J.}\ \bibnamefont {Kordyasz}}, \bibinfo {author} {\bibfnamefont
  {T.}~\bibnamefont {Kozik}}, \bibinfo {author} {\bibfnamefont
  {A.}~\bibnamefont {Le~F\`evre}}, \bibinfo {author} {\bibfnamefont
  {N.}~\bibnamefont {Le~Neindre}}, \bibinfo {author} {\bibfnamefont
  {I.}~\bibnamefont {Lombardo}}, \bibinfo {author} {\bibfnamefont
  {O.}~\bibnamefont {Lopez}}, \bibinfo {author} {\bibfnamefont
  {T.}~\bibnamefont {Marchi}}, \bibinfo {author} {\bibfnamefont
  {P.}~\bibnamefont {Marini}}, \bibinfo {author} {\bibfnamefont {S.~H.}\
  \bibnamefont {Nam}}, \bibinfo {author} {\bibfnamefont {A.}~\bibnamefont
  {Ono}}, \bibinfo {author} {\bibfnamefont {J.}~\bibnamefont {Park}}, \bibinfo
  {author} {\bibfnamefont {M.}~\bibnamefont {P\^arlog}}, \bibinfo {author}
  {\bibfnamefont {S.}~\bibnamefont {Piantelli}}, \bibinfo {author}
  {\bibfnamefont {A.}~\bibnamefont {Rebillard-Souli\'e}}, \bibinfo {author}
  {\bibfnamefont {G.}~\bibnamefont {Verde}},\ and\ \bibinfo {author}
  {\bibfnamefont {E.}~\bibnamefont {Vient}} (\bibinfo {collaboration} {INDRA
  and INDRA-FAZIA Collaborations}),\ }\href
  {https://doi.org/10.1103/PhysRevC.109.064605} {\bibfield  {journal} {\bibinfo
   {journal} {Phys. Rev. C}\ }\textbf {\bibinfo {volume} {109}},\ \bibinfo
  {pages} {064605} (\bibinfo {year} {2024})}\BibitemShut {NoStop}%
\bibitem [{\citenamefont {Mallik}\ \emph {et~al.}(2021)\citenamefont {Mallik},
  \citenamefont {Gulminelli},\ and\ \citenamefont {Gruyer}}]{Mallik2022}%
  \BibitemOpen
  \bibfield  {author} {\bibinfo {author} {\bibfnamefont {S.}~\bibnamefont
  {Mallik}}, \bibinfo {author} {\bibfnamefont {F.}~\bibnamefont {Gulminelli}},\
  and\ \bibinfo {author} {\bibfnamefont {D.}~\bibnamefont {Gruyer}},\ }\href
  {https://doi.org/10.1088/1361-6471/ac3473} {\bibfield  {journal} {\bibinfo
  {journal} {J. Phys. G: Nucl. Part. Phys.}\ }\textbf {\bibinfo {volume}
  {49}},\ \bibinfo {pages} {015102} (\bibinfo {year} {2021})}\BibitemShut
  {NoStop}%
\bibitem [{\citenamefont {Piantelli}\ \emph {et~al.}(2019)\citenamefont
  {Piantelli}, \citenamefont {Olmi}, \citenamefont {Maurenzig}, \citenamefont
  {Ono}, \citenamefont {Bini}, \citenamefont {Casini}, \citenamefont
  {Pasquali}, \citenamefont {Mangiarotti}, \citenamefont {Poggi}, \citenamefont
  {Stefanini}, \citenamefont {Barlini}, \citenamefont {Camaiani}, \citenamefont
  {Ciampi}, \citenamefont {Frosin}, \citenamefont {Ottanelli},\ and\
  \citenamefont {Valdr\'e}}]{Piantelli2019}%
  \BibitemOpen
  \bibfield  {author} {\bibinfo {author} {\bibfnamefont {S.}~\bibnamefont
  {Piantelli}}, \bibinfo {author} {\bibfnamefont {A.}~\bibnamefont {Olmi}},
  \bibinfo {author} {\bibfnamefont {P.~R.}\ \bibnamefont {Maurenzig}}, \bibinfo
  {author} {\bibfnamefont {A.}~\bibnamefont {Ono}}, \bibinfo {author}
  {\bibfnamefont {M.}~\bibnamefont {Bini}}, \bibinfo {author} {\bibfnamefont
  {G.}~\bibnamefont {Casini}}, \bibinfo {author} {\bibfnamefont
  {G.}~\bibnamefont {Pasquali}}, \bibinfo {author} {\bibfnamefont
  {A.}~\bibnamefont {Mangiarotti}}, \bibinfo {author} {\bibfnamefont
  {G.}~\bibnamefont {Poggi}}, \bibinfo {author} {\bibfnamefont {A.~A.}\
  \bibnamefont {Stefanini}}, \bibinfo {author} {\bibfnamefont {S.}~\bibnamefont
  {Barlini}}, \bibinfo {author} {\bibfnamefont {A.}~\bibnamefont {Camaiani}},
  \bibinfo {author} {\bibfnamefont {C.}~\bibnamefont {Ciampi}}, \bibinfo
  {author} {\bibfnamefont {C.}~\bibnamefont {Frosin}}, \bibinfo {author}
  {\bibfnamefont {P.}~\bibnamefont {Ottanelli}},\ and\ \bibinfo {author}
  {\bibfnamefont {S.}~\bibnamefont {Valdr\'e}},\ }\href
  {https://doi.org/10.1103/PhysRevC.99.064616} {\bibfield  {journal} {\bibinfo
  {journal} {Phys. Rev. C}\ }\textbf {\bibinfo {volume} {99}},\ \bibinfo
  {pages} {064616} (\bibinfo {year} {2019})}\BibitemShut {NoStop}%
\bibitem [{\citenamefont {Li}\ \emph {et~al.}(2018)\citenamefont {Li},
  \citenamefont {Zhang}, \citenamefont {Li}, \citenamefont {Wang},
  \citenamefont {Cui},\ and\ \citenamefont {Winkelbauer}}]{Li2018}%
  \BibitemOpen
  \bibfield  {author} {\bibinfo {author} {\bibfnamefont {L.}~\bibnamefont
  {Li}}, \bibinfo {author} {\bibfnamefont {Y.}~\bibnamefont {Zhang}}, \bibinfo
  {author} {\bibfnamefont {Z.}~\bibnamefont {Li}}, \bibinfo {author}
  {\bibfnamefont {N.}~\bibnamefont {Wang}}, \bibinfo {author} {\bibfnamefont
  {Y.}~\bibnamefont {Cui}},\ and\ \bibinfo {author} {\bibfnamefont
  {J.}~\bibnamefont {Winkelbauer}},\ }\href
  {https://doi.org/10.1103/PhysRevC.97.044606} {\bibfield  {journal} {\bibinfo
  {journal} {Phys. Rev. C}\ }\textbf {\bibinfo {volume} {97}},\ \bibinfo
  {pages} {044606} (\bibinfo {year} {2018})}\BibitemShut {NoStop}%
\bibitem [{\citenamefont {Frankland}\ \emph {et~al.}(2021)\citenamefont
  {Frankland}, \citenamefont {Gruyer}, \citenamefont {Bonnet}, \citenamefont
  {Borderie}, \citenamefont {Bougault}, \citenamefont {Chbihi}, \citenamefont
  {Ducret}, \citenamefont {Durand}, \citenamefont {Fable}, \citenamefont
  {Henri}, \citenamefont {Lemari\'e}, \citenamefont {Le~Neindre}, \citenamefont
  {Lombardo}, \citenamefont {Lopez}, \citenamefont {Manduci}, \citenamefont
  {P\^arlog}, \citenamefont {Quicray}, \citenamefont {Verde}, \citenamefont
  {Vient},\ and\ \citenamefont {Vigilante}}]{Frankland2021}%
  \BibitemOpen
  \bibfield  {author} {\bibinfo {author} {\bibfnamefont {J.~D.}\ \bibnamefont
  {Frankland}}, \bibinfo {author} {\bibfnamefont {D.}~\bibnamefont {Gruyer}},
  \bibinfo {author} {\bibfnamefont {E.}~\bibnamefont {Bonnet}}, \bibinfo
  {author} {\bibfnamefont {B.}~\bibnamefont {Borderie}}, \bibinfo {author}
  {\bibfnamefont {R.}~\bibnamefont {Bougault}}, \bibinfo {author}
  {\bibfnamefont {A.}~\bibnamefont {Chbihi}}, \bibinfo {author} {\bibfnamefont
  {J.~E.}\ \bibnamefont {Ducret}}, \bibinfo {author} {\bibfnamefont
  {D.}~\bibnamefont {Durand}}, \bibinfo {author} {\bibfnamefont
  {Q.}~\bibnamefont {Fable}}, \bibinfo {author} {\bibfnamefont
  {M.}~\bibnamefont {Henri}}, \bibinfo {author} {\bibfnamefont
  {J.}~\bibnamefont {Lemari\'e}}, \bibinfo {author} {\bibfnamefont
  {N.}~\bibnamefont {Le~Neindre}}, \bibinfo {author} {\bibfnamefont
  {I.}~\bibnamefont {Lombardo}}, \bibinfo {author} {\bibfnamefont
  {O.}~\bibnamefont {Lopez}}, \bibinfo {author} {\bibfnamefont
  {L.}~\bibnamefont {Manduci}}, \bibinfo {author} {\bibfnamefont
  {M.}~\bibnamefont {P\^arlog}}, \bibinfo {author} {\bibfnamefont
  {J.}~\bibnamefont {Quicray}}, \bibinfo {author} {\bibfnamefont
  {G.}~\bibnamefont {Verde}}, \bibinfo {author} {\bibfnamefont
  {E.}~\bibnamefont {Vient}},\ and\ \bibinfo {author} {\bibfnamefont
  {M.}~\bibnamefont {Vigilante}} (\bibinfo {collaboration} {INDRA
  Collaboration}),\ }\href {https://doi.org/10.1103/PhysRevC.104.034609}
  {\bibfield  {journal} {\bibinfo  {journal} {Phys. Rev. C}\ }\textbf {\bibinfo
  {volume} {104}},\ \bibinfo {pages} {034609} (\bibinfo {year}
  {2021})}\BibitemShut {NoStop}%
\bibitem [{\citenamefont {Cavata}\ \emph {et~al.}(1990)\citenamefont {Cavata},
  \citenamefont {Demoulins}, \citenamefont {Gosset}, \citenamefont {Lemaire},
  \citenamefont {L'H\^ote}, \citenamefont {Poitou},\ and\ \citenamefont
  {Valette}}]{Cavata1990}%
  \BibitemOpen
  \bibfield  {author} {\bibinfo {author} {\bibfnamefont {C.}~\bibnamefont
  {Cavata}}, \bibinfo {author} {\bibfnamefont {M.}~\bibnamefont {Demoulins}},
  \bibinfo {author} {\bibfnamefont {J.}~\bibnamefont {Gosset}}, \bibinfo
  {author} {\bibfnamefont {M.-C.}\ \bibnamefont {Lemaire}}, \bibinfo {author}
  {\bibfnamefont {D.}~\bibnamefont {L'H\^ote}}, \bibinfo {author}
  {\bibfnamefont {J.}~\bibnamefont {Poitou}},\ and\ \bibinfo {author}
  {\bibfnamefont {O.}~\bibnamefont {Valette}},\ }\href
  {https://doi.org/10.1103/PhysRevC.42.1760} {\bibfield  {journal} {\bibinfo
  {journal} {Phys. Rev. C}\ }\textbf {\bibinfo {volume} {42}},\ \bibinfo
  {pages} {1760} (\bibinfo {year} {1990})}\BibitemShut {NoStop}%
\bibitem [{\citenamefont {Das}\ \emph {et~al.}(2018)\citenamefont {Das},
  \citenamefont {Giacalone}, \citenamefont {Monard},\ and\ \citenamefont
  {Ollitrault}}]{Das2018}%
  \BibitemOpen
  \bibfield  {author} {\bibinfo {author} {\bibfnamefont {S.~J.}\ \bibnamefont
  {Das}}, \bibinfo {author} {\bibfnamefont {G.}~\bibnamefont {Giacalone}},
  \bibinfo {author} {\bibfnamefont {P.-A.}\ \bibnamefont {Monard}},\ and\
  \bibinfo {author} {\bibfnamefont {J.-Y.}\ \bibnamefont {Ollitrault}},\ }\href
  {https://doi.org/10.1103/PhysRevC.97.014905} {\bibfield  {journal} {\bibinfo
  {journal} {Phys. Rev. C}\ }\textbf {\bibinfo {volume} {97}},\ \bibinfo
  {pages} {014905} (\bibinfo {year} {2018})}\BibitemShut {NoStop}%
\bibitem [{\citenamefont {Rogly}\ \emph {et~al.}(2018)\citenamefont {Rogly},
  \citenamefont {Giacalone},\ and\ \citenamefont {Ollitrault}}]{Rogly2018}%
  \BibitemOpen
  \bibfield  {author} {\bibinfo {author} {\bibfnamefont {R.}~\bibnamefont
  {Rogly}}, \bibinfo {author} {\bibfnamefont {G.}~\bibnamefont {Giacalone}},\
  and\ \bibinfo {author} {\bibfnamefont {J.-Y.}\ \bibnamefont {Ollitrault}},\
  }\href {https://doi.org/10.1103/PhysRevC.98.024902} {\bibfield  {journal}
  {\bibinfo  {journal} {Phys. Rev. C}\ }\textbf {\bibinfo {volume} {98}},\
  \bibinfo {pages} {024902} (\bibinfo {year} {2018})}\BibitemShut {NoStop}%
\bibitem [{\citenamefont {Chen}\ \emph {et~al.}(2023)\citenamefont {Chen},
  \citenamefont {Li}, \citenamefont {Cui}, \citenamefont {Yang}, \citenamefont
  {Li},\ and\ \citenamefont {Zhang}}]{Chen2023}%
  \BibitemOpen
  \bibfield  {author} {\bibinfo {author} {\bibfnamefont {X.}~\bibnamefont
  {Chen}}, \bibinfo {author} {\bibfnamefont {L.}~\bibnamefont {Li}}, \bibinfo
  {author} {\bibfnamefont {Y.}~\bibnamefont {Cui}}, \bibinfo {author}
  {\bibfnamefont {J.}~\bibnamefont {Yang}}, \bibinfo {author} {\bibfnamefont
  {Z.}~\bibnamefont {Li}},\ and\ \bibinfo {author} {\bibfnamefont
  {Y.}~\bibnamefont {Zhang}},\ }\href
  {https://doi.org/10.1103/PhysRevC.108.034613} {\bibfield  {journal} {\bibinfo
   {journal} {Phys. Rev. C}\ }\textbf {\bibinfo {volume} {108}},\ \bibinfo
  {pages} {034613} (\bibinfo {year} {2023})}\BibitemShut {NoStop}%
\bibitem [{\citenamefont {Pouthas}\ \emph {et~al.}(1995)\citenamefont
  {Pouthas}, \citenamefont {Borderie}, \citenamefont {Dayras}, \citenamefont
  {Plagnol}, \citenamefont {Rivet}, \citenamefont {Saint-Laurent},
  \citenamefont {Steckmeyer}, \citenamefont {Auger}, \citenamefont {Bacri},
  \citenamefont {Barbey}, \citenamefont {Barbier}, \citenamefont {Benkirane},
  \citenamefont {Benlliure}, \citenamefont {Berthier}, \citenamefont
  {Bougamont}, \citenamefont {Bourgault}, \citenamefont {Box}, \citenamefont
  {Bzyl}, \citenamefont {Cahan}, \citenamefont {Cassagnou}, \citenamefont
  {Charlet}, \citenamefont {Charvet}, \citenamefont {Chbihi}, \citenamefont
  {Clerc}, \citenamefont {Copinet}, \citenamefont {Cussol}, \citenamefont
  {Engrand}, \citenamefont {Gautier}, \citenamefont {Huguet}, \citenamefont
  {Jouniaux}, \citenamefont {Laville}, \citenamefont {{Le Botlan}},
  \citenamefont {Leconte}, \citenamefont {Legrain}, \citenamefont {Lelong},
  \citenamefont {{Le Guay}}, \citenamefont {Martina}, \citenamefont {Mazur},
  \citenamefont {Mosrin}, \citenamefont {Olivier}, \citenamefont {Passerieux},
  \citenamefont {Pierre}, \citenamefont {Piquet}, \citenamefont {Plaige},
  \citenamefont {Pollacco}, \citenamefont {Raine}, \citenamefont {Richard},
  \citenamefont {Ropert}, \citenamefont {Spitaels}, \citenamefont {Stab},
  \citenamefont {Sznajderman}, \citenamefont {Tassan-got}, \citenamefont
  {Tillier}, \citenamefont {Tripon}, \citenamefont {Vallerand}, \citenamefont
  {Volant}, \citenamefont {Volkov}, \citenamefont {Wieleczko},\ and\
  \citenamefont {Wittwer}}]{Pouthas1995}%
  \BibitemOpen
  \bibfield  {author} {\bibinfo {author} {\bibfnamefont {J.}~\bibnamefont
  {Pouthas}}, \bibinfo {author} {\bibfnamefont {B.}~\bibnamefont {Borderie}},
  \bibinfo {author} {\bibfnamefont {R.}~\bibnamefont {Dayras}}, \bibinfo
  {author} {\bibfnamefont {E.}~\bibnamefont {Plagnol}}, \bibinfo {author}
  {\bibfnamefont {M.}~\bibnamefont {Rivet}}, \bibinfo {author} {\bibfnamefont
  {F.}~\bibnamefont {Saint-Laurent}}, \bibinfo {author} {\bibfnamefont
  {J.}~\bibnamefont {Steckmeyer}}, \bibinfo {author} {\bibfnamefont
  {G.}~\bibnamefont {Auger}}, \bibinfo {author} {\bibfnamefont
  {C.}~\bibnamefont {Bacri}}, \bibinfo {author} {\bibfnamefont
  {S.}~\bibnamefont {Barbey}}, \bibinfo {author} {\bibfnamefont
  {A.}~\bibnamefont {Barbier}}, \bibinfo {author} {\bibfnamefont
  {A.}~\bibnamefont {Benkirane}}, \bibinfo {author} {\bibfnamefont
  {J.}~\bibnamefont {Benlliure}}, \bibinfo {author} {\bibfnamefont
  {B.}~\bibnamefont {Berthier}}, \bibinfo {author} {\bibfnamefont
  {E.}~\bibnamefont {Bougamont}}, \bibinfo {author} {\bibfnamefont
  {P.}~\bibnamefont {Bourgault}}, \bibinfo {author} {\bibfnamefont
  {P.}~\bibnamefont {Box}}, \bibinfo {author} {\bibfnamefont {R.}~\bibnamefont
  {Bzyl}}, \bibinfo {author} {\bibfnamefont {B.}~\bibnamefont {Cahan}},
  \bibinfo {author} {\bibfnamefont {Y.}~\bibnamefont {Cassagnou}}, \bibinfo
  {author} {\bibfnamefont {D.}~\bibnamefont {Charlet}}, \bibinfo {author}
  {\bibfnamefont {J.}~\bibnamefont {Charvet}}, \bibinfo {author} {\bibfnamefont
  {A.}~\bibnamefont {Chbihi}}, \bibinfo {author} {\bibfnamefont
  {T.}~\bibnamefont {Clerc}}, \bibinfo {author} {\bibfnamefont
  {N.}~\bibnamefont {Copinet}}, \bibinfo {author} {\bibfnamefont
  {D.}~\bibnamefont {Cussol}}, \bibinfo {author} {\bibfnamefont
  {M.}~\bibnamefont {Engrand}}, \bibinfo {author} {\bibfnamefont
  {J.}~\bibnamefont {Gautier}}, \bibinfo {author} {\bibfnamefont
  {Y.}~\bibnamefont {Huguet}}, \bibinfo {author} {\bibfnamefont
  {O.}~\bibnamefont {Jouniaux}}, \bibinfo {author} {\bibfnamefont
  {J.}~\bibnamefont {Laville}}, \bibinfo {author} {\bibfnamefont
  {P.}~\bibnamefont {{Le Botlan}}}, \bibinfo {author} {\bibfnamefont
  {A.}~\bibnamefont {Leconte}}, \bibinfo {author} {\bibfnamefont
  {R.}~\bibnamefont {Legrain}}, \bibinfo {author} {\bibfnamefont
  {P.}~\bibnamefont {Lelong}}, \bibinfo {author} {\bibfnamefont
  {M.}~\bibnamefont {{Le Guay}}}, \bibinfo {author} {\bibfnamefont
  {L.}~\bibnamefont {Martina}}, \bibinfo {author} {\bibfnamefont
  {C.}~\bibnamefont {Mazur}}, \bibinfo {author} {\bibfnamefont
  {P.}~\bibnamefont {Mosrin}}, \bibinfo {author} {\bibfnamefont
  {L.}~\bibnamefont {Olivier}}, \bibinfo {author} {\bibfnamefont
  {J.}~\bibnamefont {Passerieux}}, \bibinfo {author} {\bibfnamefont
  {S.}~\bibnamefont {Pierre}}, \bibinfo {author} {\bibfnamefont
  {B.}~\bibnamefont {Piquet}}, \bibinfo {author} {\bibfnamefont
  {E.}~\bibnamefont {Plaige}}, \bibinfo {author} {\bibfnamefont
  {E.}~\bibnamefont {Pollacco}}, \bibinfo {author} {\bibfnamefont
  {B.}~\bibnamefont {Raine}}, \bibinfo {author} {\bibfnamefont
  {A.}~\bibnamefont {Richard}}, \bibinfo {author} {\bibfnamefont
  {J.}~\bibnamefont {Ropert}}, \bibinfo {author} {\bibfnamefont
  {C.}~\bibnamefont {Spitaels}}, \bibinfo {author} {\bibfnamefont
  {L.}~\bibnamefont {Stab}}, \bibinfo {author} {\bibfnamefont {D.}~\bibnamefont
  {Sznajderman}}, \bibinfo {author} {\bibfnamefont {L.}~\bibnamefont
  {Tassan-got}}, \bibinfo {author} {\bibfnamefont {J.}~\bibnamefont {Tillier}},
  \bibinfo {author} {\bibfnamefont {M.}~\bibnamefont {Tripon}}, \bibinfo
  {author} {\bibfnamefont {P.}~\bibnamefont {Vallerand}}, \bibinfo {author}
  {\bibfnamefont {C.}~\bibnamefont {Volant}}, \bibinfo {author} {\bibfnamefont
  {P.}~\bibnamefont {Volkov}}, \bibinfo {author} {\bibfnamefont
  {J.}~\bibnamefont {Wieleczko}},\ and\ \bibinfo {author} {\bibfnamefont
  {G.}~\bibnamefont {Wittwer}},\ }\href
  {https://doi.org/https://doi.org/10.1016/0168-9002(94)01543-0} {\bibfield
  {journal} {\bibinfo  {journal} {Nucl. Instr. and Meth. in Phys. Res. A}\
  }\textbf {\bibinfo {volume} {357}},\ \bibinfo {pages} {418} (\bibinfo {year}
  {1995})}\BibitemShut {NoStop}%
\bibitem [{\citenamefont {Pouthas}\ \emph {et~al.}(1996)\citenamefont
  {Pouthas}, \citenamefont {Bertaut}, \citenamefont {Borderie}, \citenamefont
  {Bourgault}, \citenamefont {Cahan}, \citenamefont {Carles}, \citenamefont
  {Charlet}, \citenamefont {Cussol}, \citenamefont {Dayras}, \citenamefont
  {Engrand}, \citenamefont {Jouniaux}, \citenamefont {{Le Botlan}},
  \citenamefont {Leconte}, \citenamefont {Lelong}, \citenamefont {Martina},
  \citenamefont {Mosrin}, \citenamefont {Olivier}, \citenamefont {Passerieux},
  \citenamefont {Piquet}, \citenamefont {Plagnol}, \citenamefont {Plaige},
  \citenamefont {Raine}, \citenamefont {Richard}, \citenamefont
  {Saint-Laurent}, \citenamefont {Spitaels}, \citenamefont {Tillier},
  \citenamefont {Tripon}, \citenamefont {Vallerand}, \citenamefont {Volkov},\
  and\ \citenamefont {Wittwer}}]{Pouthas1996}%
  \BibitemOpen
  \bibfield  {author} {\bibinfo {author} {\bibfnamefont {J.}~\bibnamefont
  {Pouthas}}, \bibinfo {author} {\bibfnamefont {A.}~\bibnamefont {Bertaut}},
  \bibinfo {author} {\bibfnamefont {B.}~\bibnamefont {Borderie}}, \bibinfo
  {author} {\bibfnamefont {P.}~\bibnamefont {Bourgault}}, \bibinfo {author}
  {\bibfnamefont {B.}~\bibnamefont {Cahan}}, \bibinfo {author} {\bibfnamefont
  {G.}~\bibnamefont {Carles}}, \bibinfo {author} {\bibfnamefont
  {D.}~\bibnamefont {Charlet}}, \bibinfo {author} {\bibfnamefont
  {D.}~\bibnamefont {Cussol}}, \bibinfo {author} {\bibfnamefont
  {R.}~\bibnamefont {Dayras}}, \bibinfo {author} {\bibfnamefont
  {M.}~\bibnamefont {Engrand}}, \bibinfo {author} {\bibfnamefont
  {O.}~\bibnamefont {Jouniaux}}, \bibinfo {author} {\bibfnamefont
  {P.}~\bibnamefont {{Le Botlan}}}, \bibinfo {author} {\bibfnamefont
  {A.}~\bibnamefont {Leconte}}, \bibinfo {author} {\bibfnamefont
  {P.}~\bibnamefont {Lelong}}, \bibinfo {author} {\bibfnamefont
  {L.}~\bibnamefont {Martina}}, \bibinfo {author} {\bibfnamefont
  {P.}~\bibnamefont {Mosrin}}, \bibinfo {author} {\bibfnamefont
  {L.}~\bibnamefont {Olivier}}, \bibinfo {author} {\bibfnamefont
  {J.}~\bibnamefont {Passerieux}}, \bibinfo {author} {\bibfnamefont
  {B.}~\bibnamefont {Piquet}}, \bibinfo {author} {\bibfnamefont
  {E.}~\bibnamefont {Plagnol}}, \bibinfo {author} {\bibfnamefont
  {E.}~\bibnamefont {Plaige}}, \bibinfo {author} {\bibfnamefont
  {B.}~\bibnamefont {Raine}}, \bibinfo {author} {\bibfnamefont
  {A.}~\bibnamefont {Richard}}, \bibinfo {author} {\bibfnamefont
  {F.}~\bibnamefont {Saint-Laurent}}, \bibinfo {author} {\bibfnamefont
  {C.}~\bibnamefont {Spitaels}}, \bibinfo {author} {\bibfnamefont
  {J.}~\bibnamefont {Tillier}}, \bibinfo {author} {\bibfnamefont
  {M.}~\bibnamefont {Tripon}}, \bibinfo {author} {\bibfnamefont
  {P.}~\bibnamefont {Vallerand}}, \bibinfo {author} {\bibfnamefont
  {P.}~\bibnamefont {Volkov}},\ and\ \bibinfo {author} {\bibfnamefont
  {G.}~\bibnamefont {Wittwer}},\ }\href
  {https://doi.org/https://doi.org/10.1016/0168-9002(95)00770-9} {\bibfield
  {journal} {\bibinfo  {journal} {Nucl. Instr. and Meth. in Phys. Res. A}\
  }\textbf {\bibinfo {volume} {369}},\ \bibinfo {pages} {222} (\bibinfo {year}
  {1996})}\BibitemShut {NoStop}%
\bibitem [{\citenamefont {Cussol}\ \emph {et~al.}(2002)\citenamefont {Cussol},
  \citenamefont {Lefort}, \citenamefont {P\'eter}, \citenamefont {Auger},
  \citenamefont {Bacri}, \citenamefont {Bocage}, \citenamefont {Borderie},
  \citenamefont {Bougault}, \citenamefont {Brou}, \citenamefont {Buchet},
  \citenamefont {Charvet}, \citenamefont {Chbihi}, \citenamefont {Colin},
  \citenamefont {Dayras}, \citenamefont {Demeyer}, \citenamefont {Dor\'e},
  \citenamefont {Durand}, \citenamefont {Eudes}, \citenamefont {de~Filippo},
  \citenamefont {Frankland}, \citenamefont {Galichet}, \citenamefont
  {Genouin-Duhamel}, \citenamefont {Gerlic}, \citenamefont {Germain},
  \citenamefont {Gourio}, \citenamefont {Guinet}, \citenamefont {Lautesse},
  \citenamefont {Laville}, \citenamefont {Lecolley}, \citenamefont
  {Le~F\`evre}, \citenamefont {Legrain}, \citenamefont {Le~Neindre},
  \citenamefont {Lopez}, \citenamefont {Louvel}, \citenamefont {Maskay},
  \citenamefont {Nalpas}, \citenamefont {N'Guyen}, \citenamefont {Parlog},
  \citenamefont {Plagnol}, \citenamefont {Politi}, \citenamefont {Rahmani},
  \citenamefont {Reposeur}, \citenamefont {Rivet}, \citenamefont {Rosato},
  \citenamefont {Saint-Laurent}, \citenamefont {Salou}, \citenamefont
  {Steckmeyer}, \citenamefont {Stern}, \citenamefont {Tabacaru}, \citenamefont
  {Tamain}, \citenamefont {Tassan-Got}, \citenamefont {Tirel}, \citenamefont
  {Vient}, \citenamefont {Volant},\ and\ \citenamefont
  {Wieleczko}}]{Cussol2002}%
  \BibitemOpen
  \bibfield  {author} {\bibinfo {author} {\bibfnamefont {D.}~\bibnamefont
  {Cussol}}, \bibinfo {author} {\bibfnamefont {T.}~\bibnamefont {Lefort}},
  \bibinfo {author} {\bibfnamefont {J.}~\bibnamefont {P\'eter}}, \bibinfo
  {author} {\bibfnamefont {G.}~\bibnamefont {Auger}}, \bibinfo {author}
  {\bibfnamefont {C.~O.}\ \bibnamefont {Bacri}}, \bibinfo {author}
  {\bibfnamefont {F.}~\bibnamefont {Bocage}}, \bibinfo {author} {\bibfnamefont
  {B.}~\bibnamefont {Borderie}}, \bibinfo {author} {\bibfnamefont
  {R.}~\bibnamefont {Bougault}}, \bibinfo {author} {\bibfnamefont
  {R.}~\bibnamefont {Brou}}, \bibinfo {author} {\bibfnamefont {P.}~\bibnamefont
  {Buchet}}, \bibinfo {author} {\bibfnamefont {J.~L.}\ \bibnamefont {Charvet}},
  \bibinfo {author} {\bibfnamefont {A.}~\bibnamefont {Chbihi}}, \bibinfo
  {author} {\bibfnamefont {J.}~\bibnamefont {Colin}}, \bibinfo {author}
  {\bibfnamefont {R.}~\bibnamefont {Dayras}}, \bibinfo {author} {\bibfnamefont
  {A.}~\bibnamefont {Demeyer}}, \bibinfo {author} {\bibfnamefont
  {D.}~\bibnamefont {Dor\'e}}, \bibinfo {author} {\bibfnamefont
  {D.}~\bibnamefont {Durand}}, \bibinfo {author} {\bibfnamefont
  {P.}~\bibnamefont {Eudes}}, \bibinfo {author} {\bibfnamefont
  {E.}~\bibnamefont {de~Filippo}}, \bibinfo {author} {\bibfnamefont {J.~D.}\
  \bibnamefont {Frankland}}, \bibinfo {author} {\bibfnamefont {E.}~\bibnamefont
  {Galichet}}, \bibinfo {author} {\bibfnamefont {E.}~\bibnamefont
  {Genouin-Duhamel}}, \bibinfo {author} {\bibfnamefont {E.}~\bibnamefont
  {Gerlic}}, \bibinfo {author} {\bibfnamefont {M.}~\bibnamefont {Germain}},
  \bibinfo {author} {\bibfnamefont {D.}~\bibnamefont {Gourio}}, \bibinfo
  {author} {\bibfnamefont {D.}~\bibnamefont {Guinet}}, \bibinfo {author}
  {\bibfnamefont {P.}~\bibnamefont {Lautesse}}, \bibinfo {author}
  {\bibfnamefont {J.~L.}\ \bibnamefont {Laville}}, \bibinfo {author}
  {\bibfnamefont {J.~F.}\ \bibnamefont {Lecolley}}, \bibinfo {author}
  {\bibfnamefont {A.}~\bibnamefont {Le~F\`evre}}, \bibinfo {author}
  {\bibfnamefont {R.}~\bibnamefont {Legrain}}, \bibinfo {author} {\bibfnamefont
  {N.}~\bibnamefont {Le~Neindre}}, \bibinfo {author} {\bibfnamefont
  {O.}~\bibnamefont {Lopez}}, \bibinfo {author} {\bibfnamefont
  {M.}~\bibnamefont {Louvel}}, \bibinfo {author} {\bibfnamefont {A.~M.}\
  \bibnamefont {Maskay}}, \bibinfo {author} {\bibfnamefont {L.}~\bibnamefont
  {Nalpas}}, \bibinfo {author} {\bibfnamefont {A.~D.}\ \bibnamefont {N'Guyen}},
  \bibinfo {author} {\bibfnamefont {M.}~\bibnamefont {Parlog}}, \bibinfo
  {author} {\bibfnamefont {E.}~\bibnamefont {Plagnol}}, \bibinfo {author}
  {\bibfnamefont {G.}~\bibnamefont {Politi}}, \bibinfo {author} {\bibfnamefont
  {A.}~\bibnamefont {Rahmani}}, \bibinfo {author} {\bibfnamefont
  {T.}~\bibnamefont {Reposeur}}, \bibinfo {author} {\bibfnamefont {M.~F.}\
  \bibnamefont {Rivet}}, \bibinfo {author} {\bibfnamefont {E.}~\bibnamefont
  {Rosato}}, \bibinfo {author} {\bibfnamefont {F.}~\bibnamefont
  {Saint-Laurent}}, \bibinfo {author} {\bibfnamefont {S.}~\bibnamefont
  {Salou}}, \bibinfo {author} {\bibfnamefont {J.~C.}\ \bibnamefont
  {Steckmeyer}}, \bibinfo {author} {\bibfnamefont {M.}~\bibnamefont {Stern}},
  \bibinfo {author} {\bibfnamefont {G.}~\bibnamefont {Tabacaru}}, \bibinfo
  {author} {\bibfnamefont {B.}~\bibnamefont {Tamain}}, \bibinfo {author}
  {\bibfnamefont {L.}~\bibnamefont {Tassan-Got}}, \bibinfo {author}
  {\bibfnamefont {O.}~\bibnamefont {Tirel}}, \bibinfo {author} {\bibfnamefont
  {E.}~\bibnamefont {Vient}}, \bibinfo {author} {\bibfnamefont
  {C.}~\bibnamefont {Volant}},\ and\ \bibinfo {author} {\bibfnamefont {J.~P.}\
  \bibnamefont {Wieleczko}} (\bibinfo {collaboration} {INDRA Collaboration}),\
  }\href {https://doi.org/10.1103/PhysRevC.65.044604} {\bibfield  {journal}
  {\bibinfo  {journal} {Phys. Rev. C}\ }\textbf {\bibinfo {volume} {65}},\
  \bibinfo {pages} {044604} (\bibinfo {year} {2002})}\BibitemShut {NoStop}%
\bibitem [{\citenamefont {Bougault}\ \emph {et~al.}(2014)\citenamefont
  {Bougault}, \citenamefont {Poggi}, \citenamefont {Barlini}, \citenamefont
  {Borderie}, \citenamefont {Casini}, \citenamefont {Chbihi}, \citenamefont
  {{Le Neindre}}, \citenamefont {P\^{a}rlog}, \citenamefont {Pasquali},
  \citenamefont {Piantelli}, \citenamefont {Sosin}, \citenamefont {Ademard},
  \citenamefont {Alba}, \citenamefont {Anastasio}, \citenamefont {Barbey},
  \citenamefont {Bardelli}, \citenamefont {Bini}, \citenamefont {Boiano},
  \citenamefont {Boisjoli}, \citenamefont {Bonnet}, \citenamefont {Borcea},
  \citenamefont {Bougard}, \citenamefont {Brulin}, \citenamefont {Bruno},
  \citenamefont {Carboni}, \citenamefont {Cassese}, \citenamefont {Cassese},
  \citenamefont {Cinausero}, \citenamefont {Ciolacu}, \citenamefont {Cruceru},
  \citenamefont {Cruceru}, \citenamefont {D'Aquino}, \citenamefont {Fazio},
  \citenamefont {Degerlier}, \citenamefont {Desrues}, \citenamefont {Meo},
  \citenamefont {Dueñas}, \citenamefont {Edelbruck}, \citenamefont {Energico},
  \citenamefont {Falorsi}, \citenamefont {Frankland}, \citenamefont {Galichet},
  \citenamefont {Gasior}, \citenamefont {Gramegna}, \citenamefont {Giordano},
  \citenamefont {Gruyer}, \citenamefont {Grzeszczuk}, \citenamefont {Guerzoni},
  \citenamefont {Hamrita}, \citenamefont {Huss}, \citenamefont {Kajetanowicz},
  \citenamefont {Korcyl}, \citenamefont {Kordyasz}, \citenamefont {Kozik},
  \citenamefont {Kulig}, \citenamefont {Lavergne}, \citenamefont {Legou\'{e}e},
  \citenamefont {Lopez}, \citenamefont {\L{}ukasik}, \citenamefont {Maiolino},
  \citenamefont {Marchi}, \citenamefont {Marini}, \citenamefont {Martel},
  \citenamefont {Masone}, \citenamefont {Meoli}, \citenamefont {Merrer},
  \citenamefont {Morelli}, \citenamefont {Negoita}, \citenamefont {Olmi},
  \citenamefont {Ordine}, \citenamefont {Paduano}, \citenamefont {Pain},
  \citenamefont {Pa\l{}ka}, \citenamefont {Passeggio}, \citenamefont {Pastore},
  \citenamefont {Paw\l{}owski}, \citenamefont {Petcu}, \citenamefont
  {Petrascu}, \citenamefont {Piasecki}, \citenamefont {Pontoriere},
  \citenamefont {Rauly}, \citenamefont {Rivet}, \citenamefont {Rocco},
  \citenamefont {Rosato}, \citenamefont {Roscilli}, \citenamefont {Scarlini},
  \citenamefont {Salomon}, \citenamefont {Santonocito}, \citenamefont
  {Seredov}, \citenamefont {Serra}, \citenamefont {Sierpowski}, \citenamefont
  {Spadaccini}, \citenamefont {Spitaels}, \citenamefont {Stefanini},
  \citenamefont {Tobia}, \citenamefont {Tortone}, \citenamefont {Twar\'{o}g},
  \citenamefont {Valdr\'{e}}, \citenamefont {Vanzanella}, \citenamefont
  {Vanzanella}, \citenamefont {Vient}, \citenamefont {Vigilante}, \citenamefont
  {Vitiello}, \citenamefont {Wanlin}, \citenamefont {Wieloch},\ and\
  \citenamefont {Zipper}}]{Bougault2014}%
  \BibitemOpen
  \bibfield  {author} {\bibinfo {author} {\bibfnamefont {R.}~\bibnamefont
  {Bougault}}, \bibinfo {author} {\bibfnamefont {G.}~\bibnamefont {Poggi}},
  \bibinfo {author} {\bibfnamefont {S.}~\bibnamefont {Barlini}}, \bibinfo
  {author} {\bibfnamefont {B.}~\bibnamefont {Borderie}}, \bibinfo {author}
  {\bibfnamefont {G.}~\bibnamefont {Casini}}, \bibinfo {author} {\bibfnamefont
  {A.}~\bibnamefont {Chbihi}}, \bibinfo {author} {\bibfnamefont
  {N.}~\bibnamefont {{Le Neindre}}}, \bibinfo {author} {\bibfnamefont
  {M.}~\bibnamefont {P\^{a}rlog}}, \bibinfo {author} {\bibfnamefont
  {G.}~\bibnamefont {Pasquali}}, \bibinfo {author} {\bibfnamefont
  {S.}~\bibnamefont {Piantelli}}, \bibinfo {author} {\bibfnamefont
  {Z.}~\bibnamefont {Sosin}}, \bibinfo {author} {\bibfnamefont
  {G.}~\bibnamefont {Ademard}}, \bibinfo {author} {\bibfnamefont
  {R.}~\bibnamefont {Alba}}, \bibinfo {author} {\bibfnamefont {A.}~\bibnamefont
  {Anastasio}}, \bibinfo {author} {\bibfnamefont {S.}~\bibnamefont {Barbey}},
  \bibinfo {author} {\bibfnamefont {L.}~\bibnamefont {Bardelli}}, \bibinfo
  {author} {\bibfnamefont {M.}~\bibnamefont {Bini}}, \bibinfo {author}
  {\bibfnamefont {A.}~\bibnamefont {Boiano}}, \bibinfo {author} {\bibfnamefont
  {M.}~\bibnamefont {Boisjoli}}, \bibinfo {author} {\bibfnamefont
  {E.}~\bibnamefont {Bonnet}}, \bibinfo {author} {\bibfnamefont
  {R.}~\bibnamefont {Borcea}}, \bibinfo {author} {\bibfnamefont
  {B.}~\bibnamefont {Bougard}}, \bibinfo {author} {\bibfnamefont
  {G.}~\bibnamefont {Brulin}}, \bibinfo {author} {\bibfnamefont
  {M.}~\bibnamefont {Bruno}}, \bibinfo {author} {\bibfnamefont
  {S.}~\bibnamefont {Carboni}}, \bibinfo {author} {\bibfnamefont
  {C.}~\bibnamefont {Cassese}}, \bibinfo {author} {\bibfnamefont
  {F.}~\bibnamefont {Cassese}}, \bibinfo {author} {\bibfnamefont
  {M.}~\bibnamefont {Cinausero}}, \bibinfo {author} {\bibfnamefont
  {L.}~\bibnamefont {Ciolacu}}, \bibinfo {author} {\bibfnamefont
  {I.}~\bibnamefont {Cruceru}}, \bibinfo {author} {\bibfnamefont
  {M.}~\bibnamefont {Cruceru}}, \bibinfo {author} {\bibfnamefont
  {B.}~\bibnamefont {D'Aquino}}, \bibinfo {author} {\bibfnamefont {B.~D.}\
  \bibnamefont {Fazio}}, \bibinfo {author} {\bibfnamefont {M.}~\bibnamefont
  {Degerlier}}, \bibinfo {author} {\bibfnamefont {P.}~\bibnamefont {Desrues}},
  \bibinfo {author} {\bibfnamefont {P.~D.}\ \bibnamefont {Meo}}, \bibinfo
  {author} {\bibfnamefont {J.~A.}\ \bibnamefont {Dueñas}}, \bibinfo {author}
  {\bibfnamefont {P.}~\bibnamefont {Edelbruck}}, \bibinfo {author}
  {\bibfnamefont {S.}~\bibnamefont {Energico}}, \bibinfo {author}
  {\bibfnamefont {M.}~\bibnamefont {Falorsi}}, \bibinfo {author} {\bibfnamefont
  {J.~D.}\ \bibnamefont {Frankland}}, \bibinfo {author} {\bibfnamefont
  {E.}~\bibnamefont {Galichet}}, \bibinfo {author} {\bibfnamefont
  {K.}~\bibnamefont {Gasior}}, \bibinfo {author} {\bibfnamefont
  {F.}~\bibnamefont {Gramegna}}, \bibinfo {author} {\bibfnamefont
  {R.}~\bibnamefont {Giordano}}, \bibinfo {author} {\bibfnamefont
  {D.}~\bibnamefont {Gruyer}}, \bibinfo {author} {\bibfnamefont
  {A.}~\bibnamefont {Grzeszczuk}}, \bibinfo {author} {\bibfnamefont
  {M.}~\bibnamefont {Guerzoni}}, \bibinfo {author} {\bibfnamefont
  {H.}~\bibnamefont {Hamrita}}, \bibinfo {author} {\bibfnamefont
  {C.}~\bibnamefont {Huss}}, \bibinfo {author} {\bibfnamefont {M.}~\bibnamefont
  {Kajetanowicz}}, \bibinfo {author} {\bibfnamefont {K.}~\bibnamefont
  {Korcyl}}, \bibinfo {author} {\bibfnamefont {A.}~\bibnamefont {Kordyasz}},
  \bibinfo {author} {\bibfnamefont {T.}~\bibnamefont {Kozik}}, \bibinfo
  {author} {\bibfnamefont {P.}~\bibnamefont {Kulig}}, \bibinfo {author}
  {\bibfnamefont {L.}~\bibnamefont {Lavergne}}, \bibinfo {author}
  {\bibfnamefont {E.}~\bibnamefont {Legou\'{e}e}}, \bibinfo {author}
  {\bibfnamefont {O.}~\bibnamefont {Lopez}}, \bibinfo {author} {\bibfnamefont
  {J.}~\bibnamefont {\L{}ukasik}}, \bibinfo {author} {\bibfnamefont
  {C.}~\bibnamefont {Maiolino}}, \bibinfo {author} {\bibfnamefont
  {T.}~\bibnamefont {Marchi}}, \bibinfo {author} {\bibfnamefont
  {P.}~\bibnamefont {Marini}}, \bibinfo {author} {\bibfnamefont
  {I.}~\bibnamefont {Martel}}, \bibinfo {author} {\bibfnamefont
  {V.}~\bibnamefont {Masone}}, \bibinfo {author} {\bibfnamefont
  {A.}~\bibnamefont {Meoli}}, \bibinfo {author} {\bibfnamefont
  {Y.}~\bibnamefont {Merrer}}, \bibinfo {author} {\bibfnamefont
  {L.}~\bibnamefont {Morelli}}, \bibinfo {author} {\bibfnamefont
  {F.}~\bibnamefont {Negoita}}, \bibinfo {author} {\bibfnamefont
  {A.}~\bibnamefont {Olmi}}, \bibinfo {author} {\bibfnamefont {A.}~\bibnamefont
  {Ordine}}, \bibinfo {author} {\bibfnamefont {G.}~\bibnamefont {Paduano}},
  \bibinfo {author} {\bibfnamefont {C.}~\bibnamefont {Pain}}, \bibinfo {author}
  {\bibfnamefont {M.}~\bibnamefont {Pa\l{}ka}}, \bibinfo {author}
  {\bibfnamefont {G.}~\bibnamefont {Passeggio}}, \bibinfo {author}
  {\bibfnamefont {G.}~\bibnamefont {Pastore}}, \bibinfo {author} {\bibfnamefont
  {P.}~\bibnamefont {Paw\l{}owski}}, \bibinfo {author} {\bibfnamefont
  {M.}~\bibnamefont {Petcu}}, \bibinfo {author} {\bibfnamefont
  {H.}~\bibnamefont {Petrascu}}, \bibinfo {author} {\bibfnamefont
  {E.}~\bibnamefont {Piasecki}}, \bibinfo {author} {\bibfnamefont
  {G.}~\bibnamefont {Pontoriere}}, \bibinfo {author} {\bibfnamefont
  {E.}~\bibnamefont {Rauly}}, \bibinfo {author} {\bibfnamefont {M.~F.}\
  \bibnamefont {Rivet}}, \bibinfo {author} {\bibfnamefont {R.}~\bibnamefont
  {Rocco}}, \bibinfo {author} {\bibfnamefont {E.}~\bibnamefont {Rosato}},
  \bibinfo {author} {\bibfnamefont {L.}~\bibnamefont {Roscilli}}, \bibinfo
  {author} {\bibfnamefont {E.}~\bibnamefont {Scarlini}}, \bibinfo {author}
  {\bibfnamefont {F.}~\bibnamefont {Salomon}}, \bibinfo {author} {\bibfnamefont
  {D.}~\bibnamefont {Santonocito}}, \bibinfo {author} {\bibfnamefont
  {V.}~\bibnamefont {Seredov}}, \bibinfo {author} {\bibfnamefont
  {S.}~\bibnamefont {Serra}}, \bibinfo {author} {\bibfnamefont
  {D.}~\bibnamefont {Sierpowski}}, \bibinfo {author} {\bibfnamefont
  {G.}~\bibnamefont {Spadaccini}}, \bibinfo {author} {\bibfnamefont
  {C.}~\bibnamefont {Spitaels}}, \bibinfo {author} {\bibfnamefont {A.~A.}\
  \bibnamefont {Stefanini}}, \bibinfo {author} {\bibfnamefont {G.}~\bibnamefont
  {Tobia}}, \bibinfo {author} {\bibfnamefont {G.}~\bibnamefont {Tortone}},
  \bibinfo {author} {\bibfnamefont {T.}~\bibnamefont {Twar\'{o}g}}, \bibinfo
  {author} {\bibfnamefont {S.}~\bibnamefont {Valdr\'{e}}}, \bibinfo {author}
  {\bibfnamefont {A.}~\bibnamefont {Vanzanella}}, \bibinfo {author}
  {\bibfnamefont {E.}~\bibnamefont {Vanzanella}}, \bibinfo {author}
  {\bibfnamefont {E.}~\bibnamefont {Vient}}, \bibinfo {author} {\bibfnamefont
  {M.}~\bibnamefont {Vigilante}}, \bibinfo {author} {\bibfnamefont
  {G.}~\bibnamefont {Vitiello}}, \bibinfo {author} {\bibfnamefont
  {E.}~\bibnamefont {Wanlin}}, \bibinfo {author} {\bibfnamefont
  {A.}~\bibnamefont {Wieloch}},\ and\ \bibinfo {author} {\bibfnamefont
  {W.}~\bibnamefont {Zipper}},\ }\href
  {https://doi.org/https://doi.org/10.1140/epja/i2014-14047-4} {\bibfield
  {journal} {\bibinfo  {journal} {Eur. Phys. J. A}\ }\textbf {\bibinfo {volume}
  {50}},\ \bibinfo {pages} {47} (\bibinfo {year} {2014})}\BibitemShut {NoStop}%
\bibitem [{\citenamefont {Valdr\'{e}}\ \emph {et~al.}(2019)\citenamefont
  {Valdr\'{e}}, \citenamefont {Casini}, \citenamefont {{Le Neindre}},
  \citenamefont {Bini}, \citenamefont {Boiano}, \citenamefont {Borderie},
  \citenamefont {Edelbruck}, \citenamefont {Poggi}, \citenamefont {Salomon},
  \citenamefont {Tortone}, \citenamefont {Alba}, \citenamefont {Barlini},
  \citenamefont {Bonnet}, \citenamefont {Bougard}, \citenamefont {Bougault},
  \citenamefont {Brulin}, \citenamefont {Bruno}, \citenamefont {Buccola},
  \citenamefont {Camaiani}, \citenamefont {Chbihi}, \citenamefont {Ciampi},
  \citenamefont {Cicerchia}, \citenamefont {Cinausero}, \citenamefont
  {Dell’Aquila}, \citenamefont {Desrues}, \citenamefont {Dueñas},
  \citenamefont {Fabris}, \citenamefont {Falorsi}, \citenamefont {Frankland},
  \citenamefont {Frosin}, \citenamefont {Galichet}, \citenamefont {Giordano},
  \citenamefont {Gramegna}, \citenamefont {Grassi}, \citenamefont {Gruyer},
  \citenamefont {Guerzoni}, \citenamefont {Henri}, \citenamefont
  {Kajetanowicz}, \citenamefont {Korcyl}, \citenamefont {Kordyasz},
  \citenamefont {Kozik}, \citenamefont {Lecomte}, \citenamefont {Lombardo},
  \citenamefont {Lopez}, \citenamefont {Maiolino}, \citenamefont {Mantovani},
  \citenamefont {Marchi}, \citenamefont {Margotti}, \citenamefont {Merrer},
  \citenamefont {Morelli}, \citenamefont {Olmi}, \citenamefont {Ordine},
  \citenamefont {Ottanelli}, \citenamefont {Pain}, \citenamefont {Pałka},
  \citenamefont {Pârlog}, \citenamefont {Pasquali}, \citenamefont {Pastore},
  \citenamefont {Piantelli}, \citenamefont {{de Préaumont}}, \citenamefont
  {Revenko}, \citenamefont {Richard}, \citenamefont {Rivet}, \citenamefont
  {Ropert}, \citenamefont {Rosato}, \citenamefont {Saillant}, \citenamefont
  {Santonocito}, \citenamefont {Scarlini}, \citenamefont {Serra}, \citenamefont
  {Soulet}, \citenamefont {Spadaccini}, \citenamefont {Stefanini},
  \citenamefont {Tobia}, \citenamefont {Upadhyaya}, \citenamefont {Vanzanella},
  \citenamefont {Verde}, \citenamefont {Vient}, \citenamefont {Vigilante},
  \citenamefont {Wanlin}, \citenamefont {Wittwer},\ and\ \citenamefont
  {Zucchini}}]{Valdre2019}%
  \BibitemOpen
  \bibfield  {author} {\bibinfo {author} {\bibfnamefont {S.}~\bibnamefont
  {Valdr\'{e}}}, \bibinfo {author} {\bibfnamefont {G.}~\bibnamefont {Casini}},
  \bibinfo {author} {\bibfnamefont {N.}~\bibnamefont {{Le Neindre}}}, \bibinfo
  {author} {\bibfnamefont {M.}~\bibnamefont {Bini}}, \bibinfo {author}
  {\bibfnamefont {A.}~\bibnamefont {Boiano}}, \bibinfo {author} {\bibfnamefont
  {B.}~\bibnamefont {Borderie}}, \bibinfo {author} {\bibfnamefont
  {P.}~\bibnamefont {Edelbruck}}, \bibinfo {author} {\bibfnamefont
  {G.}~\bibnamefont {Poggi}}, \bibinfo {author} {\bibfnamefont
  {F.}~\bibnamefont {Salomon}}, \bibinfo {author} {\bibfnamefont
  {G.}~\bibnamefont {Tortone}}, \bibinfo {author} {\bibfnamefont
  {R.}~\bibnamefont {Alba}}, \bibinfo {author} {\bibfnamefont {S.}~\bibnamefont
  {Barlini}}, \bibinfo {author} {\bibfnamefont {E.}~\bibnamefont {Bonnet}},
  \bibinfo {author} {\bibfnamefont {B.}~\bibnamefont {Bougard}}, \bibinfo
  {author} {\bibfnamefont {R.}~\bibnamefont {Bougault}}, \bibinfo {author}
  {\bibfnamefont {G.}~\bibnamefont {Brulin}}, \bibinfo {author} {\bibfnamefont
  {M.}~\bibnamefont {Bruno}}, \bibinfo {author} {\bibfnamefont
  {A.}~\bibnamefont {Buccola}}, \bibinfo {author} {\bibfnamefont
  {A.}~\bibnamefont {Camaiani}}, \bibinfo {author} {\bibfnamefont
  {A.}~\bibnamefont {Chbihi}}, \bibinfo {author} {\bibfnamefont
  {C.}~\bibnamefont {Ciampi}}, \bibinfo {author} {\bibfnamefont
  {M.}~\bibnamefont {Cicerchia}}, \bibinfo {author} {\bibfnamefont
  {M.}~\bibnamefont {Cinausero}}, \bibinfo {author} {\bibfnamefont
  {D.}~\bibnamefont {Dell’Aquila}}, \bibinfo {author} {\bibfnamefont
  {P.}~\bibnamefont {Desrues}}, \bibinfo {author} {\bibfnamefont
  {J.}~\bibnamefont {Dueñas}}, \bibinfo {author} {\bibfnamefont
  {D.}~\bibnamefont {Fabris}}, \bibinfo {author} {\bibfnamefont
  {M.}~\bibnamefont {Falorsi}}, \bibinfo {author} {\bibfnamefont
  {J.}~\bibnamefont {Frankland}}, \bibinfo {author} {\bibfnamefont
  {C.}~\bibnamefont {Frosin}}, \bibinfo {author} {\bibfnamefont
  {E.}~\bibnamefont {Galichet}}, \bibinfo {author} {\bibfnamefont
  {R.}~\bibnamefont {Giordano}}, \bibinfo {author} {\bibfnamefont
  {F.}~\bibnamefont {Gramegna}}, \bibinfo {author} {\bibfnamefont
  {L.}~\bibnamefont {Grassi}}, \bibinfo {author} {\bibfnamefont
  {D.}~\bibnamefont {Gruyer}}, \bibinfo {author} {\bibfnamefont
  {M.}~\bibnamefont {Guerzoni}}, \bibinfo {author} {\bibfnamefont
  {M.}~\bibnamefont {Henri}}, \bibinfo {author} {\bibfnamefont
  {M.}~\bibnamefont {Kajetanowicz}}, \bibinfo {author} {\bibfnamefont
  {K.}~\bibnamefont {Korcyl}}, \bibinfo {author} {\bibfnamefont
  {A.}~\bibnamefont {Kordyasz}}, \bibinfo {author} {\bibfnamefont
  {T.}~\bibnamefont {Kozik}}, \bibinfo {author} {\bibfnamefont
  {P.}~\bibnamefont {Lecomte}}, \bibinfo {author} {\bibfnamefont
  {I.}~\bibnamefont {Lombardo}}, \bibinfo {author} {\bibfnamefont
  {O.}~\bibnamefont {Lopez}}, \bibinfo {author} {\bibfnamefont
  {C.}~\bibnamefont {Maiolino}}, \bibinfo {author} {\bibfnamefont
  {G.}~\bibnamefont {Mantovani}}, \bibinfo {author} {\bibfnamefont
  {T.}~\bibnamefont {Marchi}}, \bibinfo {author} {\bibfnamefont
  {A.}~\bibnamefont {Margotti}}, \bibinfo {author} {\bibfnamefont
  {Y.}~\bibnamefont {Merrer}}, \bibinfo {author} {\bibfnamefont
  {L.}~\bibnamefont {Morelli}}, \bibinfo {author} {\bibfnamefont
  {A.}~\bibnamefont {Olmi}}, \bibinfo {author} {\bibfnamefont {A.}~\bibnamefont
  {Ordine}}, \bibinfo {author} {\bibfnamefont {P.}~\bibnamefont {Ottanelli}},
  \bibinfo {author} {\bibfnamefont {C.}~\bibnamefont {Pain}}, \bibinfo {author}
  {\bibfnamefont {M.}~\bibnamefont {Pałka}}, \bibinfo {author} {\bibfnamefont
  {M.}~\bibnamefont {Pârlog}}, \bibinfo {author} {\bibfnamefont
  {G.}~\bibnamefont {Pasquali}}, \bibinfo {author} {\bibfnamefont
  {G.}~\bibnamefont {Pastore}}, \bibinfo {author} {\bibfnamefont
  {S.}~\bibnamefont {Piantelli}}, \bibinfo {author} {\bibfnamefont
  {H.}~\bibnamefont {{de Préaumont}}}, \bibinfo {author} {\bibfnamefont
  {R.}~\bibnamefont {Revenko}}, \bibinfo {author} {\bibfnamefont
  {A.}~\bibnamefont {Richard}}, \bibinfo {author} {\bibfnamefont
  {M.}~\bibnamefont {Rivet}}, \bibinfo {author} {\bibfnamefont
  {J.}~\bibnamefont {Ropert}}, \bibinfo {author} {\bibfnamefont
  {E.}~\bibnamefont {Rosato}}, \bibinfo {author} {\bibfnamefont
  {F.}~\bibnamefont {Saillant}}, \bibinfo {author} {\bibfnamefont
  {D.}~\bibnamefont {Santonocito}}, \bibinfo {author} {\bibfnamefont
  {E.}~\bibnamefont {Scarlini}}, \bibinfo {author} {\bibfnamefont
  {S.}~\bibnamefont {Serra}}, \bibinfo {author} {\bibfnamefont
  {C.}~\bibnamefont {Soulet}}, \bibinfo {author} {\bibfnamefont
  {G.}~\bibnamefont {Spadaccini}}, \bibinfo {author} {\bibfnamefont
  {A.}~\bibnamefont {Stefanini}}, \bibinfo {author} {\bibfnamefont
  {G.}~\bibnamefont {Tobia}}, \bibinfo {author} {\bibfnamefont
  {S.}~\bibnamefont {Upadhyaya}}, \bibinfo {author} {\bibfnamefont
  {A.}~\bibnamefont {Vanzanella}}, \bibinfo {author} {\bibfnamefont
  {G.}~\bibnamefont {Verde}}, \bibinfo {author} {\bibfnamefont
  {E.}~\bibnamefont {Vient}}, \bibinfo {author} {\bibfnamefont
  {M.}~\bibnamefont {Vigilante}}, \bibinfo {author} {\bibfnamefont
  {E.}~\bibnamefont {Wanlin}}, \bibinfo {author} {\bibfnamefont
  {G.}~\bibnamefont {Wittwer}},\ and\ \bibinfo {author} {\bibfnamefont
  {A.}~\bibnamefont {Zucchini}},\ }\href
  {https://doi.org/https://doi.org/10.1016/j.nima.2019.03.082} {\bibfield
  {journal} {\bibinfo  {journal} {Nucl. Instr. and Meth. in Phys. Res. A}\
  }\textbf {\bibinfo {volume} {930}},\ \bibinfo {pages} {27} (\bibinfo {year}
  {2019})}\BibitemShut {NoStop}%
\bibitem [{\citenamefont {Casini}\ and\ \citenamefont {{Le
  Neindre}}(2022)}]{Casini2022}%
  \BibitemOpen
  \bibfield  {author} {\bibinfo {author} {\bibfnamefont {G.}~\bibnamefont
  {Casini}}\ and\ \bibinfo {author} {\bibfnamefont {N.}~\bibnamefont {{Le
  Neindre}}},\ }\href {https://doi.org/10.1080/10619127.2022.2133502}
  {\bibfield  {journal} {\bibinfo  {journal} {Nuclear Physics News}\ }\textbf
  {\bibinfo {volume} {32}},\ \bibinfo {pages} {24} (\bibinfo {year}
  {2022})}\BibitemShut {NoStop}%
\bibitem [{\citenamefont {Wittwer}(2004)}]{CENTRUM_ref}%
  \BibitemOpen
  \bibfield  {author} {\bibinfo {author} {\bibfnamefont {G.}~\bibnamefont
  {Wittwer}},\ }\href@noop {} {\bibinfo {title} {Clock event number transmitter
  receiver universal module, user’s manual}},\ \bibinfo {howpublished}
  {GANIL} (\bibinfo {year} {2004})\BibitemShut {NoStop}%
\bibitem [{\citenamefont {Frankland}\ \emph {et~al.}(2022)\citenamefont
  {Frankland}, \citenamefont {{Le Neindre}}, \citenamefont {Henri},
  \citenamefont {Chbihi}, \citenamefont {Fable}, \citenamefont {Gouyet},
  \citenamefont {Revenko}, \citenamefont {Beaudouin}, \citenamefont {Bougault},
  \citenamefont {Bourgault}, \citenamefont {Genard}, \citenamefont {Godefroid},
  \citenamefont {Gruyer}, \citenamefont {Lemarié}, \citenamefont {Leterrier},
  \citenamefont {Lopez}, \citenamefont {Loubeau}, \citenamefont
  {Marie-Saillenfest}, \citenamefont {Nicolle}, \citenamefont {Noury},
  \citenamefont {Perronnel}, \citenamefont {Prieur}, \citenamefont
  {Rebillard-Soulié}, \citenamefont {Rousseau}, \citenamefont {Saillant},\
  and\ \citenamefont {Vient}}]{Frankland2022}%
  \BibitemOpen
  \bibfield  {author} {\bibinfo {author} {\bibfnamefont {J.~D.}\ \bibnamefont
  {Frankland}}, \bibinfo {author} {\bibfnamefont {N.}~\bibnamefont {{Le
  Neindre}}}, \bibinfo {author} {\bibfnamefont {M.}~\bibnamefont {Henri}},
  \bibinfo {author} {\bibfnamefont {A.}~\bibnamefont {Chbihi}}, \bibinfo
  {author} {\bibfnamefont {Q.}~\bibnamefont {Fable}}, \bibinfo {author}
  {\bibfnamefont {C.}~\bibnamefont {Gouyet}}, \bibinfo {author} {\bibfnamefont
  {R.}~\bibnamefont {Revenko}}, \bibinfo {author} {\bibfnamefont
  {Q.}~\bibnamefont {Beaudouin}}, \bibinfo {author} {\bibfnamefont
  {R.}~\bibnamefont {Bougault}}, \bibinfo {author} {\bibfnamefont
  {P.}~\bibnamefont {Bourgault}}, \bibinfo {author} {\bibfnamefont
  {T.}~\bibnamefont {Genard}}, \bibinfo {author} {\bibfnamefont
  {V.}~\bibnamefont {Godefroid}}, \bibinfo {author} {\bibfnamefont
  {D.}~\bibnamefont {Gruyer}}, \bibinfo {author} {\bibfnamefont
  {J.}~\bibnamefont {Lemarié}}, \bibinfo {author} {\bibfnamefont
  {L.}~\bibnamefont {Leterrier}}, \bibinfo {author} {\bibfnamefont
  {O.}~\bibnamefont {Lopez}}, \bibinfo {author} {\bibfnamefont
  {E.}~\bibnamefont {Loubeau}}, \bibinfo {author} {\bibfnamefont
  {F.}~\bibnamefont {Marie-Saillenfest}}, \bibinfo {author} {\bibfnamefont
  {C.}~\bibnamefont {Nicolle}}, \bibinfo {author} {\bibfnamefont
  {F.}~\bibnamefont {Noury}}, \bibinfo {author} {\bibfnamefont
  {J.}~\bibnamefont {Perronnel}}, \bibinfo {author} {\bibfnamefont
  {M.}~\bibnamefont {Prieur}}, \bibinfo {author} {\bibfnamefont
  {A.}~\bibnamefont {Rebillard-Soulié}}, \bibinfo {author} {\bibfnamefont
  {L.}~\bibnamefont {Rousseau}}, \bibinfo {author} {\bibfnamefont
  {F.}~\bibnamefont {Saillant}},\ and\ \bibinfo {author} {\bibfnamefont
  {E.}~\bibnamefont {Vient}},\ }\href
  {https://doi.org/10.1393/ncc/i2022-22043-6} {\bibfield  {journal} {\bibinfo
  {journal} {Nuovo Cim. C}\ }\textbf {\bibinfo {volume} {45}},\ \bibinfo
  {pages} {43} (\bibinfo {year} {2022})}\BibitemShut {NoStop}%
\bibitem [{\citenamefont {Vient}\ \emph {et~al.}(2018)\citenamefont {Vient},
  \citenamefont {Manduci}, \citenamefont {Legou\'ee}, \citenamefont {Augey},
  \citenamefont {Bonnet}, \citenamefont {Borderie}, \citenamefont {Bougault},
  \citenamefont {Chbihi}, \citenamefont {Dell'Aquila}, \citenamefont {Fable},
  \citenamefont {Francalanza}, \citenamefont {Frankland}, \citenamefont
  {Galichet}, \citenamefont {Gruyer}, \citenamefont {Guinet}, \citenamefont
  {Henri}, \citenamefont {La~Commara}, \citenamefont {Lehaut}, \citenamefont
  {Le~Neindre}, \citenamefont {Lombardo}, \citenamefont {Lopez}, \citenamefont
  {Marini}, \citenamefont {P\^arlog}, \citenamefont {Rivet}, \citenamefont
  {Rosato}, \citenamefont {Roy}, \citenamefont {St-Onge}, \citenamefont
  {Spadaccini}, \citenamefont {Verde},\ and\ \citenamefont
  {Vigilante}}]{Vient2018}%
  \BibitemOpen
  \bibfield  {author} {\bibinfo {author} {\bibfnamefont {E.}~\bibnamefont
  {Vient}}, \bibinfo {author} {\bibfnamefont {L.}~\bibnamefont {Manduci}},
  \bibinfo {author} {\bibfnamefont {E.}~\bibnamefont {Legou\'ee}}, \bibinfo
  {author} {\bibfnamefont {L.}~\bibnamefont {Augey}}, \bibinfo {author}
  {\bibfnamefont {E.}~\bibnamefont {Bonnet}}, \bibinfo {author} {\bibfnamefont
  {B.}~\bibnamefont {Borderie}}, \bibinfo {author} {\bibfnamefont
  {R.}~\bibnamefont {Bougault}}, \bibinfo {author} {\bibfnamefont
  {A.}~\bibnamefont {Chbihi}}, \bibinfo {author} {\bibfnamefont
  {D.}~\bibnamefont {Dell'Aquila}}, \bibinfo {author} {\bibfnamefont
  {Q.}~\bibnamefont {Fable}}, \bibinfo {author} {\bibfnamefont
  {L.}~\bibnamefont {Francalanza}}, \bibinfo {author} {\bibfnamefont {J.~D.}\
  \bibnamefont {Frankland}}, \bibinfo {author} {\bibfnamefont {E.}~\bibnamefont
  {Galichet}}, \bibinfo {author} {\bibfnamefont {D.}~\bibnamefont {Gruyer}},
  \bibinfo {author} {\bibfnamefont {D.}~\bibnamefont {Guinet}}, \bibinfo
  {author} {\bibfnamefont {M.}~\bibnamefont {Henri}}, \bibinfo {author}
  {\bibfnamefont {M.}~\bibnamefont {La~Commara}}, \bibinfo {author}
  {\bibfnamefont {G.}~\bibnamefont {Lehaut}}, \bibinfo {author} {\bibfnamefont
  {N.}~\bibnamefont {Le~Neindre}}, \bibinfo {author} {\bibfnamefont
  {I.}~\bibnamefont {Lombardo}}, \bibinfo {author} {\bibfnamefont
  {O.}~\bibnamefont {Lopez}}, \bibinfo {author} {\bibfnamefont
  {P.}~\bibnamefont {Marini}}, \bibinfo {author} {\bibfnamefont
  {M.}~\bibnamefont {P\^arlog}}, \bibinfo {author} {\bibfnamefont {M.~F.}\
  \bibnamefont {Rivet}}, \bibinfo {author} {\bibfnamefont {E.}~\bibnamefont
  {Rosato}}, \bibinfo {author} {\bibfnamefont {R.}~\bibnamefont {Roy}},
  \bibinfo {author} {\bibfnamefont {P.}~\bibnamefont {St-Onge}}, \bibinfo
  {author} {\bibfnamefont {G.}~\bibnamefont {Spadaccini}}, \bibinfo {author}
  {\bibfnamefont {G.}~\bibnamefont {Verde}},\ and\ \bibinfo {author}
  {\bibfnamefont {M.}~\bibnamefont {Vigilante}},\ }\href
  {https://doi.org/10.1103/PhysRevC.98.044612} {\bibfield  {journal} {\bibinfo
  {journal} {Phys. Rev. C}\ }\textbf {\bibinfo {volume} {98}},\ \bibinfo
  {pages} {044612} (\bibinfo {year} {2018})}\BibitemShut {NoStop}%
\bibitem [{\citenamefont {Sihver}\ \emph {et~al.}(2014)\citenamefont {Sihver},
  \citenamefont {Lantz},\ and\ \citenamefont {Kohama}}]{KoxShenSihver2014}%
  \BibitemOpen
  \bibfield  {author} {\bibinfo {author} {\bibfnamefont {L.}~\bibnamefont
  {Sihver}}, \bibinfo {author} {\bibfnamefont {M.}~\bibnamefont {Lantz}},\ and\
  \bibinfo {author} {\bibfnamefont {A.}~\bibnamefont {Kohama}},\ }\href
  {https://doi.org/10.1103/PhysRevC.89.067602} {\bibfield  {journal} {\bibinfo
  {journal} {Phys. Rev. C}\ }\textbf {\bibinfo {volume} {89}},\ \bibinfo
  {pages} {067602} (\bibinfo {year} {2014})}\BibitemShut {NoStop}%
\bibitem [{\citenamefont {Lombardo}\ \emph {et~al.}(2011)\citenamefont
  {Lombardo}, \citenamefont {Agodi}, \citenamefont {Amorini}, \citenamefont
  {Anzalone}, \citenamefont {Auditore}, \citenamefont {Berceanu}, \citenamefont
  {Cardella}, \citenamefont {Cavallaro}, \citenamefont {Chatterjee},
  \citenamefont {De~Filippo}, \citenamefont {Geraci}, \citenamefont {Giuliani},
  \citenamefont {Grassi}, \citenamefont {Han}, \citenamefont {La~Guidara},
  \citenamefont {Loria}, \citenamefont {Lanzalone}, \citenamefont {Maiolino},
  \citenamefont {Pagano}, \citenamefont {Papa}, \citenamefont {Pirrone},
  \citenamefont {Politi}, \citenamefont {Porto}, \citenamefont {Rizzo},
  \citenamefont {Russotto}, \citenamefont {Trifir\`o}, \citenamefont
  {Trimarchi}, \citenamefont {Verde},\ and\ \citenamefont
  {Vigilante}}]{Lombardo2011}%
  \BibitemOpen
  \bibfield  {author} {\bibinfo {author} {\bibfnamefont {I.}~\bibnamefont
  {Lombardo}}, \bibinfo {author} {\bibfnamefont {C.}~\bibnamefont {Agodi}},
  \bibinfo {author} {\bibfnamefont {F.}~\bibnamefont {Amorini}}, \bibinfo
  {author} {\bibfnamefont {A.}~\bibnamefont {Anzalone}}, \bibinfo {author}
  {\bibfnamefont {L.}~\bibnamefont {Auditore}}, \bibinfo {author}
  {\bibfnamefont {I.}~\bibnamefont {Berceanu}}, \bibinfo {author}
  {\bibfnamefont {G.}~\bibnamefont {Cardella}}, \bibinfo {author}
  {\bibfnamefont {S.}~\bibnamefont {Cavallaro}}, \bibinfo {author}
  {\bibfnamefont {M.~B.}\ \bibnamefont {Chatterjee}}, \bibinfo {author}
  {\bibfnamefont {E.}~\bibnamefont {De~Filippo}}, \bibinfo {author}
  {\bibfnamefont {E.}~\bibnamefont {Geraci}}, \bibinfo {author} {\bibfnamefont
  {G.}~\bibnamefont {Giuliani}}, \bibinfo {author} {\bibfnamefont
  {L.}~\bibnamefont {Grassi}}, \bibinfo {author} {\bibfnamefont
  {J.}~\bibnamefont {Han}}, \bibinfo {author} {\bibfnamefont {E.}~\bibnamefont
  {La~Guidara}}, \bibinfo {author} {\bibfnamefont {D.}~\bibnamefont {Loria}},
  \bibinfo {author} {\bibfnamefont {G.}~\bibnamefont {Lanzalone}}, \bibinfo
  {author} {\bibfnamefont {C.}~\bibnamefont {Maiolino}}, \bibinfo {author}
  {\bibfnamefont {A.}~\bibnamefont {Pagano}}, \bibinfo {author} {\bibfnamefont
  {M.}~\bibnamefont {Papa}}, \bibinfo {author} {\bibfnamefont {S.}~\bibnamefont
  {Pirrone}}, \bibinfo {author} {\bibfnamefont {G.}~\bibnamefont {Politi}},
  \bibinfo {author} {\bibfnamefont {F.}~\bibnamefont {Porto}}, \bibinfo
  {author} {\bibfnamefont {F.}~\bibnamefont {Rizzo}}, \bibinfo {author}
  {\bibfnamefont {P.}~\bibnamefont {Russotto}}, \bibinfo {author}
  {\bibfnamefont {A.}~\bibnamefont {Trifir\`o}}, \bibinfo {author}
  {\bibfnamefont {M.}~\bibnamefont {Trimarchi}}, \bibinfo {author}
  {\bibfnamefont {G.}~\bibnamefont {Verde}},\ and\ \bibinfo {author}
  {\bibfnamefont {M.}~\bibnamefont {Vigilante}},\ }\href
  {https://doi.org/10.1103/PhysRevC.84.024613} {\bibfield  {journal} {\bibinfo
  {journal} {Phys. Rev. C}\ }\textbf {\bibinfo {volume} {84}},\ \bibinfo
  {pages} {024613} (\bibinfo {year} {2011})}\BibitemShut {NoStop}%
\bibitem [{\citenamefont {Sun}\ \emph {et~al.}(2010)\citenamefont {Sun},
  \citenamefont {Tsang}, \citenamefont {Lynch}, \citenamefont {Verde},
  \citenamefont {Amorini}, \citenamefont {Andronenko}, \citenamefont
  {Andronenko}, \citenamefont {Cardella}, \citenamefont {Chatterje},
  \citenamefont {Danielewicz}, \citenamefont {De~Filippo}, \citenamefont
  {Dinh}, \citenamefont {Galichet}, \citenamefont {Geraci}, \citenamefont
  {Hua}, \citenamefont {La~Guidara}, \citenamefont {Lanzalone}, \citenamefont
  {Liu}, \citenamefont {Lu}, \citenamefont {Lukyanov}, \citenamefont
  {Maiolino}, \citenamefont {Pagano}, \citenamefont {Piantelli}, \citenamefont
  {Papa}, \citenamefont {Pirrone}, \citenamefont {Politi}, \citenamefont
  {Porto}, \citenamefont {Rizzo}, \citenamefont {Russotto}, \citenamefont
  {Santonocito},\ and\ \citenamefont {Zhang}}]{Sun2010}%
  \BibitemOpen
  \bibfield  {author} {\bibinfo {author} {\bibfnamefont {Z.~Y.}\ \bibnamefont
  {Sun}}, \bibinfo {author} {\bibfnamefont {M.~B.}\ \bibnamefont {Tsang}},
  \bibinfo {author} {\bibfnamefont {W.~G.}\ \bibnamefont {Lynch}}, \bibinfo
  {author} {\bibfnamefont {G.}~\bibnamefont {Verde}}, \bibinfo {author}
  {\bibfnamefont {F.}~\bibnamefont {Amorini}}, \bibinfo {author} {\bibfnamefont
  {L.}~\bibnamefont {Andronenko}}, \bibinfo {author} {\bibfnamefont
  {M.}~\bibnamefont {Andronenko}}, \bibinfo {author} {\bibfnamefont
  {G.}~\bibnamefont {Cardella}}, \bibinfo {author} {\bibfnamefont
  {M.}~\bibnamefont {Chatterje}}, \bibinfo {author} {\bibfnamefont
  {P.}~\bibnamefont {Danielewicz}}, \bibinfo {author} {\bibfnamefont
  {E.}~\bibnamefont {De~Filippo}}, \bibinfo {author} {\bibfnamefont
  {P.}~\bibnamefont {Dinh}}, \bibinfo {author} {\bibfnamefont {E.}~\bibnamefont
  {Galichet}}, \bibinfo {author} {\bibfnamefont {E.}~\bibnamefont {Geraci}},
  \bibinfo {author} {\bibfnamefont {H.}~\bibnamefont {Hua}}, \bibinfo {author}
  {\bibfnamefont {E.}~\bibnamefont {La~Guidara}}, \bibinfo {author}
  {\bibfnamefont {G.}~\bibnamefont {Lanzalone}}, \bibinfo {author}
  {\bibfnamefont {H.}~\bibnamefont {Liu}}, \bibinfo {author} {\bibfnamefont
  {F.}~\bibnamefont {Lu}}, \bibinfo {author} {\bibfnamefont {S.}~\bibnamefont
  {Lukyanov}}, \bibinfo {author} {\bibfnamefont {C.}~\bibnamefont {Maiolino}},
  \bibinfo {author} {\bibfnamefont {A.}~\bibnamefont {Pagano}}, \bibinfo
  {author} {\bibfnamefont {S.}~\bibnamefont {Piantelli}}, \bibinfo {author}
  {\bibfnamefont {M.}~\bibnamefont {Papa}}, \bibinfo {author} {\bibfnamefont
  {S.}~\bibnamefont {Pirrone}}, \bibinfo {author} {\bibfnamefont
  {G.}~\bibnamefont {Politi}}, \bibinfo {author} {\bibfnamefont
  {F.}~\bibnamefont {Porto}}, \bibinfo {author} {\bibfnamefont
  {F.}~\bibnamefont {Rizzo}}, \bibinfo {author} {\bibfnamefont
  {P.}~\bibnamefont {Russotto}}, \bibinfo {author} {\bibfnamefont
  {D.}~\bibnamefont {Santonocito}},\ and\ \bibinfo {author} {\bibfnamefont
  {Y.~X.}\ \bibnamefont {Zhang}},\ }\href
  {https://doi.org/10.1103/PhysRevC.82.051603} {\bibfield  {journal} {\bibinfo
  {journal} {Phys. Rev. C}\ }\textbf {\bibinfo {volume} {82}},\ \bibinfo
  {pages} {051603(R)} (\bibinfo {year} {2010})}\BibitemShut {NoStop}%
\bibitem [{\citenamefont {Margueron}\ \emph
  {et~al.}(2018{\natexlab{b}})\citenamefont {Margueron}, \citenamefont
  {Hoffmann~Casali},\ and\ \citenamefont {Gulminelli}}]{Margueron2018I}%
  \BibitemOpen
  \bibfield  {author} {\bibinfo {author} {\bibfnamefont {J.}~\bibnamefont
  {Margueron}}, \bibinfo {author} {\bibfnamefont {R.}~\bibnamefont
  {Hoffmann~Casali}},\ and\ \bibinfo {author} {\bibfnamefont {F.}~\bibnamefont
  {Gulminelli}},\ }\href {https://doi.org/10.1103/PhysRevC.97.025805}
  {\bibfield  {journal} {\bibinfo  {journal} {Phys. Rev. C}\ }\textbf {\bibinfo
  {volume} {97}},\ \bibinfo {pages} {025805} (\bibinfo {year}
  {2018}{\natexlab{b}})}\BibitemShut {NoStop}%
\bibitem [{\citenamefont {Ono}\ \emph {et~al.}(1992)\citenamefont {Ono},
  \citenamefont {Horiuchi}, \citenamefont {Maruyama},\ and\ \citenamefont
  {Ohnishi}}]{Ono1992}%
  \BibitemOpen
  \bibfield  {author} {\bibinfo {author} {\bibfnamefont {A.}~\bibnamefont
  {Ono}}, \bibinfo {author} {\bibfnamefont {H.}~\bibnamefont {Horiuchi}},
  \bibinfo {author} {\bibfnamefont {T.}~\bibnamefont {Maruyama}},\ and\
  \bibinfo {author} {\bibfnamefont {A.}~\bibnamefont {Ohnishi}},\ }\href
  {https://doi.org/10.1143/ptp/87.5.1185} {\bibfield  {journal} {\bibinfo
  {journal} {Prog. Theor. Phys.}\ }\textbf {\bibinfo {volume} {87}},\ \bibinfo
  {pages} {1185} (\bibinfo {year} {1992})}\BibitemShut {NoStop}%
\bibitem [{\citenamefont {Charity}(2010)}]{Charity2010}%
  \BibitemOpen
  \bibfield  {author} {\bibinfo {author} {\bibfnamefont {R.~J.}\ \bibnamefont
  {Charity}},\ }\href {https://doi.org/10.1103/PhysRevC.82.014610} {\bibfield
  {journal} {\bibinfo  {journal} {Phys. Rev. C}\ }\textbf {\bibinfo {volume}
  {82}},\ \bibinfo {pages} {014610} (\bibinfo {year} {2010})}\BibitemShut
  {NoStop}%
\end{thebibliography}
 \bibliographystyle{apsrev4-2.bst}

\providecommand{\noopsort}[1]{}

\end{document}